\documentclass[12pt,a4paper]{article} 
\usepackage{etex}
\pdfoutput=1
\usepackage{jheppub}
\usepackage{epstopdf}
\usepackage{graphicx}
\usepackage{epsfig}
\usepackage{dcolumn}  
\usepackage{bm}    
\usepackage{amssymb} 
\usepackage{amsmath,bm}
\usepackage{amsfonts}    
\usepackage{slashed}  
\usepackage{youngtab}
\usepackage[mathscr]{euscript}
\usepackage{epsfig}
\usepackage[utf8]{inputenc}
\usepackage{tikz}
\usetikzlibrary{tikzmark,calc,,arrows,shapes,decorations.pathreplacing}
\usepackage{xcolor}
\usepackage{mathrsfs}
\usepackage{verbatim}
\usepackage{cases}
\usepackage{bbding}
\usepackage{pifont}
\usepackage{ulem}
\usepackage{dcolumn}
\usepackage{graphicx}
\usepackage{pgf}
\usepackage{pstricks}
\usepackage{pst-node}
\usepackage[all]{xy}

\usepackage{titlesec}

\titleclass{\subsubsubsection}{straight}[\subsection]

\newcounter{subsubsubsection}[subsubsection]
\renewcommand\thesubsubsubsection{\thesubsubsection.\arabic{subsubsubsection}}
\titleformat{\subsubsubsection}
  {\normalfont\normalsize\bfseries}{\thesubsubsubsection}{1em}{}
\titlespacing*{\subsubsubsection}
{0pt}{3.25ex plus 1ex minus .2ex}{1.5ex plus .2ex}

\makeatletter
\renewcommand\paragraph{\@startsection{paragraph}{5}{\z@}%
  {3.25ex \@plus1ex \@minus.2ex}%
  {-1em}%
  {\normalfont\normalsize\bfseries}}
\renewcommand\subparagraph{\@startsection{subparagraph}{6}{\parindent}%
  {3.25ex \@plus1ex \@minus .2ex}%
  {-1em}%
  {\normalfont\normalsize\bfseries}}
\def\toclevel@subsubsubsection{4}
\def\toclevel@paragraph{5}
\def\toclevel@paragraph{6}
\def\l@subsubsubsection{\@dottedtocline{4}{7em}{4em}}
\def\l@paragraph{\@dottedtocline{5}{10em}{5em}}
\def\l@subparagraph{\@dottedtocline{6}{14em}{6em}}
\makeatother

\usepackage{young}
\usepackage{graphicx,rotating,booktabs}

\usepackage{varwidth}

\newdimen\tableauside\tableauside=1.0ex
\newdimen\tableaurule\tableaurule=0.4pt
\newdimen\tableaustep
\def\phantomhrule#1{\hbox{\vbox to0pt{\hrule height\tableaurule width#1\vss}}}
\def\phantomvrule#1{\vbox{\hbox to0pt{\vrule width\tableaurule height#1\hss}}}
\def\sqr{\vbox{%
		\phantomhrule\tableaustep
		\hbox{\phantomvrule\tableaustep\kern\tableaustep\phantomvrule\tableaustep}%
		\hbox{\vbox{\phantomhrule\tableauside}\kern-\tableaurule}}}
\def\squares#1{\hbox{\count0=#1\noindent\loop\sqr
		\advance\count0 by-1 \ifnum\count0>0\repeat}}
\def\tableau#1{\vcenter{\offinterlineskip
		\tableaustep=\tableauside\advance\tableaustep by-\tableaurule
		\kern\normallineskip\hbox
		{\kern\normallineskip\vbox
			{\gettableau#1 0 }%
			\kern\normallineskip\kern\tableaurule}%
		\kern\normallineskip\kern\tableaurule}}
\def\gettableau#1 {\ifnum#1=0\let\next=\null\else
	{{\tiny\yng(1)}}s{#1}\let\next=\gettableau\fi\next}

\renewcommand{\(}{\left(}
\renewcommand{\)}{\right)}

\newcommand{\includeCroppedPdf}[2][]{%
    \IfFileExists{./#2-crop.pdf}{}{%
        \immediate\write18{pdfcrop #2 #2-crop.pdf}}%
    \includegraphics[#1]{#2-crop.pdf}}

\newcommand{\be}{ \begin{equation}}
\newcommand{\ee}{\end{equation}}
\newcommand{\bea}[1]{\begin{eqnarray}\label{#1} }
\newcommand{\eea}{\end{eqnarray}}

\def\ZZZ{{\hskip-3pt\hbox{ Z\kern-1.6mm Z}}}
\def\zzz{{\hskip-3pt\hbox{ z\kern-1mm z}}}

\newcommand{\CO}{{\cal O}}
\newcommand{\IP}{{\mathbb{P}}}
\newcommand{\IC}{{\mathbb{C}}}
\newcommand{\IR}{{\mathbb{R}}}
\newcommand{\IZ}{{\mathbb{Z}}}

\def\bal#1\eal{\begin{align}#1\end{align}}

\renewcommand{\(}{\left(}
\renewcommand{\)}{\right)}

\renewcommand{\Re}{{\rm Re}}
\renewcommand{\Im}{{\rm Im}}

\def\one{{\hbox{ 1\kern-.8mm l}}}
\def\zero{{\hbox{ 0\kern-1.5mm 0}}}

\def\e{\,{\rm e}}
\def\CN{{\mathcal{N}}}
\def\CW{{\mathcal{W}}}
\def\CY{{\mathcal{Y}}}
\def\hCY{\hat{\mathcal{Y}}}
\def\fu{{\mathfrak{u}}}
\def\fgl{{\mathfrak{gl}}}
\def\IZ{{\mathbb{Z}}}

\def\>{\rangle}
\def\<{\langle}

\title{Gluing two affine Yangians of $\mathfrak{gl}_1$
}

\author{Wei Li$^a$ and Pietro Longhi$^b$} 
\affiliation{$^a$ Institute of Theoretical Physics, Chinese Academy of Sciences\\
\hspace*{0.3cm}100190 Beijing, P.R.\ China} 
\affiliation{$^b$ Institut f\"ur Theoretische Physik, ETH Zurich, \\
\hspace*{0.3cm}CH-8093 Z\"urich, Switzerland}

\emailAdd{weili@itp.ac.cn, longhip@phys.ethz.ch}

\abstract{ 
We construct a four-parameter family of affine Yangian algebras by gluing two copies of the affine Yangian of $\mathfrak{gl}_1$. 
Our construction allows for gluing operators with arbitrary (integer or half integer) conformal dimension and arbitrary (bosonic or fermionic) statistics, which is related to the relative framing. 
The resulting family of algebras is a two-parameter generalization of the $\mathcal{N}=2$ affine Yangian, which is isomorphic to the universal enveloping algebra of $\mathfrak{u}(1)\oplus \mathcal{W}^{\mathcal{N}=2}_{\infty}[\lambda]$. 
All algebras that we construct have natural representations in terms of ``twin plane partitions'', a pair of plane partitions appropriately joined along one common leg. 
We observe that the geometry of twin plane partitions, which determines the algebra, bears striking similarities to the geometry of certain toric Calabi-Yau threefolds.
}

\begin{document}

\normalem

\setcounter{tocdepth}{3}
\maketitle

\makeatletter
\g@addto@macro\bfseries{\boldmath}
\makeatother

\section{Introduction}

There is an interesting and useful triangle of relations among the $\mathcal{W}_{1+\infty}[\lambda]$ algebra, the affine Yangian of $\mathfrak{gl}_1$, and the set of plane partitions.
\begin{equation}\label{bosonictriangle}
\xymatrix@C=1pc@R=2.2pc{
& \textrm{  affine Yangian of } \mathfrak{gl}_1
\ar[dl]_{\textrm{``iso"}} 
&  \\
\mathcal{W}_{1+\infty}[\lambda]  \ar[ur]      &&   \ar[ll]_{\textrm{irreps}}  \ar[ul]^{\qquad\qquad \qquad \qquad\qquad\qquad  \qquad \qquad \textrm{irreps}}        \textrm{plane partitions}
} 
\end{equation}
The $\mathcal{W}_{1+\infty}[\lambda]$ algebra is a family of VOAs with higher spin currents (i.e. one current per spin from $s\geq1$) parameterized by the central charge $c$ and the 't Hooft coupling $\lambda$.  
This algebra played a very important role in the holographic duality \cite{Gaberdiel:2010pz, Gaberdiel:2012uj} between Vasiliev's higher spin gravity in AdS$_3$ \cite{Vasiliev:1995dn, Vasiliev:1999ba} and the $\mathcal{W}_{N,k}$ minimal model \cite{Bais:1987zk}, as the boundary symmetry algebra in the large $N$ limit. 
Its truncation to finite $N$ appears as the symmetry algebra of various 2D CFTs, as a chiral algebra of certain 6D SCFTs \cite{Beem:2014kka}, and as a corner VOA of certain defect theories \cite{Gaiotto:2017euk}.

The affine Yangian of $\mathfrak{gl}_1$ --- denoted by $\widehat{\mathcal{Y}}(\mathfrak{gl}_1)$ --- appeared later on the scene. 
It was constructed independently, and with different formulations, by \cite{SV} and \cite{Maulik:2012wi} in the process of proving the AGT conjecture \cite{Alday:2009aq}.  
It was proposed by \cite{Prochazka:2015deb} and later proven by \cite{Gaberdiel:2017dbk} that there is an isomorphism between the affine Yangian of $\mathfrak{gl}_1$ and the universal enveloping algebra of the $\mathcal{W}_{1+\infty}[\lambda]$ algebra.\footnote{
This can be understood as the rational limit of the isomorphism between quantum toroidal algebra of $\mathfrak{gl}_1$ and $q$-deformed $\mathcal{W}_{1+\infty}[\lambda]$ algebra \cite{Miki, feigin2011, Feigin:2010qea}.}

This isomorphism is interesting because it relates two rather different algebra structures: one being a vertex operator algebra and the other a Yangian algebra.
It has also proven to be rather useful computationally \cite{Datta:2016cmw},
since the affine Yangian of $\mathfrak{gl}_1$ has a natural representation theory in terms of plane partitions \cite{feigin2012, Prochazka:2015deb}.
\footnote{
It was already proposed in \cite{feigin2012} that plane partitions are natural representations of the quantum toroidal algebra of $\mathfrak{gl}_1$, which upon taking the rational limit gives affine Yangian of $\mathfrak{gl}_1$ \cite{Tsymbaliuk:2014fvq}.}
By the isomorphism above, plane partitions also serve as a representation of $\mathcal{W}_{1+\infty}$.
This has certain advantages over the traditional coset representation, as it leaves manifest the triality symmetry of $\mathcal{W}_{1+\infty}[\lambda]$ \cite{Gaberdiel:2012ku}. 
More practically, the characters are also easy to obtain, since they are simply generating functions of plane partitions, with possibly non-trivial asymptotics along the three directions \cite{feigin2012, Prochazka:2015deb, Datta:2016cmw}.\footnote{
This is true at generic value of central charge $c$ and 't Hooft coupling $\lambda$. At special values where  null vectors arise, the $\mathcal{W}_{1+\infty}$ character counts fewer states.}  
\medskip

In addition, the triangle of relations (\ref{bosonictriangle}) turns out to be very useful for a certain  ``gluing'' construction of new VOAs. 
Constructing new VOAs is an interesting and non-trivial question in itself. 
This problem has recently gained renewed interest, due to physical setups in which new VOA structures arise \cite{Gaiotto:2017euk, Prochazka:2017qum}. 
In many cases, there are no explicit constructions of these VOAs,\footnote{
See \cite{Prochazka:2018tlo} for an construction of (the truncated version of) some of these VOAs using free field realization, which in principle also allows one to derive explicit algebraic relations.} and more generally their properties are only poorly understood.

In the attempt to supersymmetrize the triangle (\ref{bosonictriangle}) for the $\mathcal{N}=2$ supersymmetric $\mathcal{W}_{\infty}[\lambda]$ algebra, \cite{Gaberdiel:2017hcn, Gaberdiel:2018nbs} developed a construction of new affine Yangians by gluing two distinct plane partitions. 
The construction is based on the fact that  $ \mathfrak{u}(1)\oplus \mathcal{W}^{\mathcal{N}=2}_{\infty}[\lambda]$ has two commuting bosonic $\mathcal{W}_{1+\infty}$ subalgebras, and that all fermionic generators transform in irreducible representations derived from tensor products of $({\square}, \overline{{\square}})$ and $(\overline{{\square}}, {\square})$ w.r.t. the two bosonic subalgebras.
Each building block VOA, i.e. $\mathcal{W}_{1+\infty}\sim\widehat{\mathcal{Y}}(\mathfrak{gl}_1)$ has a faithful representation in terms of plane partitions. 
Fermionic generators act by modifying simultaneously the asymptotics of the two plane partitions, along one of the three directions of each side, and for this reason can be interpreted as ``gluing'' the two. 
This led to a representation of 
$\mathfrak{u}(1)\oplus \mathcal{W}^{\mathcal{N}=2}_{\infty}[\lambda]$ acting on a pair of standard ``corner'' plane partitions, subject to a relation on the asymptotics along a common ``internal'' leg.
The result of gluing two plane partitions in this way along a common direction was called ``twin plane partition".

The $\mathcal{N}=2$ supersymmetric affine Yangian was then constructed by demanding that it acts naturally on the set of twin plane partitions and that it correctly reproduces  $\mathcal{W}^{\mathcal{N}=2}_{\infty}$ charges. 
The algebra includes bosonic generators corresponding to the two copies of $\widehat{\mathcal{Y}}(\mathfrak{gl}_1)$, and in addition new fermionic generators transforming in a bi-module of the bosonic subalgebra.
The bosonic generators act independently on either copy of standard plane partitions, while the fermionic generators act on the asymptotic shapes of both plane partitions along the common leg.
Since this changes simultaneously the asymptotic boundary condition for each plane partition, the new gluing operators interact non-trivially with the bosonic sub-algebras that act locally on either plane partition. 
In addition, the new ``gluing'' generators have interesting interactions among themselves. 
The resulting supersymmetric affine Yangian should be isomorphic to the UEA of $\mathfrak{u}(1)\oplus \mathcal{W}^{\mathcal{N}=2}_{\infty}[\lambda]$.
\begin{equation}\label{susictriangle}
\xymatrix@C=1pc@R=2.2pc{
& \mathcal{N}=2 \,\,\textrm{affine Yangian of } \mathfrak{gl}_1
\ar[dl]_{\textrm{``iso"}}
&  \\
\mathcal{N}=2 \,\,\, \mathcal{W}_{\infty}[\lambda]  \ar[ur]      &&   \ar[ll]^{\textrm{irreps}}  \ar[ul]_{\textrm{irreps}}        \textrm{twin-plane-partitions}
} 
\end{equation}
\medskip

In this paper, we will generalize the gluing construction of \cite{Gaberdiel:2017hcn, Gaberdiel:2018nbs} to produce a four-parameter family of VOAs with a natural action on (suitably generalized) twin plane partitions.
In addition to $c,\lambda$, we introduce a ``shifting'' parameter $\rho \in \frac{1}{2}\mathbb{Z}_{\geq 0}$ and a ``framing'' $p$ that can take values $0$ or $\pm1$.
The shifting parameter determines the conformal dimension of the gluing operators that tie together the two copies of $\widehat{\mathcal{Y}}(\mathfrak{gl}_1)$.
The framing parameter determines the relative orientation of the two corner plane partitions, which in turn determines the statistics of the gluing operators and also fixes the relation between the affine Yangian parameters of the two corners.
Therefore, we have a four parameter family of new VOAs, characterized by:
\begin{enumerate}
\item central charge $c$ and coupling constant $\lambda$, as in the $\CN=2$ construction,
\item conformal dimension $1+\rho$ of the gluing operators,
\item relative framing $p$ of the two corners.
\end{enumerate}
In fact, since all these resulting algebras allow for further truncations, labeled by four positive integers, we actually have an eight-parameter family of new VOAs.\footnote{
For affine Yangian of $\mathfrak{gl}_1$, the truncation of the algebra \cite{Prochazka:2014gqa} can be realized by putting an obstruction to box-stacking in the plane partition \cite{Gaiotto:2017euk}. This can be easily generalized to the twin plane partition, where a generic obstruction is labeled by four non-negative integers.}
\medskip

The $\mathcal{N}=2$ affine Yangian is recovered as a special case of our family of algebras, corresponding to $\rho=1/2, p=0$.
In that case, the  gluing operators were chosen to have conformal dimension ${3}/{2}$ (corresponding to $1+\rho$), i.e.\ that of the supercharges $G^{\pm}$.
The fact that $p=0$ was also confirmed by the match with the ${\mathcal{N}}=2$ $\mathcal{W}_\infty[\lambda]$  algebra --- namely, the supercharges $G^{\pm}$ are fermionic and the 't~Hooft couplings of the two bosonic subalgebras were related by $\lambda_1+\lambda_2=1$.

We generalize these relations by allowing for general $\rho$ and $p$ initially. 
Demanding that the algebra acts naturally on twin plane partitions then imposes powerful constraints on these two parameters, leading to the conclusion that $\rho$ must be quantized by half-integers and that $p$ must be integer and valued in the range $[-1,1]$.
While these constraints would be difficult to obtain algebraically, they arise naturally from consistency of the action on twin-plane partitions.
Once these are fixed, the final algebra can be completely determined, by demanding that the action of both corner bosonic subalgebras and the newly added gluing operators have sensible, and mutually consistent, actions on the set of twin-plane partitions.
We derive all OPE relations by following a similar procedure as the one used in the $\mathcal{N}=2$ case in \cite{Gaberdiel:2018nbs}.

Another nontrivial prediction that arises from the action on twin plane partitions is that framing also determines the self-statistics of the gluing operators.
In fact their statistics switches from fermionic for $p=0$ to bosonic for $p=\pm1$. 
We derive this fact by direct counting of the low-lying configurations of twin-plane partitions in the vacuum module, with nontrivial asymptotics (on both sides) along the common leg, for $p=-1,0,1$.
These match exactly with with $q$-expansions of vacuum characters with fermionic or bosonic gluing operators, respectively for $p=0$ and $\pm1$. 
We also confirm the switch in statistics at the level of the algebra by studying the self-OPEs of gluing generators, showing that they are nilpotent on vacuum in the case $p=0$ and not otherwise.
Note that for the $\mathcal{N}=2$ case of \cite{Gaberdiel:2018nbs}, the nilpotency-on-vacuum property could have also been derived from the map to the $\mathcal{W}^{\mathcal{N}=2}_{\infty}$ algebra. Here we extend this statement to all values of $\rho$. 
\medskip

\begin{figure}[h!]
\begin{center}
\includegraphics[width=0.4\textwidth]{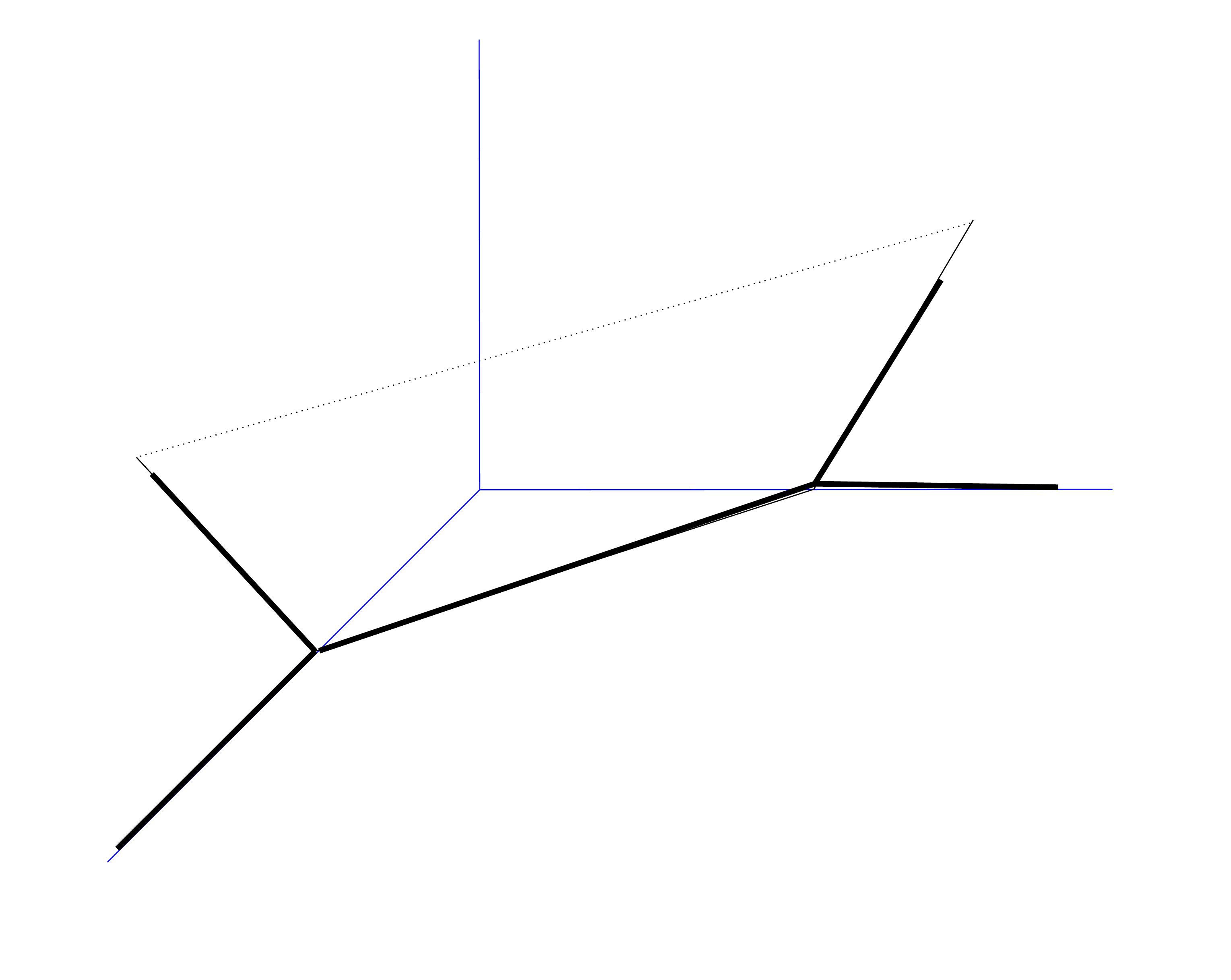}
\hspace{10pt}
\includegraphics[width=0.4\textwidth]{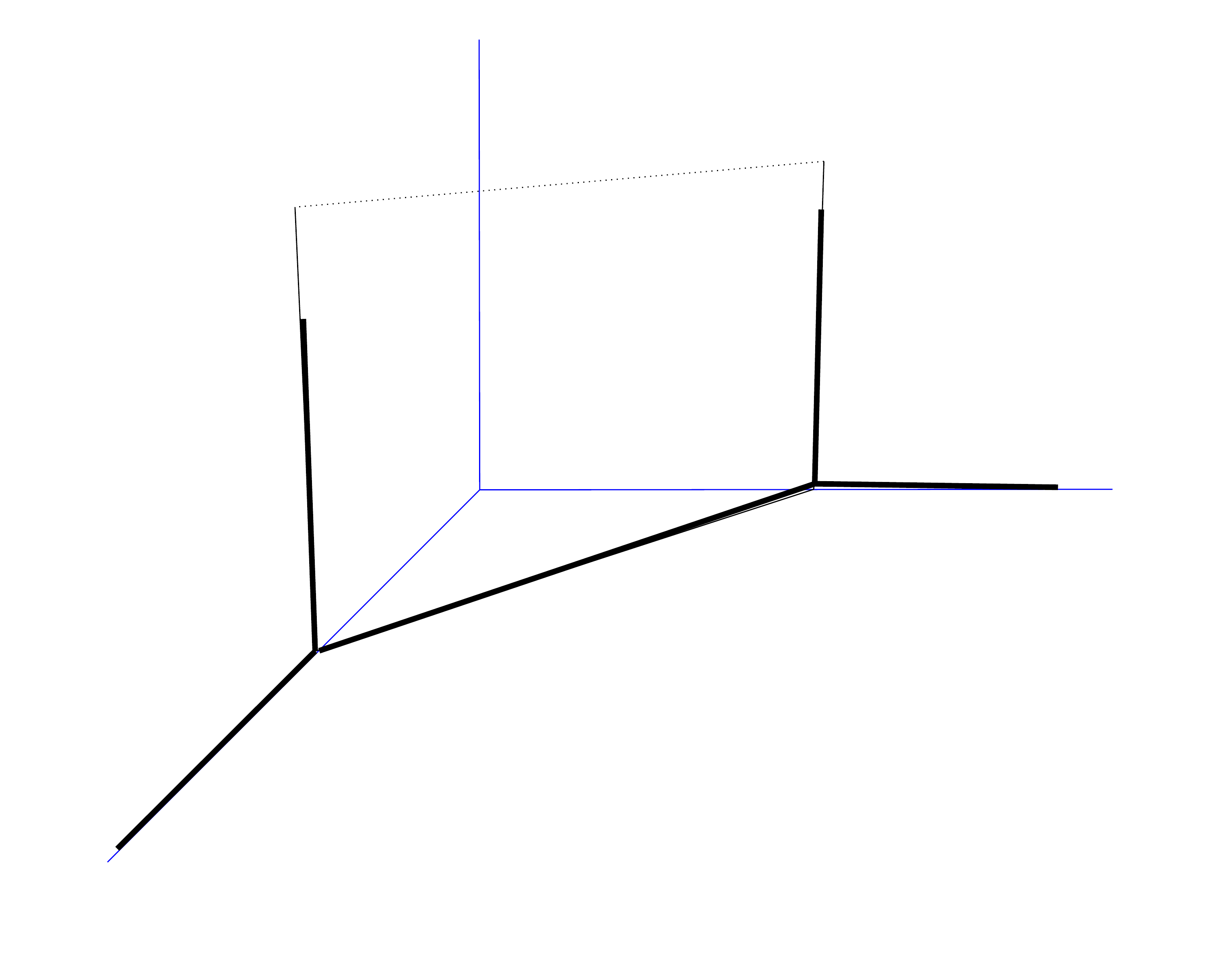}
\caption{Relative orientations of the two corners: $p=0$ on the left, $p=-1$ on the right.}
\label{fig:geometric-intuition}
\end{center}
\end{figure}

The switch in statistics of gluing operators also admits a simple geometric interpretation, which further leads to a connection to the geometry of toric Calabi-Yau threefolds obtained by gluing two copies of $\mathbb{C}^3$. 
At the level of twin plane partitions, $p$ controls the relative orientation of the axes of the two ``corners'', as sketched in Figure \ref{fig:geometric-intuition}. 
Since gluing operators act by creating infinite horizontal rows of boxes connecting the two corners,\footnote{
More precisely, they create rows of boxes on the left module, and anti-boxes on the right-module.} their action is immediately affected by the relative orientation of the walls of the room on either side.
When $p=\pm1$ the room features two parallel edges (for $p=-1$ they are the vertical ones in Figure \ref{fig:geometric-intuition}).
As a result, rows of equal length can be stacked along that direction.
Since gluing operators create rows of fixed effective length, it follows that repeated application of these operators will create a stack of rows, just like bosonic generators of a Heisenberg (sub)algebra.
On the contrary, when $p=0$ it is not possible to stack rows of equal length along any of the transverse directions, since the edges at the two corners are not parallel.
This leads to nilpotency-on-vacuum for gluing operators, and their corresponding fermionic nature.
Taking this reasoning further, one may predict that to create new rows stacked along one of the transverse directions, it is necessary to supply a certain number of extra boxes, determined by the relative slant of the two corners.
We indeed find that these heuristic considerations are independently ---and non-trivially--- predicted by self-consistency of the action of our family of VOAs on twin plane partitions.

The geometry of ``rooms'', where twin plane partitions live, bears a striking resemblance to the bases of $T^3$-fibrations of certain toric Calabi-Yau threefolds
\be\label{eq:gemetric-background}
	\CO(-1-p) \oplus \CO(-1+p) \to \IP^1 \,.
\ee
In view of this, the constraint on the range $-1\leq p\leq 1$, which arises from the demand that our algebras act on ``allowed" sets of twin plane partitions (in particular, no box sticking outside the room), can be given a natural geometric interpretation. 
Namely, in Figure~\ref{fig:geometric-intuition}, if $|p|>1$, two lines from different corners would intersect and change the topology of the room.

Another important connection to geometry arises from looking at the vacuum characters of our family of VOAs, which are respectively 
\be
\begin{split}
	(p=0):\quad& 
	\chi^{\textrm{Full}}_0(q,y) = \prod_{n=1}^{\infty} \frac{(1+y\, q^{n+\rho})^{n}(1+\frac{1}{y}\, q^{n+\rho})^{n}}{(1-q^n)^{2n}} 
	\\
	(p=\pm1):\quad&
	\chi^{\textrm{Full}}_0(q,y) 
	= 
	\prod_{n=1}^{\infty} 
	\frac{1}{(1-q^n)^{2n}  \(1-y q^{n+\rho}\)^{n}\(1-{y}^{-1}q^{n+\rho}\)^{n} } \,.
\end{split}
\ee
These characters resemble partition functions of BPS states of type IIA string theory compactified on the Calabi-Yau threefolds (\ref{eq:gemetric-background}).\footnote{
At least, in the case when $\rho=0$ they precisely coincide with (framed) Donaldson-Thomas partitions functions in maximal chambers of the moduli space of stability conditions. For other values of $\rho$ we discuss their interpretation in Section \ref{sec:BPS-interpretation}.}
This seems to suggest that our algebras are related, up to $\mathfrak{u}(1)$ factors (see later), to the 
(multi-particle) BPS algebra \cite{Kontsevich:2010px} for the corresponding theory, i.e.\ type IIA string theory compactified on (\ref{eq:gemetric-background}).\footnote{
For more recent study of related ideas see e.g.\ \cite{Rapcak:2018nsl} and reference therein.}
\medskip

The paper is organized as follows. 
In Section~\ref{sec:review} we review the building block of the gluing construction (i.e.\ the bosonic triangle (\ref{bosonictriangle})) and the gluing construction for $\mathcal{N}=2$ affine Yangian. 
In Section~\ref{sec:2parameter}, we present the two-parameter generalization of the gluing construction.
The geometric interpretation is discussed in Section \ref{sec:pqweb}. 
Section~\ref{sec:TPP} describes twin plane partitions.
In Section~\ref{sec:pole} we determine the pole structure of the actions of gluing operators on twin plane partitions and partially fix the OPEs between corner operators and gluing operators. 
In Section~\ref{sec:residue} we determine the full actions of gluing operators on twin plane partitions and fix all remaining algebraic relations.
Section~\ref{sec:summary} contains the summary of the main results and a discussion on future directions. 
We include an appendix with details on some computations.

\section{Review of affine Yangian of $\mathfrak{gl}_1$ and $\mathcal{N}=2$ gluing construction} \label{sec:review}

In this section we first review the building block of our gluing procedure, i.e.\ the triangle relation (\ref{bosonictriangle}) between $\mathcal{W}_{1+\infty}$, the affine Yangian of $\mathfrak{gl}_1$, and plane partitions. 
Then we review the construction of the $\mathcal{N}=2$ version of 
this triangle  \cite{Gaberdiel:2017hcn, Gaberdiel:2018nbs} via gluing.
The gluing construction in the current paper is a two-parameter generalization of the $\mathcal{N}=2$ gluing. 
In fact, since this generalization amounts to promoting certain quantities that used to coincide to be independent, it turns out that many of the relations that define the ${\cal N}=2$ algebra can be written in more general form by a small modification.
For this reason we will present the equations directly in the general form. 

\subsection{The affine Yangian of $\mathfrak{gl}_1$}
\subsubsection{Defining relations }

The original constructions of the affine Yangian of $\mathfrak{gl}_1$ are in the form of SH$^c$ algebra (i.e.\ the central extension of spherical degenerate double affine Hecke algebra of GL$_{n\rightarrow \infty}$) by \cite{SV} and in a (generalized) RTT formulation by \cite{Maulik:2012wi}. 
For the gluing process, we will use its reformulation by \cite{Tsymbaliuk:2014fvq},\footnote{
For the map between formulations of \cite{SV} and  \cite{Tsymbaliuk:2014fvq} see section 5.1.\ of \cite{Gaberdiel:2017dbk}; for the translation between formulations of \cite{Maulik:2012wi} and \cite{Tsymbaliuk:2014fvq}  see \cite{Prochazka:2019dvu}.} since it has a natural representation on plane partitions  (see also  \cite{Prochazka:2015deb, Gaberdiel:2017dbk} for more details).

The affine Yangian of $\mathfrak{gl}_1$ is an associative algebra, defined in terms of the following three fields 
\begin{equation}\label{generating}
e(z) = \sum_{j=0}^{\infty} \, \frac{e_j}{z^{j+1}} \ , \qquad 
f(z) = \sum_{j=0}^{\infty} \, \frac{f_j}{z^{j+1}} \ , \qquad 
\psi(z)  = 1 + \sigma_3 \, \sum_{j=0}^{\infty} \frac{\psi_j}{z^{j+1}} \ .
\end{equation}
by the following OPE-like relations \cite{Prochazka:2015deb, Gaberdiel:2017dbk} 
\begin{equation}\label{bosonicdef}
\begin{aligned}
\begin{aligned}
\psi(z)\, e(w) &  \sim  \varphi_3(\Delta)\, e(w)\, \psi(z) \\ 
\psi(z)\, f(w) & \sim \varphi_3^{-1}(\Delta)\, f(w)\, \psi(z) 
\end{aligned}
&\qquad
\begin{aligned}
e(z)\, e(w) & \sim    \varphi_3(\Delta)\, e(w)\, e(z) \\
 f(z)\, f(w) &  \sim    \varphi_3^{-1}(\Delta)\, f(w)\, f(z) 
\end{aligned}\\
[e(z)\,, f(w)]  & \sim  - \frac{1}{\sigma_3}\, \frac{\psi(z) - \psi(w)}{z-w} \ , 
\end{aligned}
\end{equation}
where $\Delta$ will henceforth denote the difference  
\begin{equation}
\Delta\equiv z-w \ ,
\end{equation}
and ``$\sim$" is understood to mean equality up to regular terms either at $z=0$ or $w=0$.\footnote{
See the discussion around eq.~(5.15) in \cite{Gaberdiel:2017dbk}}
Central to the definition is the cubic rational function $\varphi_3(u)$, defined as 
\begin{equation}\label{varphidef}
\varphi_3(u) \equiv \frac{(u+h_1) (u+h_2) (u+h_3)}{(u-h_1) (u-h_2) (u-h_3)}   \ ,
\end{equation}
with parameters $h_i$ subject to
\begin{equation}\label{sumh}
h_1+h_2+h_3=0 \ .
\end{equation}
In addition, there are Serre relations
\begin{equation}\label{Serre}
\begin{aligned}
&\sum_{\pi \in \mathcal{S}_3}\, \bigl(z_{\pi(1)} - 2 z_{\pi(2)} + z_{\pi(3)} \bigr)\, e(z_{\pi(1)})\, e(z_{\pi(2)})\, e(z_{\pi(3)}) \sim 0 \\
&\sum_{\pi \in 
\mathcal{S}_3}\, \bigl(z_{\pi(1)} - 2 z_{\pi(2)} + z_{\pi(3)} \bigr)\, f(z_{\pi(1)})\, f(z_{\pi(2)})\, f(z_{\pi(3)}) \sim 0 \ .
\end{aligned}
\end{equation}
Note that in this formulation, the definition of the affine Yangian of $\mathfrak{gl}_1$ is manifestly invariant under the permutation group $\mathcal{S}_3$ acting on the triplet $(h_1,h_2,h_3)$, as opposed to its two other formulations in \cite{SV, Maulik:2012wi}.

The defining relations above can also be translated in terms of modes $(e_j, f_j, \psi_j)$ using (\ref{generating}), for more details, see e.g.\ eq.\ (2.13)-(2.18) of \cite{Gaberdiel:2017dbk}. 
In this paper, we will mostly use the OPE-like relations in terms of fields; it is straightforward to obtain the relations in terms of modes by mode expansion.

\subsubsection{Isomorphism between affine Yangian of $\mathfrak{gl}_1$ and UEA($\mathcal{W}_{1+\infty}$)}

The affine Yangian of  $\mathfrak{gl}_1$ is isomorphic to the universal enveloping algebra of $\mathcal{W}_{1+\infty}[\lambda]$, as proposed in \cite{Prochazka:2015deb} and proven in \cite{Gaberdiel:2017dbk}.\footnote{
This can be viewed as the rational limit of the isomorphism \cite{Miki, feigin2011} between the toroidal Yangian of $\mathfrak{gl}_1$ and (the UEA of) the $q$-deformed $\mathcal{W}_{1+\infty}$ algebra.}
The map between the two sides is only in terms of modes: 
\begin{equation}\label{Wepsif}
W^{(s)}_{-1} \sim e_{s-1} \qquad \qquad W^{(s)}_{0} \sim \psi_{s} \qquad  \qquad W^{(s)}_{1} \sim f_{s-1} \ , 
\end{equation}
where we have only written leading terms here, and the ``sub-leading" terms can be obtained order by order as in \cite{Gaberdiel:2017dbk}.
A direct map between the fields of the two sides, i.e.\ between $W^{(s)}(z)$ and $\{e(z),\psi(z),f(z)\}$, is not known. 
However,  the isomorphism  can be established for the entire algebra because $\mathcal{W}_{1+\infty}[\lambda]$ only depends on two parameters (given the spectrum of one field per spin $s\geq 1$), which can be chosen to be the central charge $c$ and coupling $\lambda$, or equivalently  the coset parameters $(N, k)$. 
These are related as follows
\be\label{eq:c-and-lambda}
	c_{N,k} = (N-1)\left(1 - \frac{N(N+1)}{(N+k)(N+k+1)}\right)+1  \,,
	\qquad
	\lambda_{N,k} = \frac{N}{N+k}\,.
\ee

The map between these two parameters and the Yangian parameters $\{h_i\}$ is (see \cite{Gaberdiel:2017dbk})
\begin{equation}\label{h123}
h_1 =  -\sqrt{\frac{N+k+1}{N+k}} \ , \quad h_2 =  \sqrt{\frac{N+k}{N+k+1}} \ , \quad h_3 = \frac{1}{\sqrt{(N+k)(N+k+1)}} \ , 
\end{equation}
In addition, the central element $\psi_0$ is related to the $\mathcal{W}$ algebra parameters by  \cite{Gaberdiel:2017dbk}
\begin{equation}\label{psi0}
\psi_0 = N  \ .
\end{equation}

\subsubsection{Representation on plane partitions}\label{sec:bosonic-PP}

One important reason why the relation between $\mathcal{W}_{1+\infty}$ and the affine Yangian of $\mathfrak{gl}_1$ is useful comes from the fact that the latter has a representation theory in terms of plane partitions.
\smallskip

For a given triplet of Young diagrams $(\lambda_1, \lambda_2, \lambda_3)$, the set of plane partitions with $(\lambda_1, \lambda_2, \lambda_3)$ as asymptotic boundary conditions furnishes a representation of the affine Yangian of $\mathfrak{gl}_1$.
In particular, every plane partition $\Lambda$ is an eigenstate of $\psi(z)$:
\begin{equation}\label{ppartpsi}
\begin{aligned}
\psi(z)|\Lambda \rangle & = \bm{\Psi}_{\Lambda}(z)|\Lambda \rangle  \ ,
\end{aligned}
\end{equation}
where the eigenvalue $\bm{\Psi}_\Lambda(z)$ is defined as 
\be\label{psieig}
\bm{\Psi}_\Lambda(z) \equiv \psi_0(z) \, \prod_{ {\tiny \yng(1)} \in (\Lambda)} \varphi_3(z - h({\tiny \yng(1)}) ) \ , 
\ee
where 
\begin{equation}\label{psi0def}
\psi_0(z)\equiv 1 + \frac{\psi_0 \sigma_3}{z} 
\end{equation}
is the vacuum factor. 
Namely, each box in $\Lambda$ contributes a factor of $\varphi_3$, with variable shifted by
\be\label{hbox}
h({\tiny \yng(1)}) \equiv h_1 x_1({\tiny \yng(1)}) + h_2 x_2({\tiny \yng(1)}) +  h_3 x_3({\tiny \yng(1)}) 
\ee
with $x_i({\tiny \yng(1)})$ the $x_i$-coordinate of the box.
A plane partition configuration $\Lambda$ is in one-to-one correspondence with its eigenvalue functions $\bm{\Psi}_{\Lambda}(u)$, defined in (\ref{psieig}).

The creation operator $e(z)$ adds a box to $\Lambda$ at all possible positions (i.e.\ such that the resulting $\Lambda+\square$ is again an allowed plane partition), and the annihilation operator $f(z)$ removes a box from $\Lambda$ at all possible positions  \cite{Prochazka:2015deb,Gaberdiel:2017dbk}. These actions are encoded by the following formulae:
\begin{equation}\label{ppartef}
\begin{aligned}
e(z) | \Lambda \rangle & =  \sum_{ {\tiny \yng(1)} \in {\rm Add}(\Lambda)}\frac{\Big[ -  \frac{1}{\sigma_3} {\rm Res}_{w = h({\tiny \yng(1)})} \bm{\Psi}_{\Lambda}(w) \Big]^{\frac{1}{2}}}{ z - h({\tiny \yng(1)}) } 
| \Lambda + {\tiny \yng(1)} \rangle \ , \\
f(z) | \Lambda \rangle & =  \sum_{ {\tiny \yng(1)} \in {\rm Rem}(\Lambda)}\frac{\Big[ -  \frac{1}{\sigma_3} {\rm Res}_{w = h({\tiny \yng(1)})} \bm{\Psi}_{\Lambda}(w) \Big]^{\frac{1}{2}}}{ z - h({\tiny \yng(1)}) } | \Lambda - {\tiny \yng(1)} \rangle \ . 
\end{aligned}
\end{equation}
where ``Res" denotes the residue.
It is easy to check that with the action (\ref{ppartpsi}) and (\ref{ppartef}), the set of plane partitions $\Lambda$ (with given b.c.)  forms a faithful representation of the affine Yangian of $\mathfrak{gl}_1$ given by (\ref{bosonicdef}) and (\ref{Serre}).

Let us illustrate the action of the algebra with the first few examples. 
The charge function of the vacuum (denoted by $|\emptyset\rangle$) is
\begin{equation}
 \psi_0(u) =  1+ \frac{\sigma_3\psi_0}{u}  \ . 
\end{equation}
Applying $e(z)$ on $|\emptyset\rangle$ repeatedly generates all the states in the vacuum module.
For example,  applying $e(z)$ once generates the first descendant $|\square\rangle$ corresponding to a plane partition with one box sitting at the origin:
\begin{equation}
e(z) |\emptyset\rangle \sim \frac{1}{z} |\square\rangle \ . 
\end{equation}
The $\psi(u)$ eigenvalue is 
\begin{equation}
|\square\rangle: \qquad 
\bm{\Psi}_{\square}(u)=\psi_0(u) \cdot \varphi_3(u)\ . 
\end{equation}
Conversely, the annihilation operators $f(z)$ acts by 
\begin{equation}
f(z) |\emptyset\rangle = 0 \qquad \textrm{and} \qquad f(z) |\square\rangle \sim \frac{1}{z} |\emptyset\rangle 
\end{equation}

The irreducible representations of the affine Yangian are parametrized by three Young diagrams $(\lambda_1, \lambda_2, \lambda_3)$, which encode the asymptotics along the three directions.\footnote{
The action of the affine Yangian generators (\ref{generating}) does not modify the asymptotics.} 
For a plane partition with non-trivial asymptotics, its charge function is still given by (\ref{psieig}).
The product in (\ref{psieig}) now runs over the infinitely many boxes furnishing the non-trivial asymptotics.
In all the cases we have encountered, this infinite product is automatically regularized, due to the cancellation in the charge functions of plane partitions.

\subsection{From  $\mathcal{W}^{\mathcal{N}=2}_{\infty}$ to $\mathcal{N}=2$ affine Yangian via twin-plane partitions}

We will now review the construction of the $\mathcal{N}=2$ affine Yangian proposed in \cite{Gaberdiel:2017hcn, Gaberdiel:2018nbs}.
The goal was to generalize the bosonic triangle (\ref{bosonictriangle}). 
Unlike the bosonic case, although the $\mathcal{N}=2$ supersymmetric $\mathcal{W}_{\infty}[\lambda]$ had been known (see e.g.\ \cite{Candu:2012tr}), the corresponding $\mathcal{N}=2$ version of the 
affine Yangian of $\mathfrak{gl}_1$ was not, and neither was the relevant set of representations. 
Therefore, the construction of an $\mathcal{N}=2$ version of the triangle (\ref{bosonictriangle}) amounts to constructing the $\mathcal{N}=2$ version of the affine Yangian of $\mathfrak{gl}_1$ that is isomorphic to UEA($\mathcal{W}^{\mathcal{N}=2}_{\infty}$) and defining an appropriate set of representations upon which these two algebras act faithfully. 

The crucial hint was to study representations of $\mathcal{W}^{\mathcal{N}=2}_{\infty}$, and to interpret them in terms of plane partitions. 
In this subsection, we will review how the decomposition of $\mathcal{W}^{\mathcal{N}=2}_{\infty}$ characters suggests that the relevant representations should be a pair of plane partitions properly ``glued" among a common direction --- called twin plane partitions. 
The $\mathcal{N}=2$ affine Yangian is then constructed by demanding that it acts naturally on the set of twin plane partitions and reproduces correct $\mathcal{W}$ charges. 
The result of this approach is the $\mathcal{N}=2$ triangle (\ref{susictriangle}).

\subsubsection{Decomposition of the $\fu(1) \oplus \mathcal{W}_{\infty}^{\mathcal{N}=2}[\lambda]$ algebra}

The starting point for the construction of the $\CN=2$ affine Yangian, as proposed in \cite{Gaberdiel:2017hcn, Gaberdiel:2018nbs}, is the observation that the  $\CW^{(\CN=2)}_\infty[\lambda]$ algebra (that contains one $\mathcal{N}=2$ multiplet per spin for $s=1,2,\dots,\infty$), augmented by a free boson, contains two commuting copies of $\CW_{1+\infty}$ as a bosonic subalgebra:\footnote{
We add an extra boson to make the decomposition into left and right more symmetric. As always, the $\mathfrak{u}(1)$ field is easy to add to a $\mathcal{W}_{\infty}$ or decouple from a $\mathcal{W}_{1+\infty}$ algebra \cite{Gaberdiel:2013jpa}.}
\begin{equation}\label{eq:bos-subalg}
	\fu(1) \oplus \CW_\infty^{(\CN=2)}[\lambda]  \ \ \supset\ \  \CW_{1+\infty}[\lambda] \oplus \CW_{1+\infty}[1-\lambda]\,.	
\end{equation}
The total central charges of the l.h.s.\ is 
\begin{equation}
c^{\textrm{total}}=1+c_{\mathcal{N}=2}=c+\hat{c}
\end{equation}
where $c$ and $\hat{c}$ are the central charges of the left and right $\mathcal{W}_{1+\infty}$ algebras on the r.h.s., respectively.
In terms of $(N,k)$ parameters, the two parameters $(c,\lambda)$ of the left $\mathcal{W}_{1+\infty}$ are given by (\ref{eq:c-and-lambda}) and $(\hat{c},\hat{\lambda}=1-\lambda)$ for the right $\mathcal{W}_{1+\infty}$ by exchanging $N\leftrightarrow k$ in (\ref{eq:c-and-lambda}). 

As seen from the vacuum character of $\fu(1) \oplus \CW_\infty^{(\CN=2)}[\lambda]$
\begin{equation}\label{eq:vac-char-fermi}
	\chi^{(\CN=2)}(q,y) = \prod_{n\geq 1} \frac{
			\(1+ y q^{n+\frac{1}{2}}\)^n \(1+ y^{-1} q^{n+\frac{1}{2}}\)^n
		}{
			\(1-q^n\)^{2n}
		}\,,
\end{equation}
all the remaining fields are fermions. 
One way to characterize these additional fermionic generators is by how they transform under the bosonic subalgebra (\ref{eq:bos-subalg}). 
For this purpose, one can study how (\ref{eq:vac-char-fermi}) decomposes in terms of (characters of) representations of bosonic subalgebras. 
The denominator in (\ref{eq:vac-char-fermi})  correspond to bosonic generators accounted by (\ref{eq:bos-subalg}), whereas the numerator in (\ref{eq:vac-char-fermi}) correspond to fermionic generators.
For example, the first factor admits the following decomposition 
\begin{equation}\label{eq:char-decomp-x}
	\mathcal{W}^{\mathcal{N}=2}_{\infty}: \qquad
	\prod_{n\geq 1} \(1+ y q^{n+\frac{1}{2}}\)^n = \sum_{R} y^{|R|} \, \chi_R^{(\rm w)[\lambda]}(q) \, \chi_{R^\star}^{(\rm w)[1-\lambda]}(q)
\end{equation}
where the sum runs over all Young diagrams $R$, and 
\begin{equation}\label{eq:Neq2TPPcondition}
R^\star \equiv \overline{R^t}
\end{equation}
denotes the conjugate of the transpose of $R$.
Finally $\chi_R^{(\rm w)[\lambda]}(q)$ is the wedge part of $\mathcal{W}_{1+\infty}[\lambda]$ character for representation $R$:
\begin{equation}\label{chiwedge}
\chi_R^{[\lambda]}(q)=\chi_{\textrm{pp}}(q)\cdot\chi_R^{(\rm w)[\lambda]}(q)
\end{equation}
where $\chi_{\textrm{pp}}(q)$ is the vacuum character of $\mathcal{W}_{1+\infty}[\lambda]$; it also counts the plane partitions with trivial asymptotics and equals MacMahon function.
A similar decomposition exists for the second factor in the numerator. 

With the decomposition of the numerator, the full character (\ref{eq:vac-char-fermi}) can be decomposed into
\begin{equation}\label{N2decomp}
\begin{aligned}
\chi^{(\CN=2)}_0(q,y)&=\chi_{\rm pp}(q)^2 \left(\sum_{R} y^{|R|}\, \chi_{R}^{({\rm w})\,  [\lambda]}(q)  \chi_{R^\star}^{({\rm w})\, [1-\lambda]}(q)\right)  \left(\sum_{{S}}\frac{1}{ y^{|{S}|}}\chi_{S^\star}^{({\rm w})\,  [\lambda]}(q) \chi_{S}^{({\rm w})\, [1-\lambda]}(q)\right)
\\
&=1+ \sum_{R} y^{|{R}|}\chi_{R}^{ [\lambda]}(q) \cdot \chi_{R^\star}^{ [{1-\lambda}]}(q)+ \sum_{S} \frac{1}{y^{|S|}} \, \chi_{S^\star}^{ [\lambda]}(q) \cdot \chi_{S}^{[ {1-\lambda}]}(q)+\cdots \ ,
\end{aligned}
\end{equation}
The character analysis shows that $\fu(1)\oplus\CW_\infty^{(\CN=2)}$ can be decomposed into representations of the bosonic subalgebra (\ref{eq:bos-subalg}), with the specific property that the representation with respect to $\CW_{1+\infty}[1-\lambda]$ is the conjugate transpose (\ref{eq:Neq2TPPcondition}) of the representation with respect to $\CW_{1+\infty}[\lambda]$.
In fact, all representations appearing in the decomposition can be obtained by taking tensor powers of two ``bi-minimal'' 
representations:
\begin{equation}\label{N2fermion}
\textrm{Fermionic:}\qquad 	
	({\square},\overline {\square})
	\qquad \textrm{and}\qquad 
	(\overline {\square},{\square}) \,.
\end{equation}

\subsubsection{Twin plane partitions as representation of $\fu(1) \oplus \CW_\infty^{(\CN=2)}[\lambda]$}
\label{sec:TPPreview}

From the decomposition of the $\fu(1) \oplus \CW_\infty^{(\CN=2)}[\lambda]$ vacuum character we can deduce the relevant representations that are of plane partition type, and in turn, the building blocks of the $\mathcal{N}=2$ affine Yangian.

For the bosonic part, each copy of $\CW_{1+\infty}$ in (\ref{eq:bos-subalg}) is dual to a copy of the affine Yangian of $\fgl_1$, therefore
\begin{equation}
	\textrm{${\cal N}=2$ affine Yangian}
	 \ \ \supset \ \
	 \CY\oplus \hCY \,.
\end{equation}
The left bosonic Yangian subalgebra $\CY$ is taken to have OPEs (\ref{bosonicdef}), whereas for the right one $\hCY$ one introduces new generators $\hat e,\hat f,\hat \psi$ with OPE relations
\begin{equation}\label{bosonicdefhat}
\begin{aligned}
\begin{aligned}
\hat{\psi}(z)\, \hat{e}(w) &  \sim  \hat{\varphi}_3(\Delta)\, \hat{e}(w)\, \hat{\psi}(z) \\ 
\hat{\psi}(z)\, \hat{f}(w) & \sim \hat{\varphi}_3^{-1}(\Delta)\, \hat{f}(w)\, \hat{\psi}(z) 
\end{aligned}
&\qquad
\begin{aligned}
\hat{e}(z)\, \hat{e}(w) & \sim    \hat{\varphi}_3(\Delta)\, \hat{e}(w)\, \hat{e}(z) \\
 \hat{f}(z)\, \hat{f}(w) &  \sim    \hat{\varphi}_3^{-1}(\Delta)\, \hat{f}(w)\, \hat{f}(z) 
\end{aligned}\\
[\hat{e}(z)\,, \hat{f}(w)]  & \sim  - \frac{1}{\hat{\sigma}_3}\, \frac{\hat{\psi}(z) - \hat{\psi}(w)}{z-w} \ , 
\end{aligned}
\end{equation}
The hat over $\hat\varphi_3,\hat\sigma_3,\dots$ means that these functions are related to the un-hatted ones simply by substitution
\begin{equation}
h_i \leftrightarrow \hat{h}_i \qquad \textrm{and}\qquad \psi_0\leftrightarrow \hat{\psi}_0\ .
\end{equation} 

A priori the parameters in $\hCY$ are independent from those in $\CY$.
However in \cite{Gaberdiel:2017hcn, Gaberdiel:2018nbs}, a relation between the two set of parameters was imposed, in order to match with the $\mathcal{N}=2$ $\mathcal{W}_{\infty}$ algebra:
\begin{equation}\label{N=2relation}
\mathcal{W}^{\mathcal{N}=2}_{\infty}: \qquad h_i=\hat{h}_i \qquad \textrm{and}\qquad \psi_0+\hat{\psi}_0=-\frac{1}{h_1 h_3} \,.
\end{equation}
A quick way to see this is the following. 
Recall that the two parameters $(c,\lambda)$ and $(\hat{c},\hat{\lambda}=1-\lambda)$ of the left and right $\mathcal{W}_{1+\infty}$ algebras in (\ref{eq:bos-subalg}) are related by $N\leftrightarrow k$ in (\ref{eq:c-and-lambda}). 
On the other hand, since $\{h_i\}$ in (\ref{h123}) are invariant under $N\leftrightarrow k$,  this means that the left and right $\mathcal{W}_{1+\infty}$ algebras can have the same $\{h_i\}$ parameters. 
This has also been confirmed by checking $\mathcal{W}^{\mathcal{N}=2}$ charges of various twin plane partition configurations.
For the rest of this section, the relations (\ref{N=2relation}) will be assumed to hold. 
Nevertheless, we will keep distinct $h_i$ and $\hat h_i$ in the following, since many of the equations will generalize naturally to our new two-parameter family of algebras, where $\hat h_i$ and $h_i$ will be related in a nontrivial way.

\begin{figure}[h!]
	\centering
	\begin{tabular}{c}
	\includeCroppedPdf[width=.4\textwidth]{"xleft"}\qquad
	\includegraphics[width=.4\textwidth
		]{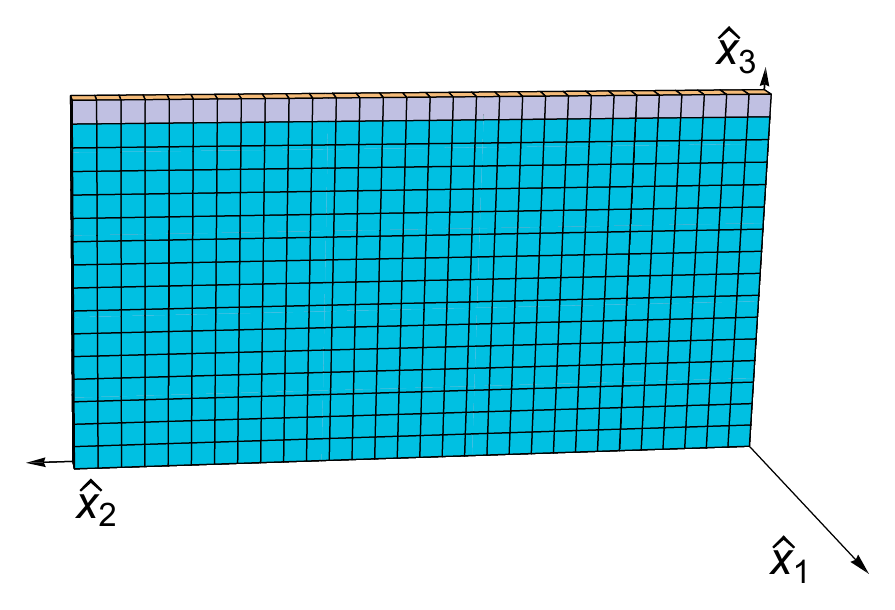}
	\end{tabular}	
	\caption{$\blacksquare=(\square,\bar{\square})$ as a twin plane partition. Left (Right): from perspective of $\mathcal{Y}$ ($\hat{\mathcal{Y}}$)
	}
	\label{figxleftright}
\end{figure}

As reviewed above in Section \ref{sec:bosonic-PP},  the affine Yangian of $\fgl_1$ admits an action on standard plane partitions, and therefore the bosonic subalgebra $\CY\oplus \hCY$ admits an action on pairs of plane partitions. We denote their coordinate systems by  $x_i$ and $\hat x_i$, respectively.
To proceed, we need to look at the fermionic generators of $\CW_\infty^{(\CN=2)}[\lambda]$.
Recall that the fermionic generators transforms as bi-representations, e.g.\ $(R, \overline{R^{t}})$, of the two bosonic $\CW_{1+\infty}[\lambda] \oplus \CW_{1+\infty}[1-\lambda]$, therefore of $\CY\oplus \hCY$.
Recall that a generic representation of a single affine Yangian of $\mathfrak{gl}_1$ is labeled by three Young diagrams $(\lambda_1,\lambda_2,\lambda_3)$, along the three directions. 
Without loss of generality, the bi-representation $(R, \overline{R^{t}})$ can be rewritten as a pair of plane partition representation
\begin{equation}
\mathcal{Y}: \quad (0,R,0) \qquad \textrm{and} \qquad \hat{\mathcal{Y}}: \quad (0,\overline{R^t},0) \,.
\end{equation}
Namely, The Young diagram $R$ is identified with the asymptotic shape of the plane partition of $\CY$ along the $x_2$-direction, while $R^\star$ with the asymptotic shape of the plane partition of $\hCY$ along the $\hat x_2$-direction.
For example, the bi-minimal  $({\square},\overline{{\square}})$ as twin-plane-partition configuration is shown in Fig. \ref{figxleftright}. More examples can be found in \cite{Gaberdiel:2018nbs}.\footnote{
We are free to view the high-wall as a ``platform'' by rotating the $\hat x_1$ and $\hat x_3$ axes. In fact, we will find that this is more natural when discussing the relation to toric Calabi-Yau geometry.
}

As explained in detail in \cite{Gaberdiel:2018nbs}, one can view the conjugate-minimal representation in terms of a ``high wall" plane partition. 
Furthermore, this high wall is situated in some sense ``behind" the boundary of the room where standard plane partitions are defined, i.e.\ at $x_{1,3}<0$ (or $\hat{x}_{1,3}<0$). 
As a consequence, the asymptotics associated to the conjugate representations can coexist with those of the regular representations: the regular representation describes the asymptotics in the quadrant with $x_{1,3}>0$ whereas the conjugate one describes the asymptotics in the quadrant with $x_{1,3}<0$.

The separate status of the $\overline{\square}$ representation from the Young diagrams built out of ${\square}$ may appear puzzling at first.
In the familiar representation theory of $A_{\mathfrak{n}}$ Lie algebras, conjugate-minimal representations can be obtained from $\mathfrak{n}$-th tensor powers of the minimal representation. 
The unusual splitting of an asymptotic Young diagram into ``minimal" and ``anti-minimal" parts is due to the lack of a well-defined counterpart of $\mathfrak{n}$ in the affine setting, which leads to the lack of a relation between the two, in contrast to the familiar case of $A_{\mathfrak{n}}$ Lie algebra representation theory.
In view of this, it is also natural that $\overline\square$ lives is located at $\hat x_1,\hat x_3<0$: this is the usual condition on the shapes of Young diagrams that forces all conjugate-minimal columns in the diagram to be on the left of shorter columns.

Pairs of plane partitions whose asymptotic shapes enjoy relation (\ref{eq:Neq2TPPcondition}) were termed \emph{twin-plane-partitions} in \cite{Gaberdiel:2018nbs}.
Due to the relation between their asymptotic shapes, we will sometimes say that the two plane partitions are ``glued'' along the  direction $x_2\sim \hat x_2$.
Twin plane partitions are characterized both by their asymptotics (both along ``internal" directions $x_2,\hat x_2$ and along ``external" ones), and by the finite configurations of boxes ${\square}$ and hatted boxes $\hat{\square}$ in their interior on either side.
The external asymptotic behavior defines the module, while interior configurations correspond to different states of a module.

\subsubsubsection{Vacuum module}

The vacuum module is defined by trivial asymptotics along external directions, but possibly nontrivial asymptotics along internal ones.

\begin{figure}[h!]
\begin{center}
\includegraphics[width=0.65\textwidth]{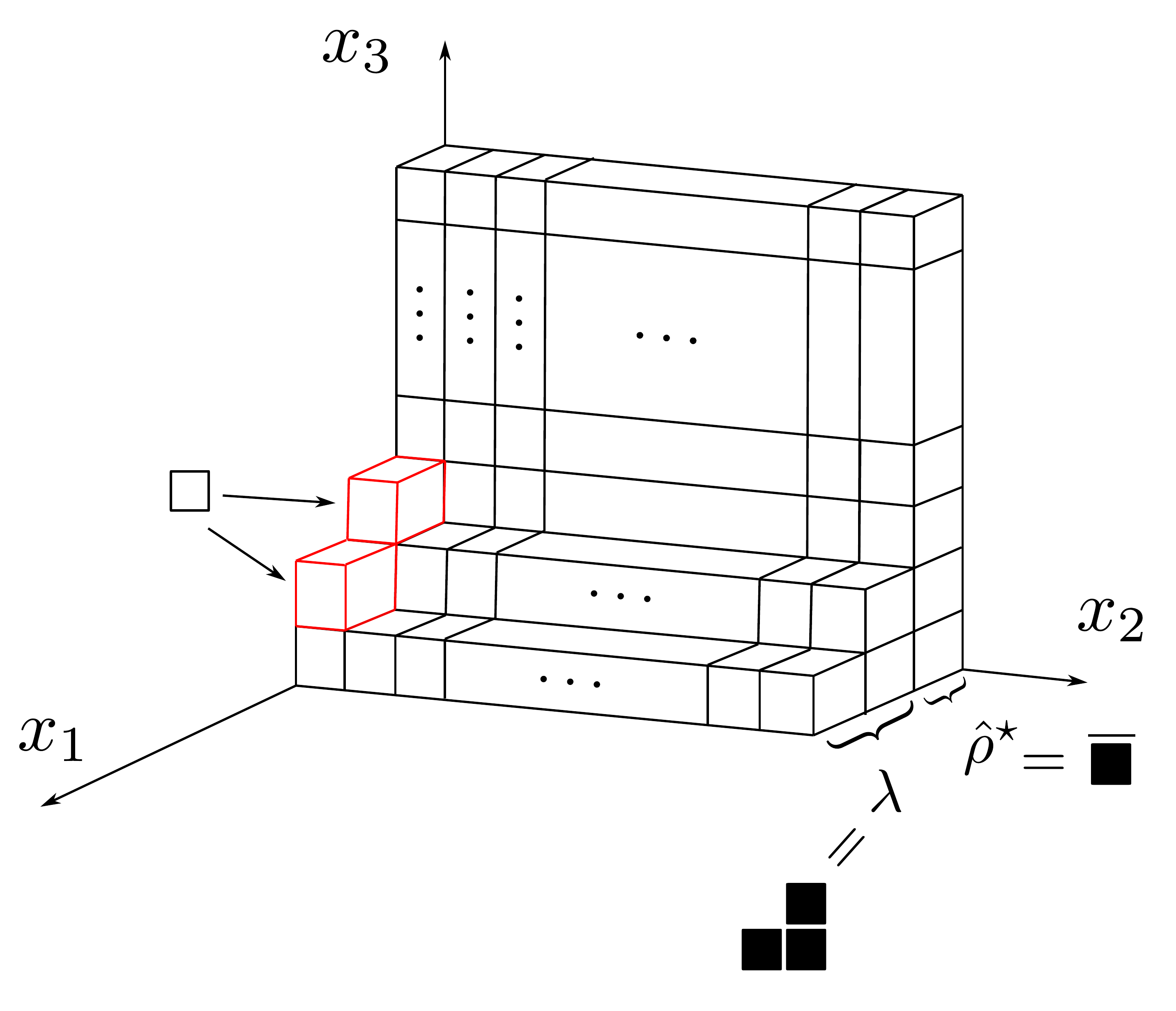}
\caption{Conventions for the description of internal asymptotics of twin-plane partitions in the vacuum module. Showing the un-hatted side only.}
\label{fig:partition-splitting}
\end{center}
\end{figure}

To characterize twin plane partitions in the vacuum module more precisely, let us introduce some notation regarding asymptotics along the $x_2$ and $\hat x_2$ internal directions.
We will use the pair 
\begin{equation}
(\lambda,\hat\rho^\star)
\end{equation} to characterize the asymptotic shape of a partition of boxes along $x_2$. $\lambda$ should be thought as a Young diagram built from copies of the minimal representation (and not to be confused with the 't Hooft coupling), whereas $\hat\rho^\star$ should be thought as its analogue for copies of the anti-minimal representation. 
Similarly, on the hatted side the conjugate representations 
\begin{equation}
(\hat\rho , \lambda^\star)
\end{equation} must characterize the asymptotic shape of a partition of hatted boxes along $\hat x_2$. The fact that these representations are the conjugates of those describing $x_2$-asymptotics follows from the analysis of the vacuum character decomposition reviewed above.
Now $\hat\rho$ should be thought as a Young diagram built from copies of the (hatted) minimal representation, whereas $\lambda^\star$ should be thought as its analogue for copies of the anti-minimal representation.
A schematic depiction of these conventions is given in Figure \ref{fig:partition-splitting}.

To summarize, in the vacuum module each twin-plane-partition configuration can be viewed as a pair of plane partitions with asymptotics 
\begin{equation}
(0, \lambda\otimes {\hat{\rho}}^{\, \star} , 0) \qquad \textrm{and} \qquad (0, {\lambda}^{\star} \otimes \hat{\rho},0) \ .
\end{equation}

\subsubsubsection{Generic module}

A generic representation is labeled by four asymptotic partitions $(\mu_1,\mu_3,\hat{\mu}_1,\hat{\mu}_3)$.
A state in this representation is a pair of plane partitions with asymptotics
\begin{equation}
(\mu_1,\lambda\otimes {\hat{\rho}}^{\, \star}, \mu_3) \qquad \textrm{and} \qquad (\hat{\mu}_1, {\lambda}^{\star}\otimes \hat{\rho},\hat{\mu}_3) \ .
\end{equation}
In this paper we will focus on twin-plane-partition configurations with trivial asymptotics along $x_1$, $x_3$, $\hat{x}_1$ and $\hat{x}_3$, i.e.\ $\mu_1=\mu_3=\hat{\mu}_1=\hat{\mu}_3=\emptyset$. These are the configurations that appear in the vacuum module.

\subsubsection{From twin plane partitions to building blocks of the glued affine Yangian}
\label{eq:heisenberg-intuition}

The three generators of the affine Yangian of $\mathfrak{gl}_1$ have the following action on single boxes ${{\square}}$ in a plane partition:
\begin{equation}\label{N2BBleft}
{{\square}}: \qquad \qquad e:\, \textrm{creation} \qquad  \psi:\, \textrm{charge}  \qquad  f:\, \textrm{annihilation} 
\end{equation}
Similarly for the generators of $\hat{\mathcal{Y}}$, we have a corresponding action on hatted single boxes $\hat\square$:
\begin{equation}\label{N2BBright}
\hat{{{\square}}}: \qquad \qquad \hat{e}:\, \textrm{creation} \qquad  \hat{\psi}:\, \textrm{charge}  \qquad  \hat{f}:\, \textrm{annihilation} 
\end{equation}

For the fermionic part, the decomposition (\ref{N2decomp}) suggests that the building blocks of the internal legs have the following asymptotic shapes\footnote{
More appropriately, we should write $\blacksquare\equiv ({\square},\hat{\overline {\square}})$ to denote that the conjugate minimal asymptotics is on the hatted side. However we will stick to the lighter notation since there is no risk of confusion on this point.}
\begin{equation}
\blacksquare\equiv ({\square},\overline {\square})
\qquad \textrm{and}\qquad 
\overline{\blacksquare}\equiv (\overline {\square},{\square})\ .
\end{equation}

Correspondingly, it is natural to introduce Yangian generators corresponding to creation, annihilation and counting of these new building blocks
\begin{equation}\label{N2BBfermion}
\begin{aligned}
\blacksquare: \qquad \qquad x:\, \textrm{creation} \qquad  P:\, \textrm{charge}  \qquad  y:\, \textrm{annihilation} \\
\overline{\blacksquare}: \qquad \qquad \bar{x}:\, \textrm{creation} \qquad  \bar{P}:\, \textrm{charge}  \qquad  \bar{y}:\, \textrm{annihilation} \\
\end{aligned}
\end{equation}
where $\{x, y, \bar{x}, \bar{y}\}$ have mode expansion: 
\begin{equation}\label{xmode}
x(z) = \sum_{r=\frac{1}{2}}^{\infty} \, \frac{x_r}{z^{r+\frac{1}{2}}} \qquad y(z) = \sum_{r=\frac{1}{2}}^{\infty} \, \frac{y_r}{z^{r+\frac{1}{2}}} \qquad \bar{x}(z) = \sum_{r=\frac{1}{2}}^{\infty} \, \frac{\bar{x}_r}{z^{r+\frac{1}{2}}} \qquad \bar{y}(z) = \sum_{r=\frac{1}{2}}^{\infty} \, \frac{\bar{y}_r}{z^{r+\frac{1}{2}}} 
\end{equation}

Just as for the bosonic case ((\ref{Wepsif}) with  (\ref{generating})), the map between the modes of $\{x,y,\bar{x},\bar{y}\}$ and the fermionic fields in $\mathcal{N}=2$ $\mathcal{W}_{\infty}$ algebra is \cite{Gaberdiel:2018nbs}:
\begin{equation}\label{Wx}
W^{(s)+}_{-3/2} \sim x_{s-\frac{1}{2}} \,, \qquad 
W^{(s)-}_{-3/2} \sim \bar{x}_{s-\frac{1}{2}} \,,\qquad
W^{(s)+}_{3/2} \sim y_{s-\frac{1}{2}} \,, \qquad W^{(s)-}_{3/2} \sim \bar{y}_{s-\frac{1}{2}}  \,,
\end{equation}
where $\sim$ means up to higher order corrections and  $W^{(s)\pm}$ are the two fermionic fields in the $\mathcal{N}=2$ multiplet whose bottom component has spin $s$, i.e. $W^{(s)\pm}$ have spin $s+\frac{1}{2}$.
(In particular, the supercharges $G^{\pm} \equiv W^{(1)\pm}$ are in the multiplet of spin $s=1$.)
To compare more directly to the bosonic map (\ref{Wepsif}) and later constructions with gluing fields of generic conformal dimension, it is more transparent to rewrite (\ref{Wx}) as
\begin{equation}\label{Vx}
V^{(s)+}_{-3/2} \sim x_{s-1} \,, \qquad 
V^{(s)-}_{-3/2} \sim \bar{x}_{s-1} \,, \qquad
V^{(s)+}_{3/2} \sim y_{s-1} \,, \qquad 
V^{(s)-}_{3/2} \sim \bar{y}_{s-1}  \,,
\end{equation}
where $V^{(s)\pm}$ has spin $s$, i.e.\ conformal dimension $s$, and $V^{(s)+}$ and $V^{(s)-}$ are conjugate to each other. 
Again, just as for the bosonic generators, the leading modes, i.e.\ the $\frac{1}{2}$ modes of  $\{x, y, \bar{x}, \bar{y}\}$ are 
the only modes that are ``outside the wedge":\footnote{
The modes ``inside the wedge" are those whose level $m$ 
(%defined by $[L_0, V^{(s)}_m]=-m V^{(s)}_m$
i.e.\ minus of eigenvalue of $L_0$) has absolute value  less than its spin: $V^{(s)}_{|m| < s}$ and they annihilate the vacuum. The modes ``outside the wedge" are $V^{(s)}_{|m|\geq s}$.}
\begin{equation}\label{Gx}
G^{+}_{-3/2} \sim x_{\frac{1}{2}} \,,\qquad  
G^{-}_{-3/2} \sim \bar{x}_{\frac{1}{2}} \,, \qquad 
G^{+}_{3/2} \sim y_{\frac{1}{2}} \,, \qquad 
G^{-}_{3/2} \sim \bar{y}_{\frac{1}{2}} \,.
\end{equation}
Namely, $x_{\frac{1}{2}}$ and $\bar{x}_{\frac{1}{2}}$ are the only modes that do not kill the vacuum. 

This suggests that the full set of generators of the $\CN=2$ affine Yangian algebra consists of those in (\ref{N2BBleft}), (\ref{N2BBright}), and (\ref{N2BBfermion}). (This idea will extend beyond the ${\cal N}=2$ case to the whole family of algebras that we construct in this paper).
Their action is easiest to describe on the ground state of the vacuum module, i.e.\ the empty configuration. 
The single-box bosonic raising operators $e,\hat e$ add single boxes (resp. hatted boxes) to the left and right plane partitions. 
Among the gluing operators, the creation operators $x(u)$ create rows (walls on the hatted side) that change the asymptotics along the internal gluing direction, thereby ``gluing" the left and right plane partitions. 
In particular, it adds a box to the asymptotic partition $\lambda$ along the $x_2$ direction, and hence simultaneously add an anti-box to the asymptotic partition $\lambda^\star$ along the $\hat{x}_2$ direction. 
Similarly, the creation operators $\bar{x}(u)$ create rows along the $\hat{x}_2$ direction from the perspective of $\hat{\rho}$, and simultaneously add walls to the asymptotics ${\hat{\rho}}^{\,\star}$  with respect to the unhatted modes.\footnote{
More accurately, $x,\bar x$ can also affect twin-plane-partitions by \emph{destroying} walls or rows, since $\square\otimes\overline\square = 1 \oplus \textrm{adj} $ includes the trivial representation. We will elaborate on this below.
}

\subsubsubsection{Charge functions of twin plane partitions}
\label{sec:psiforsingleglue}

A generic twin-plane partition from the vacuum module can be built recursively starting from the vacuum (the empty configuration), and contains the following components:
\begin{itemize}
\item A bi-representation $(\lambda, {\lambda}^{\star})$,  generated by $x$.
\item A bi-representation $({\hat{\rho}}^{\,\star}, \hat{\rho})$, generated by $\bar{x}$.
\item A collection $\mathcal{E}$ of individual $\square$s in the left plane partition, generated by $e$.
\item A collection $\hat{\mathcal{E}}$ of individual $\hat{\square}$s in the right plane partition, generated by $\hat e$.
\end{itemize}
Namely, a generic twin-plane-partition $\Lambda$ can be labeled by the quartet $(\lambda, \hat{\rho},\mathcal{E},\hat{\mathcal{E}})$.

As reviewed previously, a plane partition configuration is uniquely characterized by its charge function $\psi(u)$.
Similarly, a twin-plane-partition configuration is uniquely characterized by the charge functions $\psi(u)$ and $\hat{\psi}(u)$ of the two bosonic affine Yangians.
There are in fact four charge functions  for twin plane partitions:
\begin{equation}
(\bm{\Psi}_{\Lambda}(z) , \hat{\bm{\Psi}}_{\Lambda}(z)) \qquad \textrm{and} \qquad (\textbf{P}_{\Lambda}(z) , \bar{\textbf{P}}_{\Lambda}(z)) 
\end{equation}
defined by
\begin{equation}\label{eigendef}
\begin{aligned}
\psi(u) \, |\Lambda\rangle &= \bm{\Psi}_\Lambda(u) \, |\Lambda\rangle\ , \\
\hat{\psi}(u) \, |\Lambda\rangle &= \hat{\bm{\Psi}}_\Lambda(u) \, |\Lambda\rangle\ ,
\end{aligned}
\qquad \textrm{and}\qquad
\begin{aligned}
P(u) \, |\Lambda\rangle &= \textbf{P}_\Lambda(u) \, |\Lambda\rangle\ , \\
\bar{P}(u) \, |\Lambda\rangle &= \bar{\textbf{P}}_\Lambda(u) \, |\Lambda\rangle\ .
\end{aligned}
\end{equation}
As it turns out, the second pair of charges is not independent of the first pair.
Nevertheless, it is convenient to work with both types of charges in order to fix all the OPEs from the action of the algebra on twin plane partitions.
The charge functions have the following general form
\begin{equation}\label{XLambda}
\begin{aligned}
\bm{\Psi}_\Lambda(u) &= \psi_0(u)\, \Biggl\{
\prod_{\blacksquare \in \lambda}  \psi_{\,\blacksquare}(u) \,
\prod_{\overline{\blacksquare}\in \hat{\rho}} \psi_{\,\overline{\blacksquare}}(u) \, 
\prod_{ {\square} \in \mathcal{E}} \psi_{\,{\square}}(u) \, 
  \Biggr\} \ ,\\
\hat{\bm{\Psi}}_\Lambda(u) &= \hat{\psi}_0(u)\, \Biggl\{
\prod_{\blacksquare \in \lambda}  \hat{\psi}_{\,\blacksquare}(u) \,
\prod_{\overline{\blacksquare}\in \hat{\rho}} \hat{\psi}_{\,\overline{\blacksquare}}(u) \, 
\prod_{\hat{{\square}} \in \hat{\mathcal{E}} } \hat{\psi}_{\hat{{\square}}}(u)  \Biggr\} \ ,\\
\textbf{P}_\Lambda(u) &= P_0(u)\, \Biggl\{
\prod_{\blacksquare \in \lambda}  P_{\,\blacksquare}(u) \,
\prod_{\overline{\blacksquare}\in \hat{\rho}} P_{\,\overline{\blacksquare}}(u) \, 
\prod_{ {\square} \in \mathcal{E}} P_{\,{\square}}(u) \, 
\prod_{\hat{{\square}} \in \hat{\mathcal{E}} } P_{\hat{{\square}}}(u)  \Biggr\} \ ,\\
\bar{\textbf{P}}_\Lambda(u) &= \bar{P}_0(u)\, \Biggl\{
\prod_{\blacksquare \in \lambda}  \bar{P}_{\,\blacksquare}(u) \,
\prod_{\overline{\blacksquare}\in  \hat{\rho}} \bar{P}_{\,\overline{\blacksquare}}(u) \, 
\prod_{ {\square } \in \mathcal{E}} \bar{P}_{\,{\square}}(u) \, 
\prod_{\hat{{\square}} \in \hat{\mathcal{E}} } \bar{P}_{\hat{{\square}}}(u)  \Biggr\} \ .
\end{aligned}
\end{equation}
The various building blocks for the two charge functions can be found in Table \ref{tab2}, with the replacements $\rho=1/2, p=0$ and $S_0(u) = 1$. 

All the building blocks in (\ref{XLambda}) can be derived starting from two basic ones $\psi_{\square}(u)$ and $\hat{\psi}_{\hat{\square}}(u)$.
For example, consider the next simplest ones, $\psi_{\blacksquare}(u)$ and $\hat{\psi}_{\blacksquare}(u)$.
Consider the first $\blacksquare$ one can add, which as twin plane partition configuration is described by Fig.~\ref{figxleftright}. 
By the definition (\ref{psieig}), it has the charge function \begin{align}\label{eq:infinite-product-before}
\bm{\Psi}_{\blacksquare}(u)  & = \psi_0(u) \psi_{\blacksquare}(u) \quad\textrm{with}\quad
\psi_{\blacksquare}(u)=\prod_{{\square} \in \blacksquare} \varphi_3(u - h({\square})) =   \prod_{n=0}^{\infty} \varphi_3(u - n h_2) 
\end{align}
Evaluating the infinite product we obtain \cite{Prochazka:2015deb, Gaberdiel:2017hcn}
\begin{equation}\label{psiudef0}
 \psi_{\blacksquare}(u)=\frac{u(u +h_2)}{(u -h_1)(u -h_3)} =:
 \varphi_2(u)  
\end{equation}
One can check that (\ref{psiudef0}) reproduces the correct $\mathcal{W}$ charges for $\blacksquare$ \cite{Gaberdiel:2017hcn}.
Moreover note from this example that the in the glued algebra, the $\mathcal{S}_3$ symmetry (embodied by the $\varphi_3(u)$ function) in broken to a $\mathbb{Z}_2$ symmetry, satisfied by $\varphi_2(u)$.\footnote{
This residual $\mathbb{Z}_2$ symmetry is a special feature of $p=0$. It will be absent for general instances of our family of VOAs.}
On the other hand, the hatted charge for this configuration is 
\begin{equation}\label{conj}
\hat{\bm{\Psi}}_\blacksquare(u)=\hat{\psi}_0(u)\hat{\psi}_{\blacksquare}(u)\quad \textrm{with}\quad
\hat{\psi}_{\blacksquare}(u)=\hat \varphi_2^{-1}(-u-\hat{\sigma}_3\hat{\psi}_0)\ .
\end{equation}
Here $\hat\varphi_2$ is defined as in  (\ref{psiudef0}) with the replacement $h_i\to \hat h_i$.
See \cite{Gaberdiel:2017hcn} for a derivation of (\ref{conj}) based on the relation between $\mathcal{W}_{\infty}$ charges of a representation and those of its conjugate, namely, same for even spins and opposite for odd ones. 
See \cite{Gaberdiel:2018nbs} for a derivation of (\ref{conj}) using directly the high wall picture Fig.~\ref{figxleftright}.
In summary, both $\bm{\Psi}(u)$ and $\hat{\bm{\Psi}}(u)$ charges of the configuration $\blacksquare$ can be derived directly from naively stacking boxes, on one side only a single row and on the other a high wall. 
It is quite remarkable that the charge function of plane partitions are smart enough to regularize automatically.

The charge functions (\ref{psiudef0}) and (\ref{conj}) are for the leading $\blacksquare$, with coordinates $x_1=x_3=0$.
For $\blacksquare$ with higher $x_{1,3}$, one just shifts the variable $u$ in the charge functions by the appropriate coordinate functions, see Table \ref{tab2}, for more details see section.\ \ref{sec:TPPchargepsi}.
The conjugate ones $\hat{\bm{\Psi}}_{\blacksquare}$ and $\bm{\Psi}_{\bar{\blacksquare}}$ are obtained by $h_i \leftrightarrow \hat{h}_i$ and $\psi_0\leftrightarrow \hat{\psi}_0$. 
This way one obtains all building blocks for the charge functions $(\bm{\Psi}_{\Lambda}(u), \hat{\bm{\Psi}}_{\Lambda}(u))$.

All the remaining building blocks in (\ref{XLambda}) can be derived systematically, following the procedure to be outlined in 
section.\ \ref{sec:strategyoutline}.

\subsubsubsection{$(\psi, \hat{\psi})$ charge function of gluing operators}
\label{sec:OPEpsiGlue}

The OPE relations of the $x$ and $\bar{x}$ generators with $\psi$ and $\hat{\psi}$ can be derived by consistency with the charges of single-box operators dictated by the bosonic subalgebra. The argument goes as shown in Section \ref{sec:psiforsingleglue}, and the charges (OPE coefficients) are therefore
\begin{equation}\label{psiFBx0} 
\begin{aligned}
\psi(z) \, x(w)  &\sim  \varphi_2(\Delta) \, x(w) \,\psi(z)   \qquad
\hat{\psi}(z) \, x(w)  \sim  \hat\varphi^{-1}_2(-\Delta-\hat \sigma_3\hat{\psi}_0) \, x(w) \,\hat{\psi}(z) \\
\hat{\psi}(z) \, \bar{x}(w)  &\sim  \hat\varphi_2(\Delta) \, \bar{x}(w)\, \hat{\psi}(z)  \qquad
\psi(z) \, \bar{x}(w)  \sim  \varphi^{-1}_2(-\Delta-\sigma_3\psi_0) \, \bar{x}(w)\, \psi(z)  
\end{aligned}
\end{equation}
Although these expressions are  formally identical to those derived in \cite{Gaberdiel:2017hcn, Gaberdiel:2018nbs}, they actually differ since the relation between $\hat h_i, \hat\psi_0$ and $h_i,\psi_0$ depends on $p$ and on $\rho$ as explained above.
Here $\varphi_2(u)$ was defined in (\ref{psiudef0}), while $\hat\varphi_2$ and $\hat\sigma_3$ are obtained from $\varphi_2$ and $\sigma_3$ upon the substitution $h_i\to \hat h_i$. 
The ``$\sim$" sign means up to terms that are regular at either $z=0$ or $w=0$. 
Similarly the charges for the $y$ and $\bar{y}$ generators  are
\begin{equation}\label{psiFBy0} 
\begin{aligned}
\psi(z) \, y(w)  &\sim  \varphi_2^{-1}(\Delta) \, y(w) \,\psi(z)   \qquad
\hat{\psi}(z) \, y(w)  \sim  \hat\varphi_2(-\Delta-\hat\sigma_3\hat{\psi}_0) \, y(w) \,\hat{\psi}(z)\\
\hat{\psi}(z) \, \bar{y}(w)  &\sim \hat \varphi_2^{-1}(\Delta) \,\, \bar{y}(w) \,\hat{\psi}(z) \qquad
\psi(z) \, \bar{y}(w)  \sim  \varphi_2(-\Delta -\sigma_3 {\psi}_0 ) \,\, \bar{y}(w) \,\psi(z)  
\end{aligned}
\end{equation}
These relations are summarized in Fig.~\ref{OPEeverybody2}.

\subsection{Gluing two affine Yangians of $\mathfrak{gl}_1$ via twin-plane partitions}
\label{sec:strategyoutline}

We have now reviewed the main ideas behind the definitions of twin-plane partitions and of the enlarged set of operators  (\ref{N2BBleft}), (\ref{N2BBright})  and (\ref{N2BBfermion}) that act on them.
We will now proceed to describe the action of these operator in more detail, and explain how the algebra (i.e. the OPE coefficients) can be entirely fixed by consistency of this picture.

The definition of an action by the $\CN=2$ affine Yangian algebra on twin-plane partitions is inspired by the action of the $\fgl_1$ affine Yangian on standard plane partitions. 
It is based on two key principles.

The first one is that the action of various operators  $\mathcal{O} \in \{e,\hat e, x,\bar x,\dots\}$ on a state $|\Lambda\rangle$ corresponding to a twin-plane-partition $\Lambda$ takes the general form
\begin{equation}\label{Oform}
	\mathcal{O}(w) |\Lambda\rangle = \sum_{i} \frac{O[\Lambda \rightarrow \Lambda'_i]}{w-w^*_{i}} |\Lambda_{i}'\rangle
\end{equation}
where $ |\Lambda_{i}'\rangle$ corresponds a new twin-plane-partition that results by acting with $\mathcal{O}$ at the ``position" $w^*_i$. 
The coefficient $O[\Lambda \rightarrow \Lambda'_i]$ characterizes the residue of a charge function such as $\bm{\Psi}_\Lambda,\hat{\bm{\Psi}}_\Lambda$ or $\textbf{P}_\Lambda, \overline{\textbf{P}}_\Lambda$ depending on whether $\cal O$ is a single-box operator or a gluing operator.
Finally we sum over all possible $ |\Lambda_{i}'\rangle$. 

The second basic principle is that the set of allowed positions $w_*$ for the action of $\mathcal{O}$ on a given $\Lambda$ is determined by demanding that $\Lambda_{i}'$ is an honest twin-plane partition.
Note that these two are precisely the same principles that govern the action of the affine Yangian of $\mathfrak{gl}_1$ on the set of plane partitions. 

A remarkable property of the action on twin-plane-partitions is that these simple principles, together with self-consistency, are stringent enough to fix the pole $w^*$ as well as the the coefficients $O[\Lambda \rightarrow \Lambda'_i]$ completely. 
In turn, these characterize the algebra and allow us to compute explicit OPE relations.

Besides charge operators $(\psi(z) , \hat{\psi}(z))$ and $(P(z) , \bar{P}(z))$, the algebra includes creation and annihilation operators.
These come in two types: there are operators ($\bm{s}$) that create/kill single boxes and operators ($\bm{g}$) that create/kill non-trivial states along internal legs:
\begin{equation}
\begin{aligned}
\bm{s}: \quad e, f, \hat{e}, \hat{f}\\
\bm{g}: \quad x, y, \bar{x}, \bar{y}\\
\end{aligned}
\end{equation}
The goal is to fix the OPEs among all these operators.

The main result of \cite{Gaberdiel:2018nbs} is the proposal of a procedure to fix the poles of generators and  the residue coefficients in (\ref{Oform}), and hence the whole algebra.
Here we summarize the overall strategy, which we will once again adopt for our generalized construction, and point out future sections corresponding to each step.
\begin{enumerate}
\item\label{step1} OPEs of $(\psi(z) , \hat{\psi}(z))$ and single box operators $\bm{s}(w)$ are known. The are collected in the top and bottom parts of Fig.~\ref{OPEeverybody2}.
\begin{figure}[h!]
	\centering
	\includegraphics[trim=2cm 12cm 0cm 4cm, width=.8\textwidth]{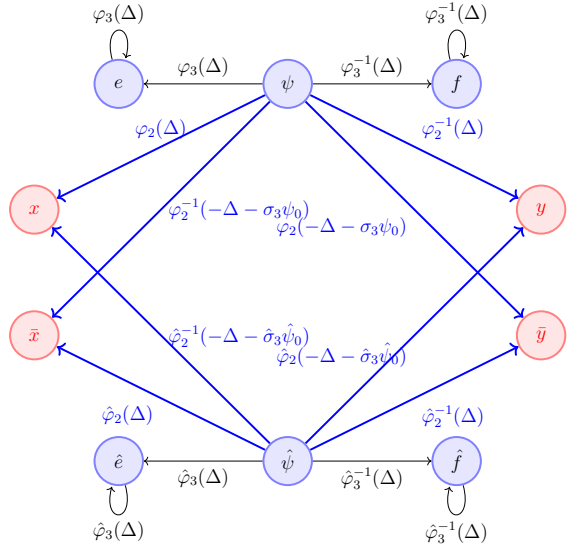}
	\caption{All generators of algebra from gluing, together with their charges with respect to $\psi(u)$ and $\hat{\psi}(u)$.}
	\label{OPEeverybody2}
\end{figure}

\item\label{step2} Determine set of ``allowed" twin plane partitions (Section~\ref{sec:TPPreview}) and compute their charge functions $(\bm{\Psi}_{\Lambda}(z) , \hat{\bm{\Psi}}_{\Lambda}(z))$.
 (Section~\ref{sec:psiforsingleglue})
 
\item\label{step3} The outcome of step-\ref{step2} immediately fixes the action of single-box operators $s$: both the allowed final states   $|\Lambda'_i\rangle$ (determined by poles $w^{*}_i$) and  the coefficients $\bm{s}[\Lambda \rightarrow \Lambda'_i]$ that appear in (\ref{Oform}).

\item\label{step4} 
The results of step-\ref{step2}, in particular the part in  Section~\ref{sec:psiforsingleglue},  also fix OPE relations between the charge function $(\psi(z) , \hat{\psi}(z))$ and  all the gluing operators $\bm{g}(w)$, which created non-trivial states along internal legs. 
See the thick blue arrows in Fig.~\ref{OPEeverybody2}. (Section~\ref{sec:OPEpsiGlue}.)

\item\label{step4} 
The results from step-\ref{step2} and step-\ref{step3} also determine, for any initial state  $|\Lambda\rangle$ and gluing operator $\bm{g}$, all allowed final states  $|\Lambda'_i\rangle$ (and therefore the associated poles $w^{*}_i$). 
Note that it is harder to fix the coefficients $\bm{g}[\Lambda \rightarrow \Lambda'_i]$.
(Section~\ref{sec:glueactiongeneric}.) 

\item\label{step5} Results from step-\ref{step4} allow us to fix almost completely the OPEs between single box operators $\{\e,f,\hat{e},\hat{f}\}$ and gluing operators $\{x,y,\bar{x},\bar{y}\}$. 
There are only two numerical constants that remain to be determined.\footnote{
Concretely, these are $a$ and $b$ in equation (\ref{Fourfunctions}) below.} See thick red arrows in  Figs.~\ref{OPEbosonicxy-conj} and \ref{figOPEbosonicfull}. 
\begin{figure}[h!]
	\centering
\includegraphics[trim=2cm 12cm 0cm 4cm, width=.8\textwidth]{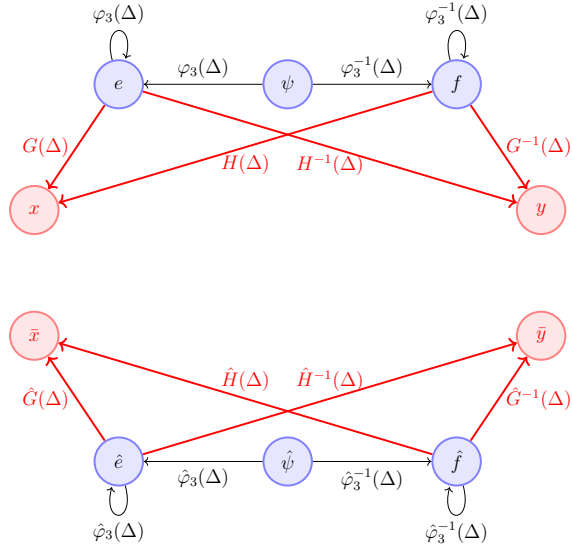}
\caption{OPEs of single box generators $(e,f)$ with $x$ and $y$, and those of the single hatted-box generators $(\hat{e},\hat{f})$ with $\bar{x}$ and $\bar{y}$. 
	}
	\label{OPEbosonicxy-conj}
\end{figure}
\begin{figure}[h!]
	\centering
\includegraphics[trim=2cm 12cm 0cm 4cm, width=.8\textwidth]{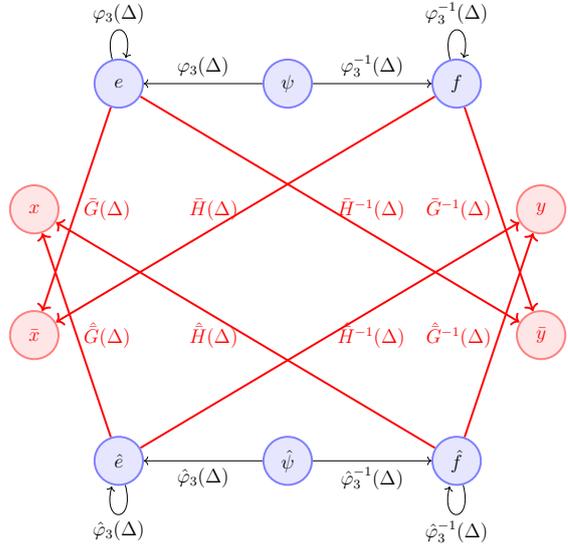}
\caption{OPEs of the single box generators $(e,f)$  with $\bar{x}$ and $\bar{y}$, and those of the single hatted-box generators $(\hat{e},\hat{f})$ with ${x}$ and ${y}$. }
	\label{figOPEbosonicfull}
\end{figure} (Section~\ref{sec:OPEcornerGluepartial}.)
\item\label{step6} To fix charge functions  $(\textbf{P}_{\Lambda}(z) , \bar{\textbf{P}}_{\Lambda}(z))$  and all the remaining OPEs:
(Section~\ref{sec:residue})
\begin{enumerate}
\item\label{step71} Fix single box contributions to $(\textbf{P}_{\Lambda}(z) , \bar{\textbf{P}}_{\Lambda}(z))$ and the remaining freedom in OPEs between all the single box operators $\{\e,f,\hat{e},\hat{f}\}$ and the gluing operators $\{x,y,\bar{x},\bar{y}\}$ left in step-\ref{step5}. (Section~\ref{sec:PboxandG}.)
\item\label{step71bis}
	The results on the single box contributions to $(\textbf{P}_{\Lambda}(z) , \bar{\textbf{P}}_{\Lambda}(z))$ from step-\ref{step71} immediately give us the OPEs between $(P, \bar{P})$ and the four single box generators $\{\e,f,\hat{e},\hat{f}\}$. See thick blue lines in  Figs.~\ref{OPEfermionicPe}.
\begin{figure}[h!]
	\centering
\includegraphics[trim=2cm 12cm 0cm 4cm, width=.8\textwidth]{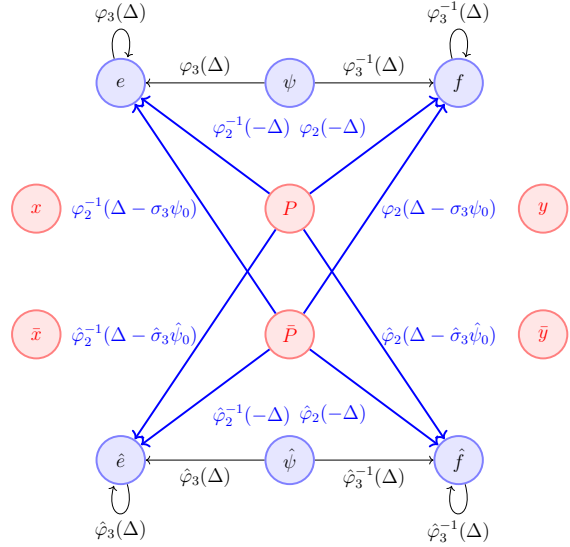}
\caption{OPEs between $(P(z), \bar{P}(z))$ and the four single box generators $\{\e,f,\hat{e},\hat{f}\}$.}
	\label{OPEfermionicPe}
\end{figure}
(Section~\ref{sec:OPEPe}.)
\item\label{step72} Use results of step-\ref{step71} to fix OPEs between $(P, \bar{P})$ and gluing operators and self-OPEs between gluing operators. See thick red lines in Figs.~\ref{OPEfermionicxy}.
\begin{figure}[h!]
	\centering
\includegraphics[trim=2cm 12cm 0cm 4cm, width=.8\textwidth]{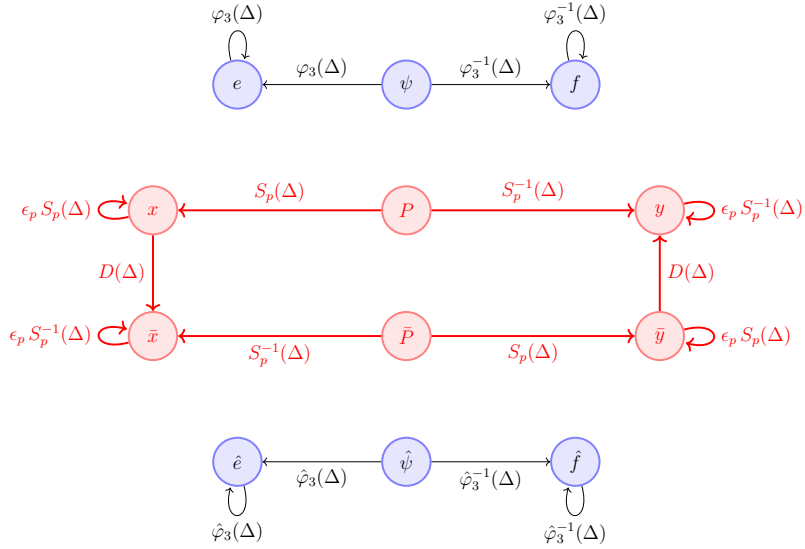}
\caption{OPEs among gluing operators. 
	}
	\label{OPEfermionicxy}
\end{figure}
(Section~\ref{sec:selfOPEglue}.)
\item\label{step73} Results from step-\ref{step71} and step-\ref{step72} allow us to fix contribution of non-trivial internal legs to $(\textbf{P}_{\Lambda}(z) , \bar{\textbf{P}}_{\Lambda}(z))$. (Section~\ref{sec:Pblackbox}.)
\item\label{step74} Results from step-\ref{step71} and step-\ref{step73} then immediately gives all the coefficients $\bm{g}[\Lambda \rightarrow \Lambda'_i]$. 
\item Result from (\ref{step74}) allows us to fix all the remaining OPEs. (Section~\ref{sec:remainingOPE}.)
\end{enumerate}
\end{enumerate}

\section{Two-parameter generalization of gluing}
\label{sec:2parameter}

In this section we describe a family of algebras constructed by a two-parameter ``gluing'' of two copies of the affine Yangian of $\fgl_1$.
The starting point of the construction is again the bosonic subalgebra $\CY\oplus \hCY$, augmented by certain gluing generators. The properties of gluing operators will be determined by self-consistency of a representation on (generalized) twin-plane-partitions.

The family of algebras that we will construct is parameterized by a \emph{shifting} modulus $\rho \in \frac{1}{2}\IZ_{\geq 0}$ and by a \emph{framing} modulus $p\in \{-1,0,1\}$. 
The $\CN=2$ affine Yangian is included in this family, and corresponds to 
\begin{equation}
\rho=1/2 \qquad \textrm{and}\qquad p=0 \,.
\end{equation}

\subsection{Moduli of the two bosonic subalgebras}

A priori, the parameters of the two bosonic Yangians $\mathcal{Y}$ and $\hat{\mathcal{Y}}$ are independent:
\begin{equation}
\begin{aligned}
&\mathcal{Y}: \qquad h_i \, (\textrm{with}\, \sum_i h_i=0) \quad \textrm{and} \quad \psi_0\\
&\hat{\mathcal{Y}}: \qquad \hat{h}_i \, (\textrm{with}\, \sum_i \hat{h}_i=0) \quad \textrm{and} \quad \hat{\psi}_0
\end{aligned}
\end{equation}
For the $\mathcal{N}=2$ affine Yangian that corresponds to $\mathcal{W}^{\mathcal{N}=2}_{\infty}$, the parameters of the two sides are related by (\ref{N=2relation}).
The two-parameter generalization corresponds to a modification of these two relations.

Now we still require that the shifted affine Yangian have two commuting subalgebras 
\begin{equation}
\mathcal{Y}\oplus \hat{\mathcal{Y}} \sim {\cal W}_{1+\infty}[\lambda] \oplus {\cal W}_{1+\infty}[\hat{\lambda}] 
\end{equation}
but with different relations between $\lambda$ and $\hat{\lambda}$. 

We shall stick to the convention of the $\CN=2$ construction in \cite{Gaberdiel:2017hcn, Gaberdiel:2018nbs} and use
$(e,\psi,f)$ to denote the  generators of the left affine Yangian $\mathcal{Y}$ and $(\hat{e},\hat{\psi},\hat{f})$ for the right one $\hat{\mathcal{Y}}$.
The left affine Yangian $\mathcal{Y}$ corresponds to $\mathcal{W}_{1+\infty}[\lambda]$ and has parameters given by (\ref{h123}) and $\psi_0=N$. 
Here we use the $(N,k)$ parametrization of $\mathcal{W}_{1+\infty}$, which is related to the central charge $c$ and 't Hooft coupling $\lambda$  as in (\ref{eq:c-and-lambda}).
The right affine Yangian $\hat{\mathcal Y}$ corresponds to $\mathcal{W}_{1+\infty}[\hat\lambda]$, whose parameters $\{\hat{h}_i\}$ and $\hat{\psi}_0$ are related to $\hat{N}$ and $\hat{k}$ via the hatted version of  (\ref{h123}). Likewise  $\hat{c}$ and $\hat{\lambda}$ are related to $\hat N$ and $\hat k$ by the hatted version of  (\ref{eq:c-and-lambda}). 
The total central of the glued algebra is
\begin{equation}
c^{\textrm{total}}=c+\hat{c}
\end{equation}
since the 
the total stress energy tensor is
\begin{equation}\label{Ttotal}
T^{\textrm{total}}(z)=T(z)+\hat{T}(z)
\end{equation}
where $T$ and $\hat{T}$ are the stress energy tensors of the bosonic $\mathcal{W}_{1+\infty}$ algebra of the left and right corners, respectively,  and $T(z)\hat{T}(w)\sim 0$.

\subsection{Shifting modulus: conformal dimension of gluing operators}
The shifting modulus $\rho$ controls the conformal dimension of the gluing generators, in the following sense. 

\subsubsection{Conformal dimension}

Let us compute the conformal dimension of the gluing operators $x$ transforming as  $({\square},\overline {\square})$ and $\bar{x}$ as $(\overline {\square},{\square})$. 
It is enough to focus on $x$, since $x$ and $\bar{x}$ have the same conformal dimension.
Recall that the conformal dimension of any representation would be (see \cite[eq. (3.5)]{Gaberdiel:2017dbk})
\be
	h = \frac{1}{2}\(\psi_2+\hat\psi_2\)
\ee
where $\psi_2,\hat\psi_2$ are the coefficients (of the eigenfunctions) of $\psi(u),\hat\psi(u)$, as defined in (\ref{generating}) and its counterpart for $\hat{\mathcal{Y}}$. 

The $x$ operator acting on the ground state creates $|\blacksquare\rangle$, whose charge is obtained by applying (\ref{XLambda}) 
\begin{equation}\label{xcharge}
	\bm{\Psi_\blacksquare}(u)=\psi_0(u)\varphi_2(u) \qquad \textrm{and}
	\qquad  
	\hat{\bm{\Psi}}_{\blacksquare}(u)=\hat{\psi}_0(u)\hat{\varphi}_2^{-1}(-u-\hat{\sigma}_3\hat{\psi}_0)\,,
\end{equation}
where 
\begin{equation}
\psi_0(u)=1+\sigma_3\frac{\psi_0}{u} \qquad \textrm{and} \qquad \hat{\psi}_0(u)=1+\hat{\sigma}_3\frac{\hat{\psi}_0}{u}\,.
\end{equation}
Expanding (\ref{xcharge}) we obtain the following mode eigenvalues  
\be
	\psi_2 = 1 - h_1 h_3\psi_0\qquad \textrm{and} \qquad 
	\hat\psi_2 = 1 - \hat h_1 \hat h_3\hat \psi_0\,,
\ee
which leads to
\begin{equation}
h_{\blacksquare}=1-\frac{h_1 h_3 \psi_0+ \hat h_1 \hat h_3 \hat{\psi}_0}{2}\,.
\end{equation}
Here we denoted by $h_{\blacksquare}$ the conformal dimension of the state $|\blacksquare\rangle$, which therefore coincides with the conformal dimension of $x$. Note that this should not be regarded as the conformal dimension of \emph{generic} rows (created at $x_{1,3}\neq 0$).
We see that (\ref{N=2relation}) would lead to $h_{\blacksquare}=\frac{3}{2}$ (i.e.\ the conformal dimension of the supercharges $G^{\pm}$). 
However, we can also obtain different  $h_{\blacksquare}$ if we modify  (\ref{N=2relation}).
\medskip

The shift parameter $\rho$ is defined by its relation to the conformal dimension of the $\blacksquare$ representation:
\begin{equation}\label{eq:rho-def}
	h_{\blacksquare} = 1+\rho \,.
\end{equation} 
Accordingly, the second relation in  (\ref{N=2relation}) is modified into
\begin{equation}\label{keyrel}
\boxed{ h_1 h_3 \psi_0+ \hat h_1 \hat h_3 \hat{\psi}_0 = - {2\rho} \ .  }
\end{equation}

The mode expansion of the gluing fields are
\begin{equation}\label{xmoderho}
x(z) = \sum_{r=\rho}^{\infty} \, \frac{x_r}{z^{r+1-\rho}} \qquad \textrm{and}\qquad y(z) = \sum_{r=\rho}^{\infty} \, \frac{y_r}{z^{r+1-\rho}} 
\end{equation}
and 
\begin{equation}\label{xbarmoderho}
\bar{x}(z) = \sum_{r=\rho}^{\infty} \, \frac{\bar{x}_r}{z^{r+1-\rho}} \qquad \textrm{and}\qquad  \bar{y}(z) = \sum_{r=\rho}^{\infty} \, \frac{\bar{y}_r}{z^{r+1-\rho}}  \,.
\end{equation}
A mode expansion for $P(z), \bar P(z)$ will be given in Section \ref{sec:P-psi-rel}.
The map between the modes of the gluing operators and the spin of fields in the $\mathcal{W}$ algebra basis (\ref{Vx}) is now modified into
\begin{equation}
V^{(s)+}_{-1-\rho} \sim x_{s-1}\,,  \qquad 
V^{(s)-}_{-1-\rho} \sim \bar{x}_{s-1} \,,  \qquad
V^{(s)+}_{1+\rho} \sim y_{s-1} \,, \qquad 
V^{(s)-}_{1+\rho} \sim \bar{y}_{s-1} \,.  
\end{equation}
The fields $V^{(s)\pm}$ have conformal dimension $s$ with respect to the total stress energy tensor (\ref{Ttotal}); and $V^{(s)+}$ and $V^{(s)-}$ are conjugate to each other.
As in the $\mathcal{N}=2$ case (\ref{Gx}), the leading modes $\{x_{\rho}, y_{\rho},\bar{x}_{\rho}, \bar{y}_{\rho}\}$ are the only modes that are ``outside the wedge":
\begin{equation}
V^{(\rho+1)+}_{-1-\rho} \sim x_{\rho} \,, \qquad 
V^{(\rho+1)-}_{-1-\rho} \sim \bar{x}_{\rho} \,,  \qquad V^{(\rho+1)+}_{1+\rho} \sim y_{\rho} \,, \qquad 
V^{(\rho+1)-}_{1+\rho} \sim \bar{y}_{\rho} \,.
\end{equation}
Namely, the lowest modes $x_{\rho}$ and $\bar{x}_{\rho}$ are the only ones that do not annihilate the vacuum.

\subsubsection{Shifted vacuum character with fermionic gluing operators}

Recall the vacuum character (\ref{eq:vac-char-fermi}) of the $\mathcal{N}=2$ affine Yangian, where the numerator was interpreted as the contribution from fermionic generators of conformal dimension $1+\frac{1}{2}$.
The shift induced by $\rho$ in the conformal dimension of these operators leads to the generalized vacuum character
\begin{equation}\label{chi0}
\chi^{\textrm{Full}}_0(q,y) = \prod_{n=1}^{\infty} \frac{(1+y\, q^{n+\rho})^{n}(1+\frac{1}{y}\, q^{n+\rho})^{n}}{(1-q^n)^{2n}} \ .
\end{equation}

Once again, we can study the decomposition of this character, to extract information on how gluing operators transform under the left and right $\mathcal{W}_{1+\infty}$ algebras. 
Plugging the character identity
\begin{equation}\label{charid}
\prod_{n=1}^{\infty} (1+y\, q^{n+\rho})^{n} = \sum_{R} y^{|R|}\chi_{R}^{({\rm w})\,  [\lambda]}(q) \cdot \chi_{R^\star}^{({\rm w})\, [\hat{\lambda}]}(q) \ ,
\end{equation}
where
\be\label{stardef}
R^{\star} \equiv \overline{R^T} 
\ee
and $\chi_R^{(\rm w)[\lambda]}(q)$ is the wedge part of $\mathcal{W}_{1+\infty}[\lambda]$ character for representation $R$ (see (\ref{chiwedge})),
into the vacuum character (\ref{chi0}), we find
\begin{equation}
\begin{aligned}
\chi^{\textrm{Full}}_0(q,y)&=\chi_{\rm pp}(q)^2 \left(\sum_{R} y^{|R|}\, \chi_{R}^{({\rm w})\,  [\lambda]}(q)  \chi_{R^\star}^{({\rm w})\, [\hat{\lambda}]}(q)\right)  \left(\sum_{{S}}\frac{1}{ y^{|{S}|}}\chi_{S^\star}^{({\rm w})\,  [\lambda]}(q) \chi_{S}^{({\rm w})\, [\hat{\lambda}]}(q)\right)\\
&=1+ \sum_{R} y^{|{R}|}\chi_{R}^{ [\lambda]}(q) \cdot \chi_{R^\star}^{ [\hat{\lambda}]}(q)+ \sum_{S} \frac{1}{y^{|S|}} \, \chi_{S^\star}^{ [\lambda]}(q) \cdot \chi_{S}^{[ \hat{\lambda}]}(q)+\cdots \ ,
\end{aligned}
\end{equation}
As in the $\mathcal{N}=2$ construction, we find that all fermionic generators come in representations of the form $(R\oplus S^\star ,R^\star \oplus S)$  under the left and right $\mathcal{W}_{1+\infty}$ algebras. 

\subsubsection{Shifted vacuum character with bosonic gluing operators}

In the vacuum character (\ref{chi0}), the gluing operators are fermionic. 
A priori, we could also have bosonic gluing operators, which gives the vacuum character 
\begin{equation}\label{eq:vac-char-bose}
	\chi^{\textrm{Full}}_0(q,y) 
	= 
	\prod_{n=1}^{\infty} 
	\frac{1}{(1-q^n)^{2n}  \(1-y q^{n+\rho}\)^{n}\(1-{y}^{-1}q^{n+\rho}\)^{n} } \ ,
\end{equation}
Now using the character identity
\begin{equation}\label{charidB}
\prod_{n=1}^{\infty} (1+y\, q^{n+\rho})^{-n} = \sum_{R} y^{|R|}\chi_{R}^{({\rm w})\,  [\lambda]}(q) \cdot \chi_{\bar{R}}^{({\rm w})\, [\hat{\lambda}]}(q) \ ,
\end{equation}
The vacuum character (\ref{eq:vac-char-bose}) is decomposed as
\begin{equation}\label{eq:char-decomp}
\begin{aligned}
	\chi^{\textrm{Full}}_0(q,y)
	&= \chi_{\rm pp}(q)^2 \left(\sum_{R} y^{|R|}\, \chi_{R}^{({\rm w})\,  [\lambda]}(q)  \chi_{\bar{R}}^{({\rm w})\, [\hat\lambda]}(q)\right) \left(\sum_{{S}}\frac{1}{ y^{|{S}|}}\chi_{\bar{S}}^{({\rm w})\,  [\lambda]}(q) \cdot \chi_{S}^{({\rm w})\, [\hat\lambda]}(q)\right)\\
&=1+ \sum_{R} y^{|{R}|}\chi_{R}^{ [\lambda]}(q) \cdot \chi_{\bar{R}}^{ [\hat\lambda]}(q)+ \sum_{S} \frac{1}{y^{|S|}} \, \chi_{\bar{S}}^{ [\lambda]}(q) \cdot \chi_{S}^{ [\hat\lambda]}(q)+\cdots \ .
\end{aligned}
\end{equation}
Thus, similar to the case with fermionic gluing operators, now the bosonic gluing operators come in representations of the form $(R\oplus \bar{S},\bar{R}\oplus S)$ under the left and right $\mathcal{W}_{1+\infty}$ algebras. 
The difference to keep in mind, compared to the fermionic case, is the absence of the \emph{transpose} for the Young diagram on the hatted side.

Here we considered the vacuum character with bosonic gluing operators without any apparent motivation.
We will see immediately that the boson/fermion nature of the gluing operators are correlated with the relative orientation of the two plane partitions.

\subsection{Framing modulus: relative orientation of two plane partitions}
\label{sec:3solutions}

The framing modulus parameterizes the relation between $h_i$ and $\hat h_i$ variables that separately characterize each of the commuting bosonic subalgebras $\CY$ and $\hCY$ respectively.
Since we will ``glue'' along the $x_2\sim \hat x_2$ direction, we set
\begin{equation}\label{h2}
	\boxed{\hat h_2 = h_2}
\end{equation}

To determine the relation between $(h_1,h_3)$ and $(\hat{h}_1,\hat{h}_3)$, we consider how to create a state with two $\blacksquare$ using the gluing operator $x(u)$, starting with $|\blacksquare\rangle$. 
Recall that in the ${\cal N}=2$ case $\hat h_i=h_i$, and it was found that $x\cdot x \, |\emptyset\rangle\sim 0$, in agreement with the fermionic nature of the gluing operators. 
This key fact emerged naturally from twin plane partitions, as follows. 
To begin with, acting with $x$ on the vacuum $|\emptyset\rangle$ creates a state with a single row of boxes $|\blacksquare\rangle$. 
Creation of a second row could take place either next to the first row along the $x_1$ direction, or along the $x_3$ direction.
However, simply applying $x$ on $|\blacksquare\rangle$ was found to annihilate the state, since the charge functions $(\psi,\hat\psi)$ of the resulting states would be incompatible with any admissible twin plane partition. 

It was found that, in order to create a second row, it was necessary to first create a ``bud'', consisting of a single box placed at position $h_i$ for $i=1$ or $3$, next to the first row. Then a second row along the $x_2$ direction could be consistently created by applying $x$ on the state $|\blacksquare+\square_{i}\rangle$ with $i=1,3$.
With this in mind, we therefore allow for the possibility that even in the more general case when $\hat h_i\neq h_i$, a bud with a certain number of $\square$'s may be necessary to create a second row.\footnote{
A detailed study of this will be performed in section~\ref{sec:minimalBud}, where we will find that except for $p=0$, buds could be asymmetric asymmetric 
between the $x_1$ and $x_3$ directions.}

Let $ s_1, s_3\in \mathbb{Z}_{\geq 0}$ denote the length of minimal buds required to add a second row, displaced in the $x_1$ and $x_3$ directions, respectively.
The transverse position where the row is created in the $(x_1,x_3)$-plane coincides with the $(x_1,x_3)$-coordinates of the pole in (\ref{Oform}) for the action of $x$ that creates the row, and is measured in integer units of $h_1, h_3$ on the unhatted side, or in integer units of $\hat h_1, \hat h_3$ on the hatted side.
The longitudinal displacement of the row (which corresponds to the length of the bud\footnote{
The  minimal $x_2$-displacement will be denoted the ``minimal bud''. In general, the bud can be longer than the minimal one.}) gives the $x_2$-coordinate of the pole and is measured in units of $h_2=\hat h_2$.
We will now derive a relation between $h_i$ and $\hat h_i$ by considering the simultaneous description of the pole of $x$ in both coordinate frames.

From the character analysis of the previous subsection, it is clear that gluing operators should transform in the $(\square,\overline\square)$ representation of the two components of the bosonic subalgebra (this is true both from the analysis of fermionic characters and of bosonic ones). 
If $x$ creates a $\square$ in the asymptotic Young diagram on the un-hatted side, it creates a $\overline \square$ in the asymptotic Young diagram on the hatted side.
Adopting conventions of \cite{Gaberdiel:2018nbs}, the transverse coordinates of the $\overline\square$'s must be \emph{negative}.\footnote{
This is natural given the definition of Young diagram. If $\square$ are stacked within the positive quadrant of the $(\hat x_1,\hat x_3)$-plane according to standard rules relating the length of a row compared to the previous one, then $\overline\square$ must be to the far left in the diagram, i.e. in a negative quadrant of the $\hat x_1,\hat x_3$-plane.}
If the $\square$ in the asymptotic Young diagram on the un-hatted side is displaced along the $x_1$ direction, there are two natural options for the displacement of $\overline\square$ in the Young diagram on he hatted side: either $-\hat h_1$ or $-\hat h_3$. 
The two are related by a change in the orientation of the volume form on the hatted room, and we fix the orientation in such a way that it is compatible with the one on the un-hatted side, i.e.\ displacing along the positive $x_1$ direction is correlated with displacing along the negative $\hat{x}_3$ direction.
This leads to the following general relation
\begin{equation}\label{chiralitySame}
h_1+s_1 \, h_2=-\hat{h}_3 \qquad \textrm{and} \qquad h_3+s_3 \, h_2=-\hat{h}_1 \,.
\end{equation}

These expressions correspond to the two possible positions of the poles of $x$ when acting on $|\blacksquare + \textrm{min.bud}\rangle$, for this  reason they include buds on the un-hatted side.
Taking the sum of these equations and using the properties $\sum_i h_i = \sum_i \hat h_i = 0$ together with the gluing condition $h_2=\hat{h}_2$ leads to
\be\label{eq:s1s3}
	s_1+s_3=2 \,.
\ee
Since $s_1 , s_3$ must be non-negative, there are only three solutions:
\begin{equation}\label{3solutions}
\begin{aligned}
(s_1,s_3)=(2,0):& \qquad\qquad \hat{h}_1=-h_3 \qquad \quad \,\,\hat{h}_3=h_3-h_2\\
(s_1,s_3)=(1,1):& \qquad\qquad \hat{h}_1=h_1 \qquad\qquad \, \hat{h}_3=h_3\\
(s_1,s_3)=(0,2):& \qquad\qquad \hat{h}_1=h_1-h_2 \qquad \hat{h}_3=-h_1
\end{aligned}
\end{equation}
which can be more efficiently expressed as
\begin{equation}\label{eq:h-hhat-p}
	\boxed{
	\hat h_1 = h_1 - p\, h_2 
	}
	\qquad \textrm{and}\qquad
	\boxed{
	\hat h_3 = h_3 + p\, h_2
	}
\end{equation} 
where 
\begin{equation}\label{eq:p-s-i-relation}
p\equiv \frac{s_3-s_1}{2}
\end{equation} 
Namely, the three cases in (\ref{3solutions}) correspond to
\begin{equation}
p=-1, 0, 1
\end{equation}
respectively.
When $p=0$ this reduces to $\hat h_i=h_i$, just like for the $\CN=2$ affine Yangian reviewed above, the other two cases are new. 
In subsection \ref{sec:statistics} we will give an alternative derivation of these three possibilities, from considerations on plane partitions and Fermi/Bose statistics of gluing operators.

\subsection{Correlation between framing and self-statistics of gluing operators}
\label{sec:self-statistics}

\subsubsection{Vacuum character expansion}
\label{sec:vacuum-char-analysis-bose}

We have argued above that there are three possible relative orientation between the two plane partitions, labeled by the framing modulus $p=0,\pm 1$.
Here we will show how the framing modulus dictates whether the gluing operators are fermionic or bosonic. 

Recall that the three cases (\ref{3solutions}) arise from three different scenarios when creating states that consist of two infinite rows $\blacksquare$.
What distinguished these three cases was the length of ``buds'' that are needed to create the next-minimal infinite row, as dictated by the general constraints of admissible twin plane partitions.
In all three cases, no bud was necessary to create the minimal row by acting with $x$ on the vacuum $|\emptyset\rangle$. 
However the next-minimal rows demand the presence of a bud, and furthermore these have different lengths $(s_1,s_3)$ depending on whether the next-minimal row is displaced along the $x_1$ direction, or along the $x_3$ direction. 
In the case $p=0$ we found $(s_1,s_3)=(1,1)$ while for $p=\pm1$ we found $(2,0)$ and $(0,2)$, see (\ref{3solutions}).

Now we show that our intuitive argument about lengths of buds implies a precise prediction for the vacuum character. 
First, since each single box in a bud contributes $1$ to the conformal dimension of a state, the difference between these three cases can be immediately seen from the $q$-expansion of their vacuum characters. (For all three cases each $x$ contributes $h=1+\rho$.)
Let $y$ be the fugacity counting the number of infinite rows $\blacksquare$, the contributions from configurations with two $\blacksquare$'s are
\begin{equation}\label{3solutionsL2}
\begin{aligned}
(s_1,s_3)=(2,0):& \qquad \qquad y^2(q^{2h+2}+q^{2h}+\cdots) \qquad \qquad (p=-1)\\
(s_1,s_3)=(1,1):& \qquad\qquad y^2(2q^{2h+1}+\cdots)\qquad \qquad \qquad(p=0)\\
(s_1,s_3)=(0,2):& \qquad\qquad y^2(q^{2h}+q^{2h+2}+\cdots)\qquad \qquad (p=1)
\end{aligned}
\end{equation}
where $h=1+\rho$ is the conformal dimension of the gluing operator\footnote{
It is important that this is the dimension of $x$, and {not} the infinite row $\blacksquare$ whose length may vary depending on its transverse position.} and  we have only written the leading terms, omitting descendants.

We immediately see that the case of $p=\pm 1$ is structurally different from the case of $p=0$,
 since they must have a different vacuum character. 
To proceed, we compare the vacuum characters in which the gluing operators are fermionic or bosonic:
\begin{eqnarray}
\chi^{\textrm{Fermion}}_0(q,y) &= &\prod_{n=1}^{\infty} \frac{(1+y\, q^{n+\rho})^{n}(1+\frac{1}{y}\, q^{n+\rho})^{n}}{(1-q^n)^{2n}} \label{chiF}\\
\chi^{\textrm{Boson}}_0(q,y) 
	&= &
	\prod_{n=1}^{\infty} 
	\frac{1}{(1-q^n)^{2n}  \(1-y q^{n+\rho}\)^{n}\(1-{y}^{-1}q^{n+\rho}\)^{n} } \label{chiB}
\end{eqnarray}
To match to the three cases (\ref{3solutionsL2}), we use the character identity 
\be\label{eq:char-id-fermionic}
\prod_{n=1}^{\infty} (1+y\, q^{n+\rho})^{n} = \sum_{R} y^{|R|}\chi_{R}^{({\rm w})\,  [\lambda]}(q) \cdot \chi_{R^\star}^{({\rm w})\, [\hat{\lambda}]}(q) 
\ee
\be\label{eq:char-id-bosonic}
\prod_{n=1}^{\infty} (1+y\, q^{n+\rho})^{-n} = \sum_{R} y^{|R|}\chi_{R}^{({\rm w})\,  [\lambda]}(q) \cdot \chi_{\bar{R}}^{({\rm w})\, [\hat{\lambda}]}(q) \ ,
\ee
and expand
\begin{equation}\label{eq:fermionic-char-buds}
\begin{split}
	& \prod_{n=1}^{\infty} (1+y\, q^{n+\rho})^{n} \\
	 = & 1 
	+ y \chi_{\blacksquare}^{({\rm w})\,  [\lambda]}(q) \cdot \chi_{\overline \blacksquare}^{({\rm w})\, [\hat{\lambda}]}(q) 
	+ y^2 
		\left(\chi_{\blacksquare\blacksquare}^{({\rm w})\,  [\lambda]}(q) \cdot \chi_{\overline{\substack{\blacksquare\\\blacksquare}}}^{({\rm w})\, [\hat{\lambda}]}(q) 
		+
		\chi_{{\substack{\blacksquare\\\blacksquare}}}^{({\rm w})\,  [\lambda]}(q) \cdot \chi_{\overline{\blacksquare\blacksquare}}^{({\rm w})\, [\hat{\lambda}]}(q) 
		\right)
		+ O(y^3)
	\\
	= & 1 
	+ y \left( q^{h} + \dots \right) 
	+ y^2 
	\left( 
		 2 q^{2h+1} + \dots 
	\right)
	+ O(y^3)
\end{split}
\end{equation}
and 
\begin{equation}\label{eq:bosonic-char-buds}
\begin{split}
	 & \prod_{n=1}^{\infty} (1+y\, q^{n+\rho})^{-n}
	 \\
	 = & 1 
	+ y \chi_{\blacksquare}^{({\rm w})\,  [\lambda]}(q) \cdot \chi_{\overline \blacksquare}^{({\rm w})\, [\hat{\lambda}]}(q) 
	+ y^2 
		\left(\chi_{\blacksquare\blacksquare}^{({\rm w})\,  [\lambda]}(q) \cdot \chi_{\overline{{\blacksquare\blacksquare}}}^{({\rm w})\, [\hat{\lambda}]}(q) 
		+
		\chi_{{\substack{\blacksquare\\\blacksquare}}}^{({\rm w})\,  [\lambda]}(q) \cdot \chi_{\overline{\substack{\blacksquare\\\blacksquare}}}^{({\rm w})\, [\hat{\lambda}]}(q) 
		\right)+ O(y^3)
	\\
	= & 1 
	+ y \left( q^{ h} + \dots \right) 
	+ y^2 
	\left( 
		 q^{2h} + q^{2h+2} + \dots 
	\right)
	+ O(y^3)
\end{split}
\end{equation}
where $h=1+\rho$ is the conformal dimension of the gluing operators, and the ellipses denote higher  powers of $q$. 
The term of interest is the leading order coefficient of $y^2$: the power $2h$ accounts for two gluing operators, and $2h+1$ accounts for two gluing operators plus a single box, etc.

Comparing (\ref{eq:fermionic-char-buds}) and (\ref{eq:bosonic-char-buds}) with (\ref{3solutionsL2}), we immediately see that the $p=0$ case, for all values of the shifting modulus $\rho$, corresponds to the \textit{fermionic} gluing operator, with vacuum character given by (\ref{chiF}) and the internal legs transforming as $(Y, \bar{Y}^{t})$; whereas the $p=\pm 1$ cases, for all values of the shifting modulus $\rho$, correspond to the \textit{bosonic} gluing operators, with vacuum character given by (\ref{chiB}) and internal legs transforming as $(Y,\bar{Y})$.

Indeed, this change in statistics is exactly what we expect. 
Let's first look at the case of $p=0$: the first application of $x$ on $|\emptyset\rangle$ acts by creating a ``minimal'' row at the origin, while the second one attempts to create a row displaced either along $x_1$, or along $x_3$. 
In either case it requires the addition of a single box to fill the whole length (see the discussion below in subsection \ref{sec:statistics}). For this reason $x\cdot x \, |\emptyset\rangle$ gives zero, reflecting the fermionic nature of this operator.

For the case of $p=\pm1$, from the minimal-bud analysis we find \emph{two distinct} configurations for the creation of the second row, encoded by the leading terms in the coefficient of $y^2$. 
The term $q^{2h}$ corresponds to two rows created by acting with $x\cdot x$ on the vacuum: these are rows along the $x_2\sim \hat x_2$ direction and stacked along $x_3\sim \hat x_1$ for $p=-1$, or along $x_1\sim \hat x_3$ for $p=1$.
The term with power $q^{2h+2}$ corresponds to the configuration obtained by acting twice with $x$ and twice with\footnote{
In fact, with any combination of two operators chosen at will from $\{e,\hat e\}$.} $e$, and this is precisely what we would expect for two rows stacked along the opposite directions, due to the relative slant of $x_1,\hat x_3$ for $p=-1$, or that of $x_3,\hat x_1$ for $p=1$  (see the discussion of Section \ref{sec:statistics}.)
Therefore, when $p=\pm1$, $x\cdot x \, |\emptyset\rangle$ no longer vanishes, in agreement with the bosonic nature of the gluing operator for these choices of the framing.

Besides these intuitive arguments, we will see below that the consistency of twin plane partitions forces exactly these types of configurations to have these conformal dimensions. 
This will provide an independent (and much more rigorous) check that our construction corresponds to fermionic gluing operators when $p= 0$ and bosonic when $p=\pm 1$.

Finally, we can further see that, since statistics can be either  fermionic or bosonic, these are the only two options, and they are realized by $p=0,\pm1$, and no other value of $p$. 
Recall that in section \ref{sec:3solutions} we showed that constraints from twin plane partitions (i.e.\ there cannot be box outside the room) allows only these three choices.
Then in section \ref{sec:pqweb}, we have seen that a map to the $(p,q)$ web also restricts us to these three choices.
Now we have yet  another reason why one should expect no other choices of framing.
Thus we conclude that our gluing construction exhausts all possible choices of framing.

\subsubsection{Gluing generators and generalized twin plane partitions}\label{eq:heisenberg-intuition-bis}

From the character decomposition, one again sees that for $p=0, \pm1$, all representations come from tensor powers of the two ``bi-minimal" building blocks,  transforming as 
\begin{itemize}
\item[$\blacksquare$]:\quad  minimal w.r.t.\ $\mathcal{Y}$ and anti-minimal w.r.t\ $\hat{\mathcal{Y}}$; 
\item[$\overline{\blacksquare}$]:\quad  anti-minimal w.r.t.\ $\mathcal{Y}$ and minimal w.r.t.\ $\hat{\mathcal{Y}}$. 
\end{itemize}
Therefore for both $p=0$ and $p=\pm1$, we introduce operators $x$ and $\bar{x}$ defined as creation operators of the two bi-minimals, with $x$ adding a box to $\lambda$, and $\bar{x}$ adding a box to $\rho$. 
The annihilation operators are  $y$ for $x$ and  $\bar{y}$ for $\bar{x}$.
These four operators, $(x,y,\bar{x},\bar{y})$, are the fermionic gluing generators for $p=0$ and bosonic ones for $p=\pm 1$.
Their transformation properties under the two copies of the affine Yangian of $\fgl_1$ are summarized by the following table
\begin{center}
\begin{tabular}{|c|c|c|}
\hline \\[-15pt]
\hbox{gluing operators} & \hbox{left ${\cal Y}$ } & \hbox{right $\hat{{\cal Y}}$} \\ \hline
&& \\[-12pt]
$x$ & \hbox{minimal} & \hbox{anti-minimal } \\
$\bar{x}$ &  \hbox{anti-minimal } & \hbox{minimal } \\
$y$ & \hbox{anti-minimal } & \hbox{minimal } \\
$\bar{y}$ &  \hbox{minimal } & \hbox{anti-minimal } \\ \hline
\end{tabular}
\end{center}
We chose to adopt the same notation as for the $\CN=2$ affine Yangian for the gluing generators, despite the fact that the algebras for $p=0$ and $p=\pm1$ are different. 
However this notational choice will turn out to be convenient, since we will be able to lay out a unified treatment of all choices of framing.

\subsection{Relative orientations of asymptotic shapes of twin-plane partitions}\label{sec:relative-orientation}

With the possibilities of framing, it is important to be careful in setting conventions for orientations of various coordinate axes of a twin plane partition. 
This is the case especially when comparing asymptotics of the left plane partition to those of the right plane partition, in order to make sense of the notion of transpose (or not transpose) conjugate representation appearing in the character decompositions (\ref{eq:char-id-fermionic}) and (\ref{eq:char-id-bosonic}). 

For a single plane partition, the three axes $x_{1,2,3}$ are equivalent. 
Correspondingly, the three parameters $h_{1,2,3}$ are also on an equal footing.\footnote{
It is only in the map to $\mathcal{W}_{N,k}$ algebra did we introduce a preferred choice that singles out the $x_3$ as the non-perturbative direction.
Namely, the truncation of the $\mathcal{W}_{1+\infty}$ down to the finite $\mathcal{W}_{1+N}$ algebra happens along the $x_3$ direction. 
This is reflected by  the fact that in the map (\ref{h123}), $h_3\rightarrow 0$ whereas $h_1$ and $h_2$ remain finite in the large $N$ limit.}
However, once we have made a choice for the symmetric/anti-symmetric direction for the left plane partition, the choice for the right one, i.e.\ whether $\hat{x}_1$ or $\hat{x}_3$ is the symmetric direction, is important, for the following reason.

Let $Y$ label the asymptotic shape of a plane partition in the left corner along the $x_2$-direction, its symmetric and anti-symmetric directions are defined referring to $x_1, x_3$ by an inessential choice of convention.  
On the other hand, $\bar{Y}$ characterizes the asymptotic shape of a plane partition from the right corner along the $\hat x_2$ leg, and it matters whether we define symmetric and anti-symmetric directions to be (respectively) $(\hat x_1, \hat x_3)$ or $(\hat x_3, \hat x_1)$. 
This is crucial because it distinguishes between $\bar{Y}$ and $\bar{Y}^t$, and therefore between bosonic and fermionic gluing operators.

From the discussion of minimal buds and their effects on statistics, we have all the necessary information to deduce how to fix the symmetric and antisymmetric axes of the right-partition's asymptotics.
In the fermionic case ($p=0$), since $h_i = \hat h_i$, the choice of symmetric axis on the right partition must coincide with that of the left partition hence $(x_1,\hat x_1)$ are symmetric axes by our choice of convention.\footnote{
Here, as in \cite{Gaberdiel:2018nbs}, we are taking $x_3$ to be the antisymmetric direction. However it would be equally fine to choose $x_1$ for the same purpose. For $p=0$ the two choices are completely democratic. In fact, there really is no ``symmetric'' axis for the fermionic case $p=0$, but only a choice of antisymmetric one.} 
Similarly the antisymmetric axis of the asymptotic shape of partitions will be $(x_3,\hat x_3)$.
When $p=-1$ we have seen that the buds have lengths $(s_1,s_3)=(2,0)$. 
This means that $x$ is ``bosonic" in the sense that repeated application of $x$ creates adjacent $\blacksquare$'s along the $x_3$ direction, since no bud is necessary.\footnote{
So far, we only discussed buds for the next-minimal rows. But the reasoning can be extended to rows created in all positions in the room. We will not discuss the details since we derive the general formula for minimal buds in later sections using the action of the algebra. Here we just use the fact that, acting with $x^k$ on the vacuum $|\emptyset\rangle$ will create $k$ rows stacked along the direction corresponding to a length-zero bud for the next-minimal row.} 
Recalling that for $p=-1$ we found $\hat h_1\sim h_3$ in (\ref{3solutions}), the symmetric directions of asymptotic Young diagrams are therefore $(x_3,\hat x_1)$, while $(x_1,\hat x_3)$ are the symmetric ones. 
The situation is simply reversed for $p=1$.\footnote{
Unlike for $p=0$, the symmetry exchanging $x_1,x_3$ is broken when $p=\pm1$, this gives rise to a distinguished symmetric direction (and an antisymmetric one). Another way to derive which direction is antisymmetric is to study the $\psi,\hat\psi$ charge functions of ``high-walls'', since the position of their poles correspond to available slots for single-boxes, and one of these slots is always atop a wall. By definition, the antisymmetric axis must coincide with the direction along which the wall is raised.}

For convenience we summarize these rules in the following table.\footnote{
These pictures are only meant to depict the orientation of asymptotic partitions. 
By convention, the horizontal axis is the symmetric axis, while the vertical one is the anti-symmetric one (as in the general theory of Young diagrams).
When applying these rules to twin-plane partitions, one should keep in mind that high-walls (corresponding to the $\overline\blacksquare$ on the hatted side) are actually placed ``behind the wall of the room'' \cite{Gaberdiel:2018nbs}. When computing the action of the algebra on twin plane partitions in later sections, this will be made explicit.}
\be
\label{eq:axes-conventions}
\begin{array}{c|c|c}
 p & 0 &  -1 \\
 \hline
\begin{array}{c}
 \textrm{self-statistics}  \\
 \textrm{of gluing operators}
 \end{array}
 &\textrm{fermionic} & \textrm{bosonic} \\ 
 \hline 
\begin{array}{c}
 \textrm{asymptotics} \\
 \textrm{along $x_2$ and $\hat{x}_2$}
 \end{array} &(Y,\bar{Y}^t)  & (Y,\bar{Y}) \\
 \hline  
\begin{array}{c}
 \textrm{symmetric} \\
 \textrm{axes}
 \end{array}
 & (x_1 , \hat x_1)& (x_3 , \hat x_1)\\
 \hline 
 \begin{array}{c}
 \textrm{anti-symmetric}  \\
 \textrm{axes}
 \end{array}
 & (x_3, \hat x_3) & (x_1, \hat x_3) \\
 \hline
 \begin{array}{c}
 \textrm{example} \\
 \textrm{(sym. = horizontal)} \\
\textrm{(antisym. = vertical)}
\end{array}& \raisebox{-30pt}{\includegraphics[width=0.38\textwidth]{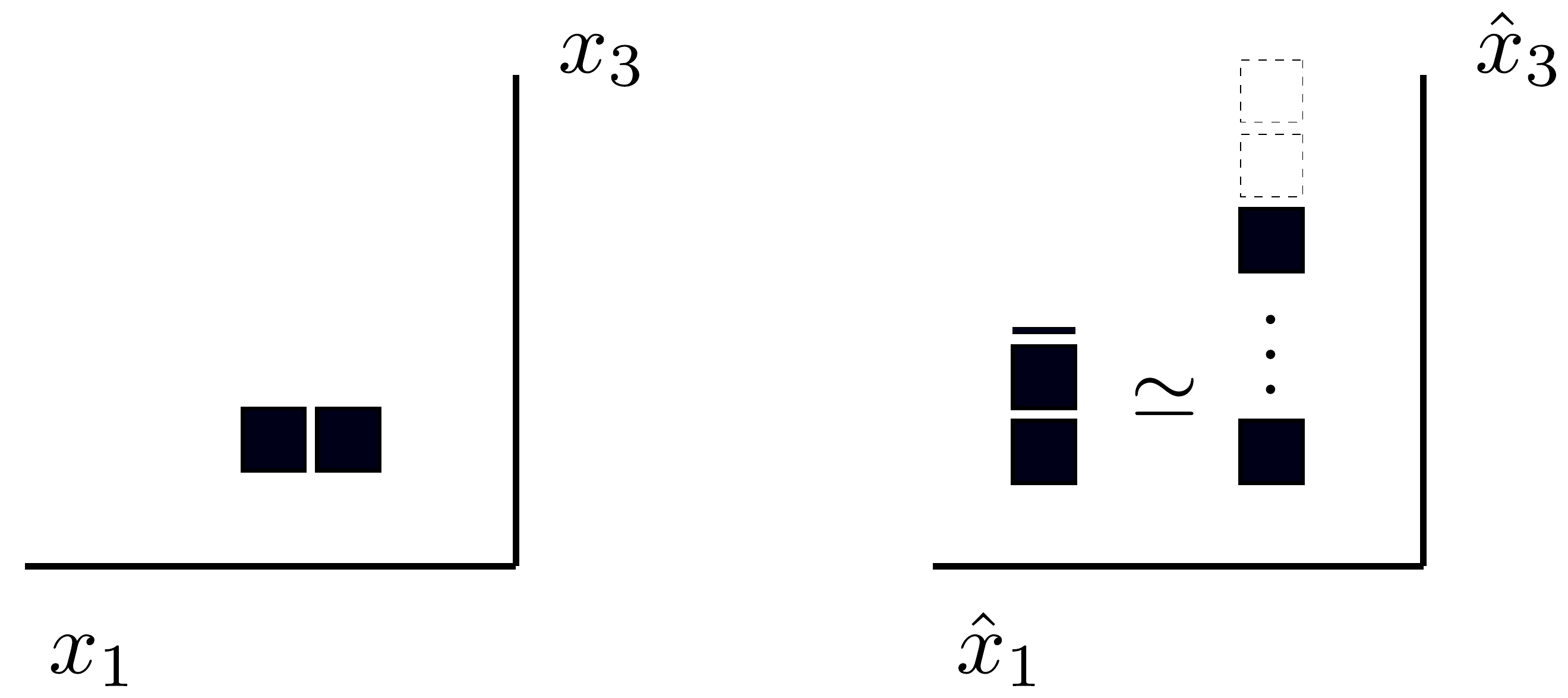}} & \raisebox{-30pt}{\includegraphics[width=0.38\textwidth]{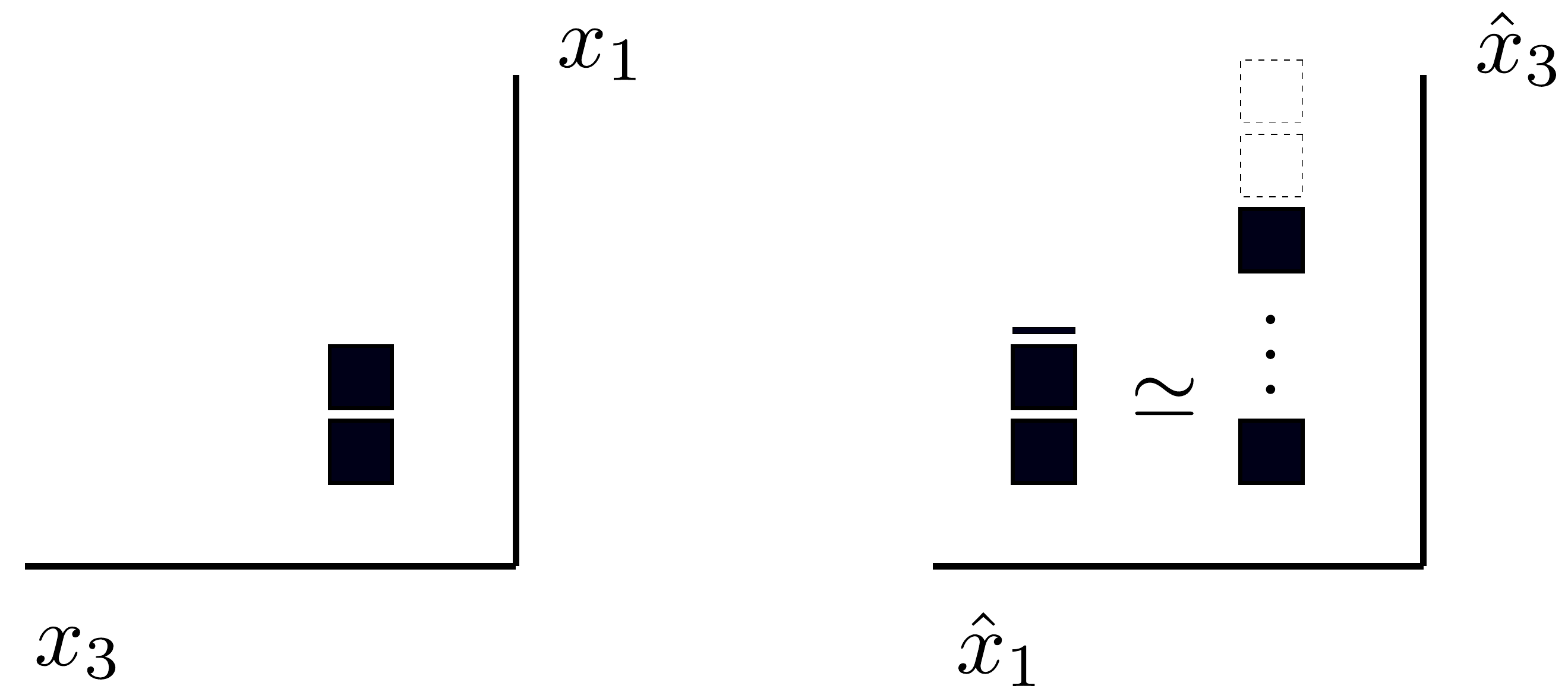}}
 \end{array}
\ee

\section{Relation to $(p,q)$ webs}\label{sec:geometry}

\label{sec:pqweb}
We will now illustrate a relation of twin plane partitions for $p=0,\pm1$ to the geometry of certain toric Calabi-Yau threefolds 
\be
	\CO(-s_3) \oplus \CO(-s_1) \to \IP^1\,,
\ee
where $s_1, s_3$ are the parameters in (\ref{3solutions}).
These geometries are known to be dual to $(p,q)$-webs of fivebranes \cite{Leung:1997tw}. 
This leads naturally to a connection to work of \cite{Prochazka:2017qum}, which conjectured that certain chiral algebras associated to this system can be obtained by gluing universal building blocks, consisting of the chiral algebra associated with a single Y-junction \cite{Gaiotto:2017euk}.
Our construction produces a family of affine Yangian algebras associated to $(p,q)$ webs with two trivalent vertices and a single internal leg. We expect that certain finite truncations of our algebra should reproduce the chiral algebras considered by \cite{Gaiotto:2017euk, Prochazka:2017qum}.

\subsection{From twin plane partitions to toric geometry}
\label{sec:geometric-meaning}

We start by observing that the relations among parameters $h_i,\hat h_i$ derived in the previous section naturally mimic geometric properties of certain toric Calabi-Yau threefolds. 

Recall that the affine Yangian from the  left (and right) corner constrain the $\{h_i\}$ (and $\{\hat{h}_i\}$) parameters to satisfy
\be\tag{a1}\label{projection1}
	\sum^3_{i=1} h_i =0 
	\qquad \textrm{and} \qquad 
	\sum^3_{i=1} \hat{h}_i =0 
\ee
In the glued algebra, the two corners shares a common $x_2$ direction, hence 
\be\tag{a2}\label{A2}
	h_2=\hat{h}_2
\ee
In addition, the minimal bud condition from the twin plane partition (\ref{chiralitySame}) further constrain the left and right parameters
\be\tag{a3}\label{A3}
	h_1+\hat{h}_3=-s_1 \, h_2  
	\qquad \textrm{and} \qquad 
	h_3+\hat{h}_1=-s_3 \, h_2 =(-2+s_1)\, h_2
\ee
with $s_{1,3}\in\mathbb{Z}$ and $s_1+s_3=2$ (see eq.\ (\ref{eq:s1s3})).
Finally, the constraint that the resulting twin plane partition cannot have buds sticking out the left and right wall demands that 
\be\tag{a4}\label{A4}
	s_1 \geq 0 
	\qquad \textrm{and} \qquad 
	s_3 =2-s_1 \geq 0 \ .
\ee
where we used (\ref{eq:s1s3}).

Now we show that these four conditions (\ref{projection1})-(\ref{A4}) on the algebra match nicely with conditions in toric geometry.
Recall that a toric Calabi-Yau threefold can be represented by a $(p,q)$ diagram in the base of a $T^2\times \mathbb{R}$ fibration.
Let us consider such a diagram with two vertices and one internal leg.
Label the $(p,q)$ charge of the two vertices by 
\begin{equation}
V_i=(p,q)_i \qquad \textrm{and} \qquad \hat{V}_i=(\hat{p},\hat{q})_i \qquad \textrm{with} \quad   i=1,2,3 \,.
\end{equation}
with $\{p_i, q_i, \hat{p}_i, \hat{q}_i\} \in \mathbb{Z}$.
The Calabi-Yau condition demands
\be\label{CalabiYau}\tag{g1}
	\sum_i V_i = 0 
	\qquad \textrm{and }\qquad 
	\sum_i\hat{V}_i=0\,,
\ee
Let the internal leg shared by the two vertices be $V_2\sim\hat{V}_2$. 
If we choose the opposite directions for the two vertices, e.g.\  all the $V_i$ pointing outwards and all $\hat{V}_i$ inwards, we have
\be\label{G2}\tag{g2}
V_2=\hat{V}_2
\ee
Furthermore, the smoothness condition demand
\begin{equation}\label{smoothness}
V_1\wedge V_2  =V_2\wedge V_3  =V_3\wedge V_1=  \hat{V}_1 \wedge \hat{V}_2 =  \hat{V}_2 \wedge \hat{V}_3 =  \hat{V}_3 \wedge \hat{V}_1 \,.
\end{equation}
Using (\ref{G2}), this can be written as
\be\label{G3}\tag{g3}
	V_1+\hat{V}_3=c_1 V_2 
	\qquad
	\textrm{and} 
	\qquad 
	V_3+\hat{V}_1=(-2-c_1)V_2 
\ee
with $c_{1}\in \mathbb{Z}$ (since $V_i$ and $\hat{V}_i$ are integer vectors). 

\begin{figure}[h!]
\begin{center}
\includegraphics[width=0.95\textwidth]{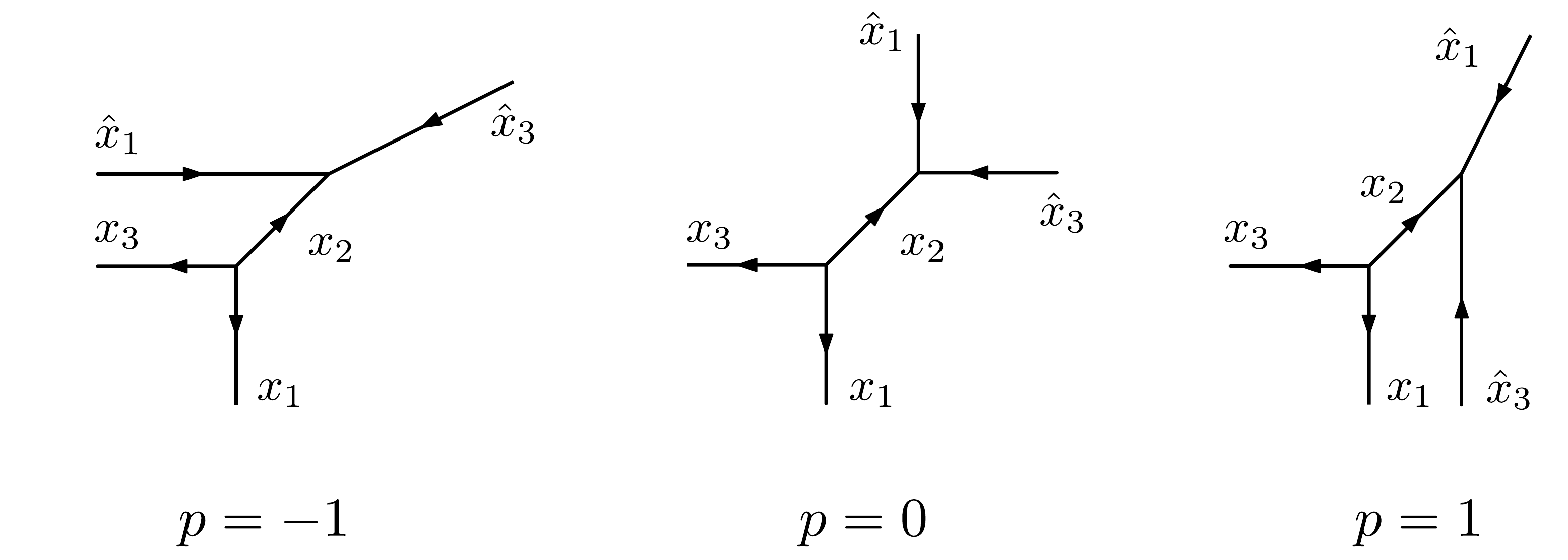}
\caption{Relation between coordinates $x_i$ and $\hat x_i$ for framing $p$. For twin plane partitions, $x_i \mapsto h_i$ and $\hat{x}_i \mapsto \hat{h}_i$. For $(p,q)$ web of toric geometry, $x_i \mapsto V_i$ and $\hat{x}_i \mapsto \hat{V}_i$.}
\label{fig:framing}
\end{center}
\end{figure}

Finally, it is convenient to take advantage of the overall SL$(2,\mathbb{Z})$ freedom to bring the vectors at one vertex in the following form:
\begin{equation}
V_1=(1,0)\,, \qquad \qquad V_2=(-1,-1)\,, \qquad \qquad V_3=(0,1)\,, 
\end{equation}
where the charge vectors point outwards.
Then the Calabi-Yau condition (\ref{CalabiYau}) and the smoothness condition (\ref{smoothness}) constrain the second vertex to have charge vectors (pointing inwards)
\begin{equation}
\hat{V}_1=(p+1,p)\,, \qquad \qquad \hat{V}_2=(-1,-1)\,, \qquad \qquad\hat{V}_3=(-p,1-p)\,,  
\end{equation}
where $p= \hat{q}_1$ and a priori $p\in\mathbb{Z}$. 
Plotting this $(p,q)$ diagram (see Figure~\ref{fig:framing}), one can easily see that in order for the external legs from the two vertices not to intersect, one need to impose 
\be\tag{g4}\label{G4}
	|p|\leq 1\,, 
\ee
as already observed by \cite{Prochazka:2017qum}.

The four constraints (\ref{projection1})-(\ref{A4}) coming from twin plane partition matches precisely to constraints (\ref{CalabiYau})-(\ref{G4}) on the choice of $(p,q)$ web for toric Calabi-Yau's!
In particular, the fact that $s_1$  and $ s_3$ have to be non-negative, i.e. there should not be box outside the room, translates to the constraint of $|p|\leq 1$ in the $(p,q)$ web, which means the legs from the two vertices shouldn't intersect. 
The end result is that the three cases in (\ref{3solutions}) precisely correspond to the three different $(p,q)$ web with two trivalent vertices shown in Figure \ref{fig:framing}. 

\subsection{From toric geometry to twin plane partitions}

Above we have seen how certain basic properties of twin plane partitions resemble some features of certain toric Calabi-Yau geometries.
Here we take the opposite perspective and show how starting from suitable toric Calabi-Yau geometries one can recover features of twin plane partitions.

Compared to the previous discussion, a salient novelty  that emerges from the geometry of toric threefolds is the 3d nature of twin plane partitions. 
From the algebraic viewpoint, the relations $\sum_i h_i = 0$ (and the hatted counterpart) effectively reduce the description of 3d plane partitions to two dimensions.
On the other hand, toric geometry offers a natural 3d perspective, which arises from changing the choice of fibration from $T^2\times \IR$ to $T^3$.

\subsubsection{$\CO(-1) \oplus \CO(-1) \to \IP^1$}

This geometry is known as the resolved conifold, and can be constructed as a gauge linear sigma model with four chiral fields and a single $U(1)$ gauge group, with the following charge assignments
\be\label{eq:conifold-charges}
	\begin{array}{ccccc}
		& z_1 & z_2 & z_3 & z_4 \\
		Q^i & 1 & 1 & -1 & -1
	\end{array}
\ee
The fact that $\sum_i Q^i = 0$ ensures that this toric threefold is Calabi-Yau.
The GIT quotient is obtained by taking the $U(1)$ quotient of the locus defined by $D$-term equations. With the charge assignments given above, these turn out to be
\be
	|z_1|^2 +|z_2|^2 - |z_3^2| -|z_4^2| = t \,.
\ee

\begin{figure}[h!]
\begin{center}
\includegraphics[width=0.48\textwidth]{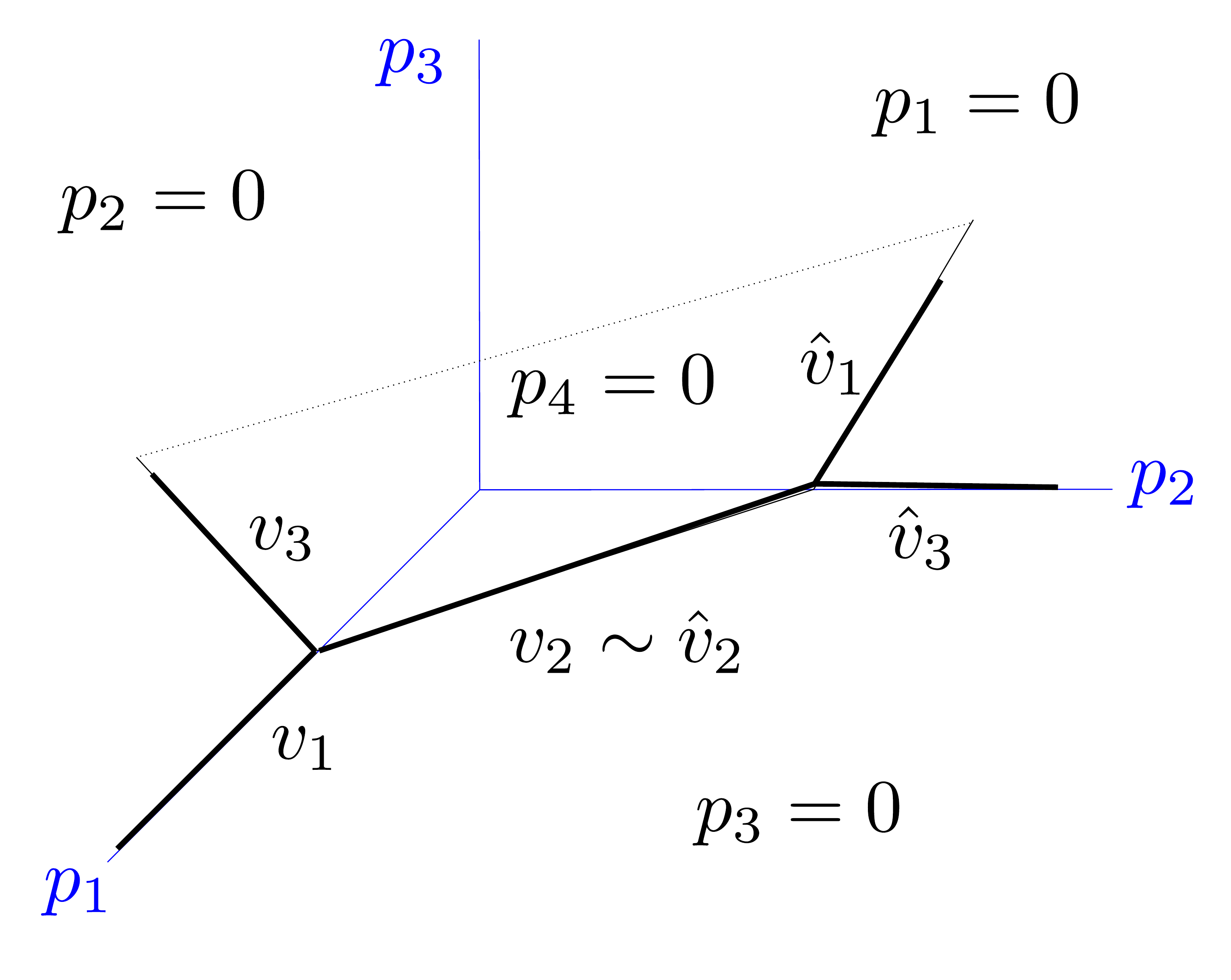}
\includegraphics[width=0.48\textwidth]{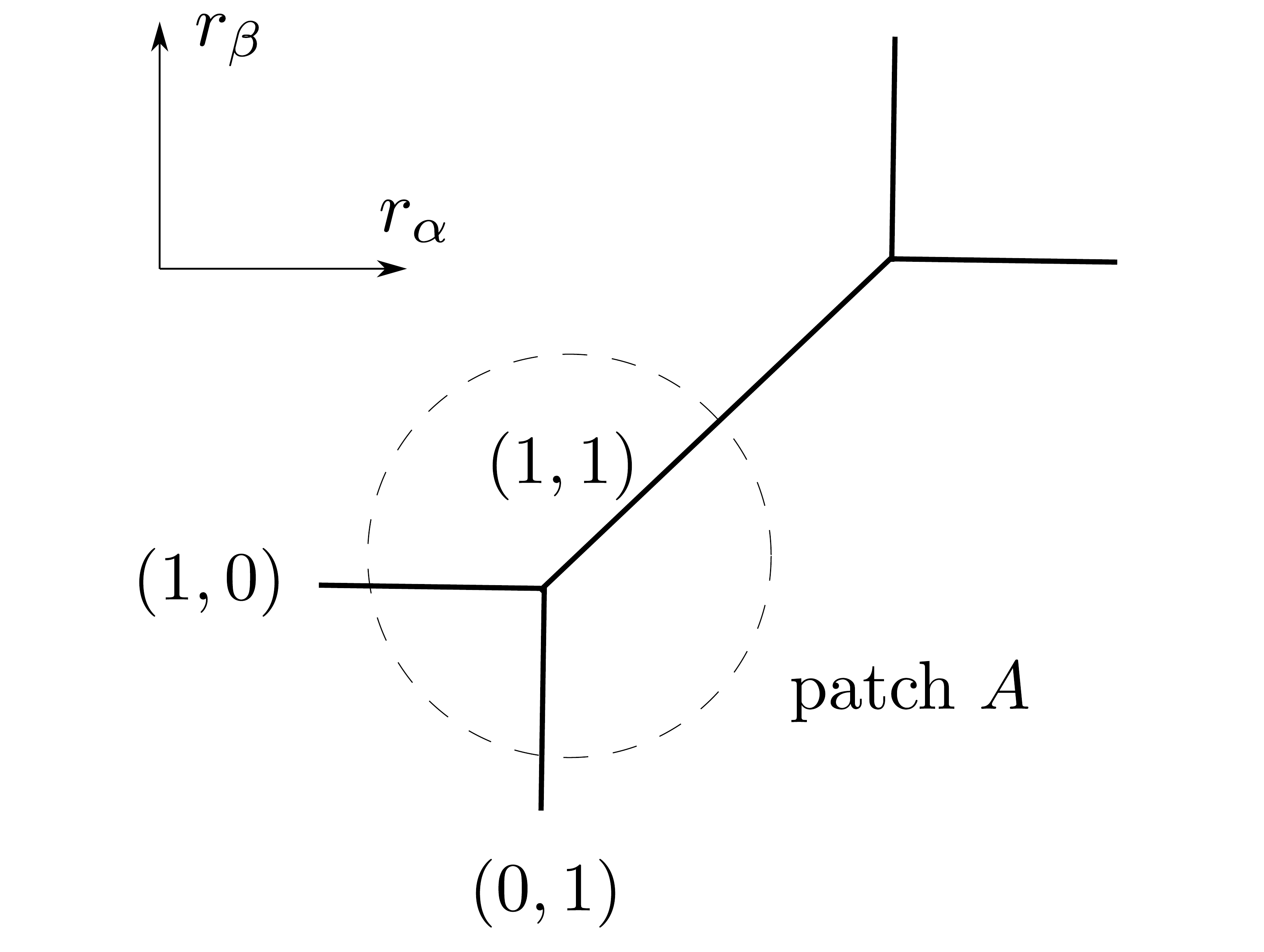}
\caption{Left: base of the $T^3$ fibration for the conifold. Right: base for the $T^2\times\IR$ fibration.}
\label{fig:conifold}
\end{center}
\end{figure}

\paragraph{$T^3$ fibration}
Thinking of the resolved conifold as a $T^3$ fibration means that we take the base to be spanned by $p_1\dots p_4$ with $p_i = |z_i|^2$ and subject to 
\be\label{eq:conifold-base}
	p_1,p_2, p_3\geq 0, \qquad p_4 = p_1+p_2 - p_3-t\geq 0
\ee
The fiber is then parametrized by the four phases of $z_i$ modulo the $U(1)$ quotient.
The first three equations in (\ref{eq:conifold-base}) single out the positive octant of $\IR^3_{p_1,p_2, p_3}$. 
Taking $t>0$, the fourth equation determines a plane, whose positive half-space intersects with $\IR^3_{>0}$ to give the base of the conifold.
The four planes intersect along {five edges} and two vertices, see Figure \ref{fig:conifold}.
The fiber shrinks at the boundaries of the base: at one of the planes $p_i=0$, the corresponding $S^1_i$ shrinks; at edges a $T^2$ shrinks; finally at the corners the whole fiber shrinks.

The vectors describing (up to a sign) the orientation of each edge are as follows 
\be
	\text{left corner} : \ \ 
	\left\{
	\begin{split}
		{v_1} &= (1,0,0) \\
		{v_2} &= (-1,1,0) \\
		{v_3} &= (1,0,1) 
	\end{split}
	\right.
	\qquad
	\text{right corner} : \ \ 
	\left\{
	\begin{split}
		\hat v_1 &= (0,1,1) \\
		\hat v_2 &= (1,-1,0) \\
		\hat v_3 &= (0,1,0)
	\end{split}
	\right.
\ee
We chose to normalize vectors in integer units, as this is suitable for a lattice discretization of space.\footnote{
Also note that $\hat v_2 = -v_2$ unlike in subsection \ref{sec:geometric-meaning}. To compare the two, one can simply flip signs for the three vectors of the right corner.}
Remarkably, this implies that $\sum v_i = \sum \hat v_i = (1,1,1)$, i.e. the diagonal dimension of boxes on the two sides agree.\footnote{
This condition ensures that one can consistently project the 3d setup to the plane transverse to this vector. The lattices arising from projection of the two corners will agree.
}

The geometry of this base shares some tantalizing features of the twin plane partition,  in particular regarding the length of ``buds'' discussed above (a rigorous derivation of these lengths will be given below through algebraic arguments).
First, the two vertices are $q_1=(t,0,0)$ and $q_2=(0,t,0)$ on the $p_1, p_2$ axes, respectively.
Their separation is 
\be
	q_2-q_1 = t \cdot v_2\,.
\ee
Therefore $t$ is the length of a row of boxes stacked along the internal edge of the base, since the size of each box is $v_2=\hat v_2$ along that direction.\footnote{
\label{foot:effective-length}
Intuitively one may expect that $t$ is related to $\rho$, since both describe the effective length of the internal leg. Indeed, the conformal dimension of a configuration depends linearly on the number of single-boxes $\square$ and $\hat\square$ \cite{Gaberdiel:2018nbs}, therefore $1+\rho$ can be viewed as the ``effective'' number of boxes in the infinite row $\blacksquare$. However the relation between $t$ and $\rho$ is somewhat nontrivial, a possible interpretation will be given in Section \ref{sec:BPS-interpretation}
} 

Next consider two new points 
\be
	q'_1=q_1+v_1 = (t+1,0,0)
	\qquad\textrm{and} \qquad
	q'_2=q_2 + \hat v_3 = (0,t+1,0)
\ee
their new distance will be
\be
	{(q'_2-q'_1) = (q_2-q_1) +  v_2} \,.
\ee
This has a straightforward interpretation.
The new row along the $v_2$ direction is placed next to the minimal one, and is displaced transversely by $(1,1,0)=v_1+\hat v_3$. 
Since the orientation of the row has an angle of more than $\pi/2$ with the planes $p_1=0$ and $p_2=0$, the displacement implies that the new row should be longer by one unit of $v_2$. 
But since the operator $x$ creates rows of length $t$, this would be disallowed as a twin plane partition (the row would not arrive at the corner on the other side, but hang away from it leaving a gap of length $v_2$). 
Instead, to create an admissible configuration one should supply a ``bud'' consisting of a single box, either at the left corner or at the right one. 
The bud together with the row created by $x$ will then result in an admissible twin plane partition. 
The overall conformal dimension should increase by the same amount as the (effective) number of boxes, which is $1+(1+\rho)$. 
This is consistent with the fact that $e(z)$ creates a single box (the bud) and has conformal dimension 1, while $x$ has dimension $1+\rho$.

From the viewpoint of the left corner, we would say that the row is displaced (transversely) by $v_1$, whereas from the viewpoint of the right corner we would say that it is displaced by $\hat v_3$. 
Above we defined $s_i$ as the length (in units of $v_2$) of the bud that is necessary to add the next-minimal row, shifted by $v_i$ with respect to the minimal-length row. 
Since we created a next-minimal row shifted along $v_1$ (as opposed to $v_3$), we have just recovered 
\be
	\boxed{s_1 = 1}
\ee
Repeating the argument with a next-minimal row displaced by $v_3$ would give 
\be
	\boxed{s_3 = 1}\,.
\ee 
We have seen that the conifold geometry viewed as a $T^3$ fibration naturally reproduces the bud structure for the framing $p=0$. Next we will see that the relation between $h_i$ and $\hat h_i$ also arises naturally.

\paragraph{$\IR\times T^2$ fibration} 
A useful way to think about toric threefolds is to introduce a decomposition into $\IC^3$ patches, one for each vertex of the toric diagram.
Let patch $A$ correspond to the vertex on the left, located at $z_2=z_3=z_4=0$ with $z_1\neq 0$. 
Changing coordinates to
\be\label{eq:conifold-patch-A}
\begin{split}
	r_{\alpha} & = |z_4|^2 - |z_1|^2 +t \,, 
	\qquad
	r_{\beta}  = |z_3|^2 - |z_1|^2 +t \,,
	\qquad
	r_{\gamma}  = \Im(z_1 z_3 z_4) \,, 
\end{split}
\ee
we work on the slice of the base generated by $(r_\alpha,r_\beta)\in \IR^2$, where the left vertex is located at the origin.
In this patch, the fiber includes a $T^2$  generated by
\be
	(z_1, z_3, z_4)\mapsto (e^{-i(\alpha+\beta)} z_1, e^{i\beta} z_3, e^{i\alpha} z_4),
\ee
as well as a real line generated by $\Re(z_1 z_3 z_4) $. 

By an overall $SL(2,\IZ)$ freedom to parametrize the torus fiber, we can fix $S^1_\alpha$ to be the $(1,0)$ cycle of $T^2$. This circle shrinks when $|z_1|^2=t+ |z_4|^2$, which by the D-term equations coincides with the coordinate axis $p_1$ (i.e. $|z_2|^2=|z_3|^2=0$).
This locus corresponds to $r_\alpha = 0,r_\beta = -|z_4|\leq 0$ in the $T^2\times \IR$ fibration and it is the edge denoted by $(0,1)$ in Figure \ref{fig:conifold}. 
Likewise we can fix $S^1_\beta$ to be the $(0,1)$ cycle, which shrinks when $r_\beta = 0,r_\alpha = -|z_3|\leq 0$. 
This structure of the base is shown in Figure \ref{fig:conifold}.

A similar analysis can be performed in patch $B$ near the other vertex, located at $(r_\alpha,r_\beta)=(t,t)$.
It is natural to change coordinates using the D-term equations, to reflect the fact that the right corner located at $z_1= z_3= z_4 = 0$
\be\label{eq:conifold-patch-B}
	r_\alpha = |z_2|^2- |z_3|^2  \,,
	\qquad
	r_\beta = |z_2|^2- |z_4|^2 \,.
\ee
From these we construct the other piece of the diagram, in the same way as for patch $A$, see Figure \ref{fig:conifold}. 

\paragraph{The relation between $h_i$ and $\hat h_i$}
Now consider a box placed near the left corner in the $T^3$ fibration. We assume its sides are described by $v_1, v_2, v_3$, consistently with the previous discussion.
In the $T^2\times \IR$ fibration the three sides of a box are described as follows.
$v_1$ is the unit vector along the $p_1=|z_1|^2$ direction, with no components along other directions. Therefore $v_1$ maps to
\be
	v_1 \quad\to \quad
	\delta |z_1|^2 = 1
	\,,\quad
	\delta |z_2|^2 = 0
	\,,\quad
	\delta |z_3|^2 = 0
	\,,\quad
	\delta |z_4|^2 = 1
	\,.
\ee
Using coordinates of the $T^2\times \IR$ fibration in patch $A$ (\ref{eq:conifold-patch-A}), this translates into
\be
	v_1 \quad\to \quad(\delta r_\alpha, \delta r_\beta) = (0,-1) = H_1\,. 
\ee
By the same argument we obtain
\be
	v_2 \quad\to \quad (1,1) = H_2
	\,,
	\qquad \qquad
	v_3 \quad\to \quad (-1,0) = H_3\,.
\ee
Similarly we define from the $\hat v_i$ the following 2d vectors in the $T^2\times \IR$ base
\be
	\hat H_1  = (0,1)
	\,,\qquad \qquad
	\hat H_2  = (-1,-1)
	\,,\qquad \qquad 
	\hat H_3  = (1,0) 
\ee
Note that these vectors satisfy
\be
\boxed{\begin{split}
	H_1+H_2+H_3 = 0
	& \qquad \qquad
	\hat H_1+\hat H_2+\hat H_3 = 0
	\\
	&-\hat H_i = H_i
\end{split}}
\ee
where the minus sign is due the fact that we have chosen $\hat{H}_i$ to point outwards from the second vertex instead of inwards (as in the convention for section.\ \ref{sec:geometric-meaning}.)
These are precisely the relations between $h_i$ and $\hat h_i$ in the case $p=0$, see (\ref{eq:h-hhat-p}).

\subsubsection{$\CO(0) \oplus \CO(-2) \to \IP^1$}

Let us now consider a close relative of the conifold, the geometry $\CO(0) \oplus \CO(-2) \to \IP^1$.
This also admits a gauged linear sigma model construction with four chiral fields and a single $U(1)$ gauge group, with the following charge assignments
\be\label{eq:conifold-charges}
	\begin{array}{ccccc}
		& z_1 & z_2 & z_3 & z_4 \\
		Q^i & 1 & 1 & 0 & -2
	\end{array}
\ee
The D-term equations are now.
\be
	|z_1|^2 +|z_2|^2  -2|z_4^2| = t\,.
\ee

\begin{figure}
\begin{center}
\includegraphics[width=0.48\textwidth]{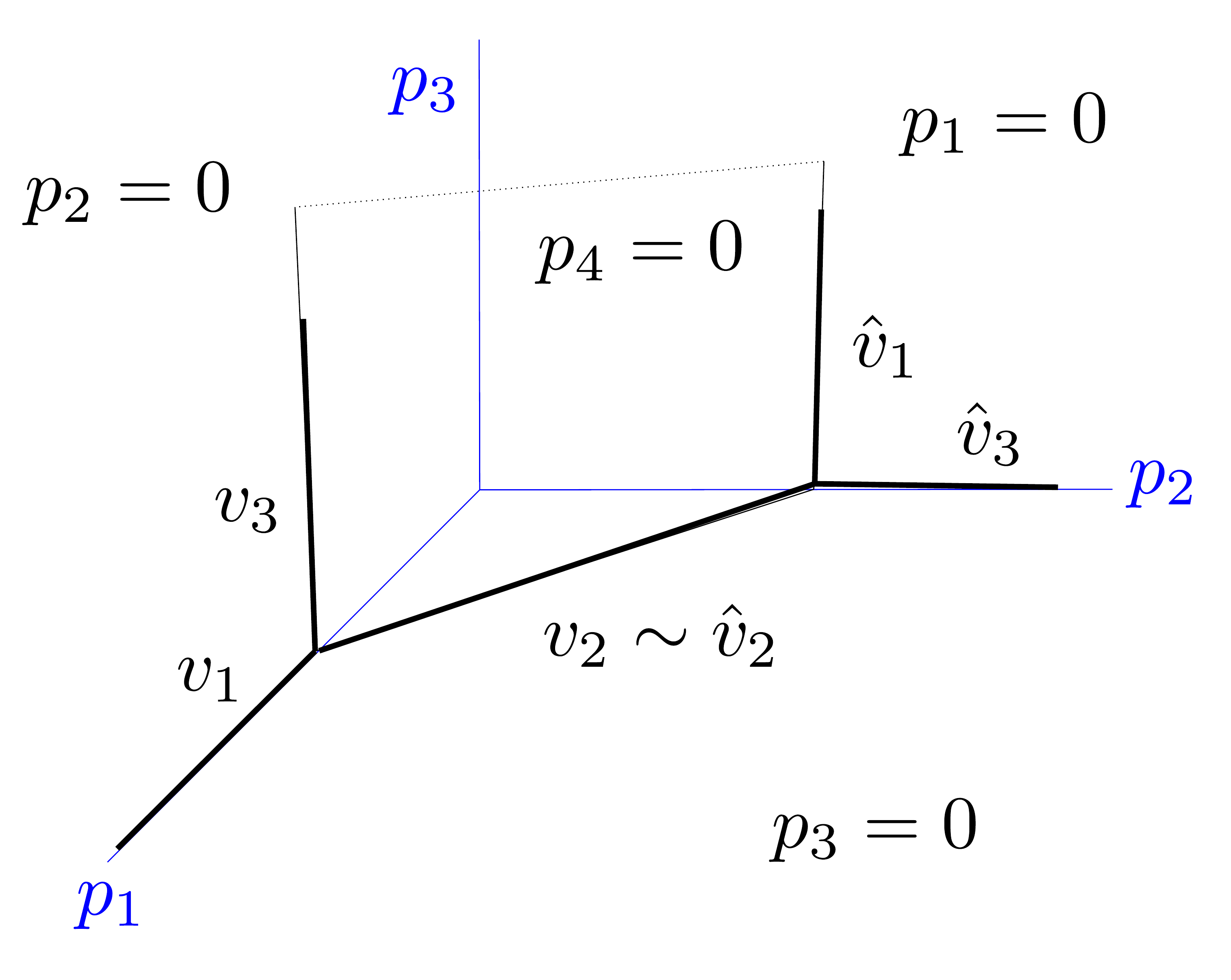}
\includegraphics[width=0.48\textwidth]{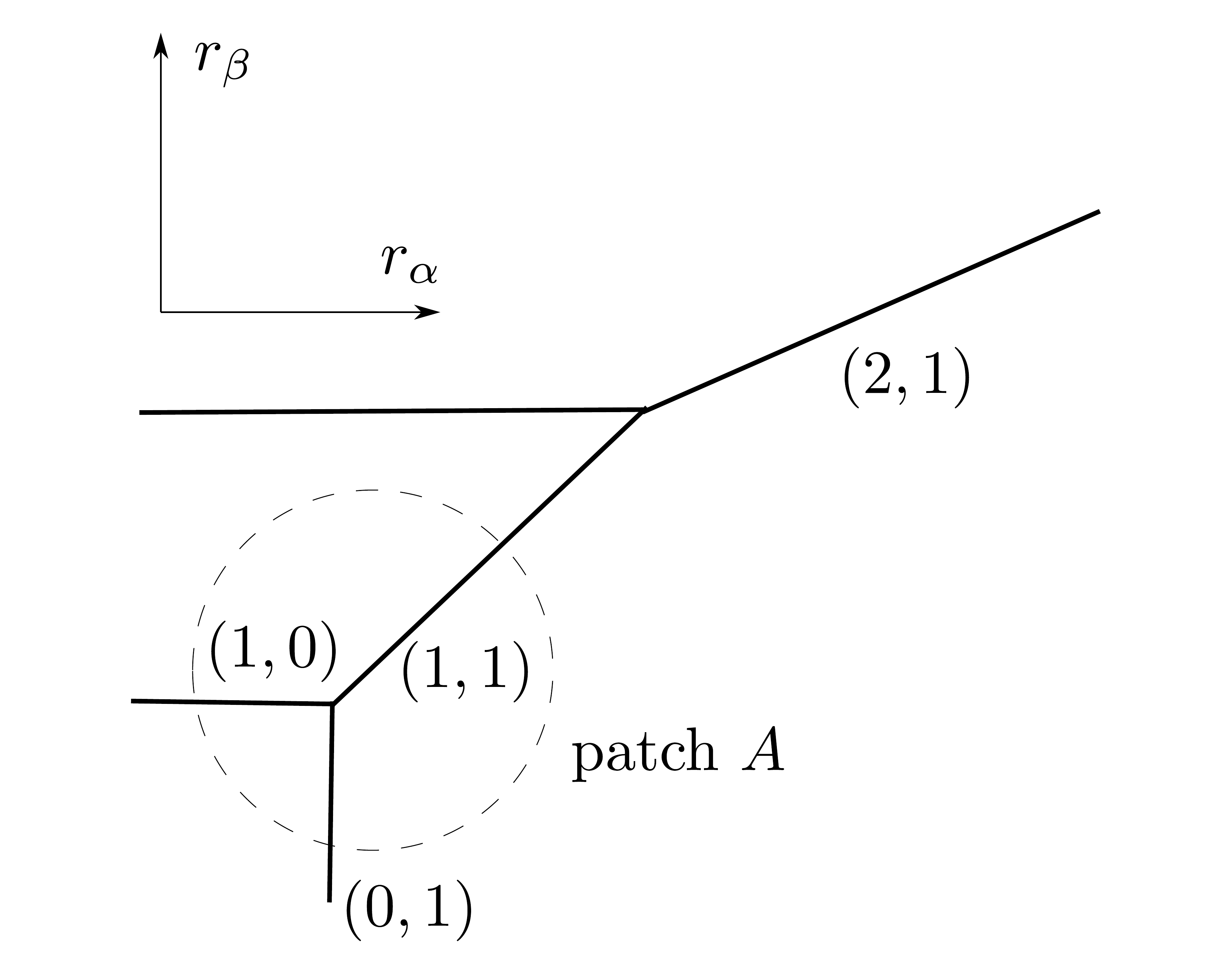}
\end{center}
\caption{Left: base of the $T^3$ fibration for $\CO(0) \oplus \CO(-2) \to \IP^1$. Right: base for the $T^2\times\IR$ fibration.}
\label{fig:O0O2}
\end{figure}

\paragraph{$T^3$ fibration}
Viewing the geometry as a $T^3$ fibration means that we take the base to be spanned by $p_1\dots p_4$ with $p_i = |z_i|^2$ and subject to 
\be
	p_1,p_2, p_3\geq 0, \qquad p_4 = \frac{1}{2}\(p_1+p_2 -t\)\geq 0
\ee
The geometry of the base is shown in Figure \ref{fig:O0O2}

The four planes intersect along five edges, whose orientations correspond to the vectors
\be
	\text{left corner} : \ \ 
	\left\{
	\begin{split}
		{v_1} &= (2,0,0) \\
		{v_2} &= (-1,1,0) \\
		{v_3} &= (0,0,1) 
	\end{split}
	\right.
	\qquad
	\text{right corner} : \ \ 
	\left\{
	\begin{split}
		\hat v_1 &= (0,0,1) \\
		\hat v_2 &= (1,-1,0) \\
		\hat v_3 &= (0,2,0)
	\end{split}
	\right.
\ee
Note the normalization of $v_1$ and $\hat v_3$ relative to $v_3,v_2=-\hat v_2,\hat v_1$, is chosen to comply with the property that $\sum_{i} v_i=\sum_{i} \hat v_i = (1,1,1)$ as in the case $p=0$.\footnote{
Again this condition ensures that one can consistently project the 3d setup to the plane transverse to this vector. The lattices arising from projection of the two corners will agree.
}

Let us now fix once again  points $q_1=(t,0,0)$ and $q_2=(0,t,0)$ corresponding to the two corners.
Their separation is\footnote{
As for the conifold, this is suggestive of a relation between $t$ and $\rho$. See footnote \ref{foot:effective-length}.
}
\be
	q_2-q_1 = t \cdot v_2\,.
\ee 

Next consider two new points 
\be
	q'_1=q_1+v_1 = (t+2,0,0)
	\qquad
	q'_2=q_2 + \hat v_3 = (0,t+2,0)\,.
\ee
Their distance is
\be
	\boxed{(q'_2-q'_1) = (q_2-q_1) +  2 v_2}
\ee
We can interpret this as follows: to create a next-minimal oriented along the $v_2$ direction, and displaced by $(2,2,0)=v_1+\hat v_3$, it will cause the row to be longer by  \emph{two} units of $v_2$. 
Repeating the arguments as for the conifold, we recognize this as the statement that creation of this next-minimal row would require the presence of a bud of length two. 
From the viewpoint of the left corner, we would say that the row is displaced (transversely) by $v_1$, whereas from the viewpoint of the right corner we would say that it is displaced by $\hat v_3$. 
Therefore we recovered
\be
	\boxed{s_1 = 2.}
\ee
Consider then creation of a next-minimal row displaced transversely along the $p_3$ direction. 
Its endpoints would be 
\be
	q''_1=q_1+v_3 = (t,0,1)
	\qquad
	q''_2=q_2 + \hat v_1 = (0,t,1)
\ee
their new distance will be
\be
	\boxed{(q''_2-q''_1) = (q_2-q_1)}
\ee
We can interpret this as follows: to create a next-minimal row displaced along $(0,0,1)=v_3=\hat v_1$, it will cause the row to have the \emph{same length} as the minimal row.
We recognize this as the statement that there is no need to insert a bud in this case (the bud would have length zero).
We recovered
\be
	\boxed{s_3 = 0.}
\ee
Taken together, these relations suggest that we are in the case of framing $p=-1$.
Next let us confirm this by reproducing the relations defining the parameters of the algebra $h_i, \hat h_i$.

\paragraph{$\IR\times T^2$ fibration} 
The left corner is located at $z_2=z_3=z_4=0$ with $z_1\neq 0$. 
Let us change coordinates to  
\be\label{eq:O0O2-patchA}
	r_{\alpha}  = |z_2|^2- |z_3|^2 
	\,,\qquad
	r_{\beta}  = \frac{1}{2}\(|z_2|^2- |z_1|^2 +t\) 
	\,,\qquad
	r_{\gamma}  = \Im(z_1 z_3 z_4) \,,
\ee
and work on the slice of the base generated by $(r_\alpha,r_\beta)\in \IR^2$.
The segment of the $p_1$ axis with $p_2=p_3=0$ and $p_1\geq t$ maps to $r_\alpha=0, r_\beta = \frac{1}{2}\(-|z_1|^2+t\) \leq 0$ (it is denoted $(0,1)$ in Figure \ref{fig:O0O2}).
The edge corresponding to $v_3$ is described instead by $p_2=0, p_1=t, p_3\geq0$ and therefore maps to $r_\alpha\leq 0, r_\beta = 0$ it is denoted $(1,0)$ in Figure \ref{fig:O0O2}.

In patch $B$ (near the other vertex) the right corner is located at $(r_\alpha,r_\beta)=(t,t)$. 
The portion of the $p_2$ axis ($p_1=p_3=0$) with $p_2\geq t$ now maps to
$r_\alpha\geq t, r_\beta = \frac{1}{2}(r_\alpha+t)$.
Now the edge with orientation $\hat v_1$ located at $p_2=t, p_4=0$ and $p_3\geq 0$ will map to the half-line $r_\alpha \leq t $ and $r_\beta = t$.
Overall this gives the rest of the diagram in Figure \ref{fig:O0O2}.

\paragraph{The relation between $h_i$ and $\hat h_i$}
Now consider a box placed near the left corner in the $T^3$ fibration. We assume its sides are described by $v_1, v_2, v_3$, consistently with the previous discussion.
In the $T^2\times \IR$ fibration the three sides of a box are described as follows.
$v_1$ is the unit vector along the $p_1=|z_1|^2$ direction, with no components along other directions. Therefore $v_1$ maps to
\be
	v_1 \quad\to \quad
	\delta |z_1|^2 = 1
	\,,\quad
	\delta |z_2|^2 = 0
	\,,\quad
	\delta |z_3|^2 = 0
	\,,\quad
	\delta |z_4|^2 = 1/2
	\,.
\ee
Using coordinates of the $T^2\times \IR$ fibration in patch $A$ (\ref{eq:O0O2-patchA}), this translates into\footnote{
Since $(\delta r_\alpha, \delta r_\beta)= (0,-1/2)$, we rescale by an overall inessential constant to achieve integral normalization. We will do the same for $\hat H_3$ below.}
\be
	v_1 \quad\to \quad(\delta r_\alpha, \delta r_\beta)\sim (0,-1) = H_1\,. 
\ee
By the same argument we obtain
\be
	v_2 \quad\to \quad (1,1) = H_2
	\,,
	\qquad \qquad
	v_3 \quad\to \quad (-1,0) = H_3\,.
\ee
Similarly we define from the $\hat v_i$ the following 2d vectors in the $T^2\times \IR$ base
\be
	\hat H_1  = (-1,0)
	\,,\qquad \qquad
	\hat H_2  = (-1,-1)
	\,,\qquad \qquad
	\hat H_3  = (2,1) 
\ee
Note that these vectors satisfy
\be
\boxed{\begin{split}
	H_1+H_2+H_3 = 0
	\qquad \qquad&
	\hat H_1+\hat H_2+\hat H_3 = 0
	\\
	-\hat H_1 = H_1+ H_2
	\qquad \qquad
	-\hat H_2 = H_2
	&
	\qquad\qquad
	-\hat H_3 =H_3-H_2\,,
\end{split}}
\ee
Again, the minus sign is due the fact that we have chosen $\hat{H}_i$ to point outwards from the second vertex instead of inwards (as in the convention for section.\ \ref{sec:geometric-meaning}.)
These are precisely the relations between $h_i$ and $\hat h_i$ in the case $p=-1$, if we flip the signs of $\hat H_i$, see (\ref{eq:h-hhat-p}).

\subsection{Statistics from geometry of plane partitions}
\label{sec:statistics}

We have argued in Section \ref{sec:self-statistics} that the gluing operators $x,\bar x,y,\bar y$ obey \emph{Fermi} statistics if $p=0$, but \emph{Bose} statistics if $p=\pm1$.
We will now show that the same conclusion can be reached from the geometric picture of the $(p,q)$ diagram.

Gluing operators act on twin-plane-partitions by creating an infinite row of boxes (or a wall, viewed from the other side) along the common $x_2\sim \hat x_2$ direction.
Since the conformal dimension is tied to the number of boxes in a plane partition, one should think of the effective length of the row (and wall) created by a gluing operator as fixed by its conformal dimension. 
However, the ``slots'' where a row may be created within the room have varying length.
It follows from these two facts that the gluing operators have different statistics depending on the choice of framing $p$.

First let us consider the case of $p=0$.  
We first apply $x$ on the vacuum configuration to create a row/wall pair along the $x_2$ axis. 
Acting again with $x$ should create a second row/wall adjacent to the former, either displaced by $h_1=\hat h_1$ along the $x_1\sim \hat x_1$ axis, or by $h_3=\hat h_3$ along the $x_3\sim \hat x_3$ axis.

But it is clear from Figure \ref{fig:conifold} that neither works, because from a 3d perspective if we shift along $x_1$ on the un-hatted side, this would correspond to a shift along $x_3$ on the hatted side, and vice-versa.\footnote{
When thinking of plane partitions as configurations of boxes in a room, the $h_i$ variables describe the size of single boxes in the three directions $x_i$ (and similarly for $\hat h_i$). 
However, since $h_1+h_2+h_3=0$, these are really the \emph{projected} lengths of a box's sides on the plane in which  we project the plane partition configuration, such as in Figure \ref{fig:framing}.
}
We should in fact shift the wall along the \emph{negative} $\hat x_3$ direction, as already explained in deriving (\ref{chiralitySame}).
Since $-\hat h_3 = h_2+h_1$ (as opposed to $-\hat h_3= h_1$) it follows that the infinite row/wall pair would not be able to fill the whole length of the room along the (displaced) $x_2$-direction, leaving a gap of length $h_2$ (or $\hat h_2$ equivalently).
Since this is not an allowed configuration for plane partitions, one concludes that $x\cdot x \, |\emptyset\rangle \sim0$. 
This nilpotency-on-vacuum nature of $x$ is consistent with its fermionic nature for $p=0$.

In contrast, for $p=-1$ the edges of the room corresponding to axes $x_3$ and $\hat x_1$ are \emph{parallel}, as is manifest both from Figure \ref{fig:O0O2} and from the identity $\hat h_1=-h_3$. Therefore acting twice with $x$ will create two rows/walls along the $x_3\sim \hat x_1$ direction, leaving no gaps along $x_2$.
In fact, acting arbitrarily many times with $x$ can keep creating rows/walls stacked along $x_3\sim \hat x_1$. The situation for $p=1$ is essentially identical upon switching $x_1\leftrightarrow x_3$ and $\hat x_1\leftrightarrow \hat x_3$.
For this reason the gluing operators $(x,y,\bar{x},\bar{y})$ behaves as a bosonic operator when $p=\pm1$.

\subsection{Relative orientation of asymptotic shapes from geometry}
Now that we have a geometric interpretation of twin plane partitions, let us briefly comment again on the relative orientation of asymptotic shapes of twin plane partitions along the common direction, discussed in Section \ref{sec:relative-orientation}.

\begin{figure}[h!]
\begin{center}
\includegraphics[width=0.85\textwidth]{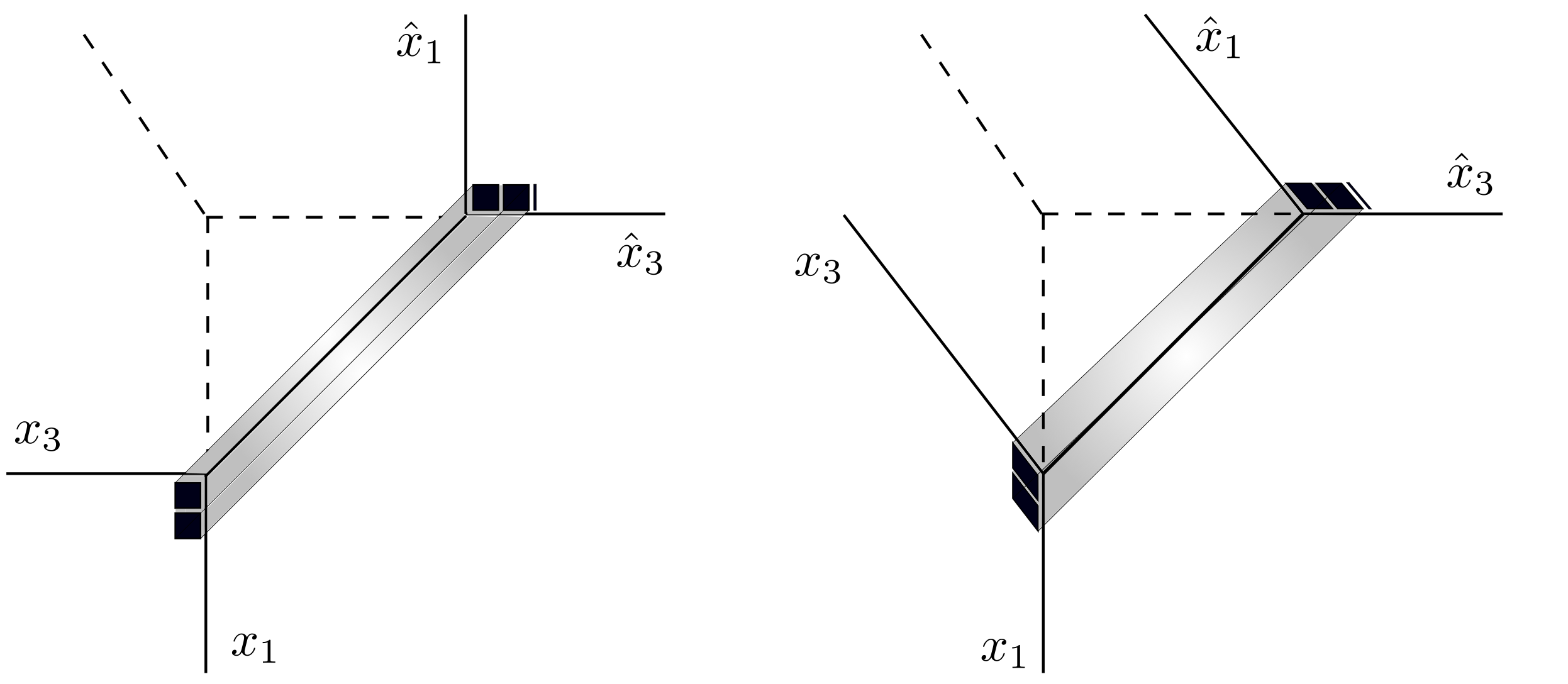}
\caption{Sketch of the box interpretation of gluing generators, where the room is viewed as the base of the $T^3$ fibration of the toric Calabi-Yau threefolds corresponding to $p=0$ and $p=-1$.}
\label{fig:gluing-compatibility}
\end{center}
\end{figure}

Note that in going from $p=0$ to $p=\pm 1$ we have two simultaneous effects: one is the change of asymptotics by a transpose, while the other is the change of vertical/horizontal axes on the (un-)hatted side. 
These effects effectively cancel each other on plane partitions, since they both correspond to a transposition of the Young diagram describing the asymptotic shape of the hatted partition.
This phenomenon is especially natural from the viewpoint of twin-plane partition asymptotics, if viewed from a 3d perspective as illustrated by Figure \ref{fig:gluing-compatibility}.  There, infinite rows of boxes on one side appear from the other side as infinite rows of anti-boxes, or high-walls. On the left we show a configuration with two rows for $p$=0. On the right we show a configuration with two rows for $p=-1$.  (cf.\ Figures \ref{fig:conifold} and \ref{fig:O0O2}) The relative orientations of asymptotic shapes along the $x_2\sim \hat x_2$ direction are derived from table (\ref{eq:axes-conventions}).

\section{Twin plane partitions for generic $p$ and $\rho$}
\label{sec:TPP}

Here we give an overview of the structure of twin plane partitions, extended to the two-parameter family of gluing parameterized by $\rho$ and $p$.

\subsection{Coordinate functions}

%Recall that for a plane partition $\Lambda$, a box in $\Lambda$ is labelled by its coordinates $x_i({\square})$, with $i=1,2,3$ and $x_i({\square})\in \mathbb{Z}_{\geq 0}$, i.e.\ the box at the origin has coordinate $(0,0,0)$. 
%We now generalize this to a coordinate system for the twin-plane-partitions. 
The coordinate system for twin-plane-partitions is a generalization of the one used for plane partitions \cite{Gaberdiel:2018nbs}; in particular, it coincides with the one for the left (resp. right) plane partitions when focusing on the left (resp. right) corner.
We use coordinates $x_i$ to label  the boxes in the left plane partition (denoted by ${\square}$), and $\hat{x}_i$ for hatted boxes  $\hat{{\square}}$ in the right plane partition. 
The coordinates for the boxes sitting at the bottom are
% (as opposed to atop a high wall) is %inherited from the $\fgl_1$ description:
\begin{equation}
\begin{aligned}
\square\textrm{ at bottom}:\qquad &x_1(\square)\,, x_2(\square)\,,x_3(\square)\,=0,1,2,3\dots\\
\hat{\square}\textrm{ at bottom}:\qquad &\hat{x}_1(\hat{\square})\,, \hat{x}_2(\hat{\square})\,,\hat{x}_3(\hat{\square})\,=0,1,2,3\dots
\end{aligned}
\end{equation}
For both left and right plane partitions, the first box in the corner has coordinate $(0,0,0)$.

For boxes sitting atop the left high-wall, the natural coordinate system is%\footnote{Relative to the conventions of \cite{Gaberdiel:2018nbs}, we take here $x_1,x_3<0$.}
\begin{equation}\label{coordtop}
\square \textrm{ on top}:\qquad \begin{cases} x_1(\square)&=-1,-2,\dots\qquad \\ x_2(\square)&=0,1,2,\dots\qquad\\
 x_3 (\square)\,&=-1,-2,\dots \ , \qquad
\end{cases}
\end{equation}
Similarly for $\hat{\square}$ atop the right wall,
\begin{equation}\label{coordhattop}
\hat{\square} \textrm{ on top}:\qquad \begin{cases} \hat{x}_1(\hat{\square})&=-1,-2,\dots\qquad \\
\hat{x}_2(\hat{\square})&=0,1,2,\dots\qquad\\
\hat{x}_3(\hat{\square})\,&=-1,-2,\dots\ . \qquad 
\end{cases}
\end{equation}
In either case we define 
\begin{align}\label{h-hath-def}
h(\square) & \equiv \left(\sum^{3}_{i=1}x_i(\square)\,h_i \right) 
- \delta_{{\rm top}} \sigma_3  \psi_0 \qquad 
\hat{h}(\hat{\square})  \equiv \left(\sum^3_{i=1}\hat{x}_i(\hat{\square})\,\hat h_i \right)
 - \delta_{{\rm top}} \hat  \sigma_3 \hat{\psi}_0  \ ,
\end{align}
where $\delta_{{\rm top}}=1$ if $\square$ (or $\hat{\square}$) sits atop the wall, and $\delta_{{\rm top}}=0$ otherwise. 
Intuitively, $\sigma_3\psi_0$ is the effective height of a wall on the unhatted side, and $\hat\sigma_3\hat\psi_0$ is the same quantity for the hatted side.
Thus $h(\square)$ and $\hat{h}(\hat{\square})$ coincide with the positions of  poles of the charge functions, where single-box operators can add or remove boxes, see Section~3.2 of \cite{Gaberdiel:2018nbs}. 
Note that  $x_{1,3}<0$ and $\hat{x}_{1,3}<0$ in eqs.~(\ref{coordtop}) and (\ref{coordhattop}), i.e. the conjugate representations should be described in terms of ``high walls" located in the quadrant with $x_{1,3}<0$ and $\hat{x}_{1,3}<0$.
\smallskip

We also need to introduce suitable coordinates for the individual rows and walls along the $x_2\sim\hat x_2$ direction.
Let us extend the notation $Y^\star$ to all three choices of framing, as follows
\be\label{eq:star-def}
	Y^\star := \left\{\begin{array}{ll}
		\overline Y^t \qquad & \qquad  (p=0)\\
		\overline Y \qquad & \qquad (p=\pm 1)
	\end{array}\right.
\ee
in agreement with the observations made in Section \ref{sec:self-statistics}.
In all three cases the asymptotic partitions along the common direction are determined by pairs of Young diagrams $(\lambda, {\lambda}^{\star})$ and $(\hat{\rho}^{\,\star},\hat{\rho})$. 
To describe the coordinate functions of  the $\blacksquare$s (or $\overline{\blacksquare}$s) in these Young diagrams, it is enough to focus on the representation $\lambda$ and $\hat{\rho}$. 
Each box in the Young diagram $\lambda$ is labeled by $\blacksquare$, and its position described by the coordinates
\begin{equation}
\blacksquare :\qquad x_1(\blacksquare)\,, x_3(\blacksquare)=0,1,2,3 \ldots \ .
\end{equation}

Since $\blacksquare$ is ``visible'' from both sides, i.e.\ $\blacksquare$ is inside the Young diagram $\lambda$ on the left and the dual ${\lambda}^{\star}$ on the right, it has  two coordinate functions simultaneously:
\begin{equation}
g(\blacksquare)\equiv x_1(\blacksquare) h_1+ x_3(\blacksquare) h_3\ ,  
\qquad
\hat{g}(\blacksquare)\equiv -x_3(\blacksquare)\hat  h_1- x_1(\blacksquare) \hat h_3+ h_2-\hat \sigma_3\hat{\psi}_0\label{ghatdefx} \ ,
\end{equation}
Note the parallel between $\hat{g}({\blacksquare})$ and $\hat{h}(\hat\square)$ when $\hat\square$ is atop the right wall.
\smallskip

Similarly, asymptotic partitions $(\hat{\rho}^{\,\star}, \hat\rho)$ along the internal leg are made of ``anti-boxes'' in the Young diagram $\hat{\rho}$, which we label by $\overline{\blacksquare}$. 
We describe their positions with  coordinates
\begin{equation}
\overline{\blacksquare} :\qquad \hat{x}_1(\overline{\blacksquare})\,, \hat{x}_3(\overline{\blacksquare})=0,1,2,3  \ldots \ .
\end{equation}
Once again these configurations are visible on both sides, and therefore have two coordinate functions
\begin{equation}
g(\overline\blacksquare)\equiv-\hat{x}_3(\overline{\blacksquare}) \, h_1- \hat{x}_1(\overline{\blacksquare}) \, h_3+ h_2-\sigma_3\psi_0
\qquad \hat{g}(\overline\blacksquare)\equiv\hat{x}_1(\overline{\blacksquare}) \, \hat h_1+ \hat{x}_3(\overline{\blacksquare}) \, \hat h_3 \ . \label{ghatdef}
\end{equation}
Note the parallel between $g(\overline{\blacksquare})$ and $h(\square)$ when $\square$ is atop the left wall.

The charge functions of infinite rows and walls feature a ``transposition'' on one side compared to the other: for example in (\ref{ghatdefx}) $x_3$ pairs with $\hat h_1$ while  $x_1$ pairs with $\hat h_3$, while a similar statement with hats exchanged applies to (\ref{ghatdef}). 
When $p=0$,  the transpose arises from the definition of $\lambda^\star$, see (\ref{eq:star-def}). For $p=\pm1$ instead it is due to the switching of symmetric and anti-symmetric axes, see (\ref{eq:axes-conventions}).

For later convenience we define another coordinate function for $\blacksquare$ and $\bar{\blacksquare}$:
\begin{align}
h(\blacksquare) & =  - x_3(\blacksquare) \hat h_1 - x_1(\blacksquare) \hat h_3 
=\hat{g}(\blacksquare) + \hat \sigma_3\hat{\psi}_0- h_2\label{hblack}\\
\hat{h}(\overline{\blacksquare}) & = - \hat{x}_3(\overline{\blacksquare})  h_1 - \hat{x}_1(\overline{\blacksquare})  h_3 =g(\overline{\blacksquare}) + \sigma_3\psi_0-h_2\label{hbarblack} \ . 
\end{align}
As explained in Section~4.5 of \cite{Gaberdiel:2018nbs},  $h(\blacksquare)$ is directly related to the pole when adding a $\blacksquare$, whereas $\hat{g}(\blacksquare)$ is in the conjugate charge function of  $\blacksquare$. 
And similarly for $\hat{h}(\overline{\blacksquare})$ and $g(\overline{\blacksquare})$, respectively. 

\begin{table}[h!]
\centering
\vspace{.10\textwidth}
 \rotatebox{90}{
 \begin{varwidth}{.89\textheight}
\resizebox{1.15\columnwidth}{!}{
\begin{tabular}{ccccc}
 & $\psi(u)$ 
 & $\hat{\psi}(u)$&$P(u)$
 & $\bar{P}(u)$ \vspace{4pt}\\ 
 \hline  \\[-10pt]
 \hbox{vac.} & $\psi_0(u)$ & $\hat{\psi}_0(u)$ & $\Bigl(1 + \frac{\sigma_3 \psi_0}{u}\Bigr) \, \Bigl( 1 - \frac{\hat \sigma_3 \hat{\psi}_0}{u} \Bigr)  $ &$
\Bigl(1 + \frac{\hat \sigma_3 \hat{\psi}_0}{u}\Bigr) \, \Bigl( 1 - \frac{\sigma_3 {\psi}_0}{u} \Bigr)  $ \vspace{4pt} \\
 \hline  \\[-10pt]
$\square$ 
& $\varphi_3(u-h(\square))$ 
& $1$
& $\varphi^{-1}_2(-u+h(\square))$ 
& $\varphi^{-1}_2(u-h(\square)-\sigma_3 \psi_0)$ \vspace{4pt}\\
\hline  \\[-10pt]
$\hat{\square}$ 
&  $1$ 
& $\hat \varphi_3(u-\hat{h}(\hat\square))$ 
& $\hat \varphi^{-1}_2(u-\hat{h}(\hat\square)-\hat \sigma_3 \hat{\psi}_0)$ 
& $\hat \varphi^{-1}_2(-u+\hat{h}(\hat\square))$  \vspace{4pt} \\
\hline  \\[-10pt]
$\blacksquare$ 
& $\varphi_2(u-g(\blacksquare))$ 
& $\hat \varphi^{-1}_2(-u+h(\blacksquare)-\hat \sigma_3 \hat{\psi}_0)$ 
& ${\displaystyle  S_p\bigl(u - g(\blacksquare) \bigr)\!\!  \prod^{m(1-p)+n(1+p)-1}_{k=0}\hat{\varphi}^{-1}_2(-u+g(\blacksquare)+k h_2)}$
& ${\displaystyle {S}_p\bigl( u- \hat g(\blacksquare) \bigr) \!\!
	\prod^{m(1-p)+n(1+p)+1+2\rho}_{k=0} \varphi^{-1}_2(u-\hat g(\blacksquare)+k h_2 )}$ \vspace{4pt} \\
\hline  \\[-10pt]
$\overline{\blacksquare}$ 
& $\varphi^{-1}_2(-u+\hat{h}(\overline{\blacksquare})-\sigma_3 \psi_0)$
& $\hat \varphi_2(u-\hat{g}(\overline{\blacksquare}))$ 
&${\displaystyle 
S_p\bigl(g(\overline\blacksquare) - u\bigr) \!\!\prod^{\hat{m}(1+p) +\hat{n}(1-p) +1+2\rho}_{k=0}  \hat \varphi^{-1}_2(u- {g}(\overline{\blacksquare})+k h_2)
}$
& ${\displaystyle {S}_p\bigl( \hat g(\overline{\blacksquare}) - u)  \bigr)\!\! \prod^{\hat{m}(1+p)+\hat{n}(1-p)-1}_{k=0} \varphi^{-1}_2(-u+\hat{g}(\overline{\blacksquare})+k h_2)}$ \vspace{4pt} \\ 
\hline
\end{tabular}
}
\caption{The eigenvalues of the different factors. Recall that $h(\blacksquare) =\hat{g}(\blacksquare) + \hat \sigma_3\hat{\psi}_0- h_2$ and $\hat{h}(\overline{\blacksquare}) =g(\overline{\blacksquare}) + \sigma_3\psi_0-h_2$
}\label{tab2}
\end{varwidth}
}
\end{table}

\subsection{Charge functions of twin plane partitions}
\label{sec:TPPcharge}
\subsubsection{$(\psi,\hat{\psi})$ charge functions}
\label{sec:TPPchargepsi}

The charge (eigen-)functions of operators  $(\psi(u),\hat{\psi}(u))$ can be factorized according to different contributions:
\begin{equation}\label{chargefunctions}
\begin{cases}
\bm{\Psi}_{\Lambda}(u) \equiv \psi_0(u) \cdot \psi_{\lambda}(u)\cdot \psi_{\hat{\rho}}(u)\cdot \psi_{\mathcal{E}}(u)\\
\hat{\bm{\Psi}}_{\Lambda}(u)\equiv \hat{\psi}_0(u) \cdot \hat{\psi}_{\lambda}(u)\cdot \hat{\psi}_{\hat{\rho}}(u)\cdot\hat{\psi}_{\hat{\mathcal{E}}}(u) 
\end{cases}
\end{equation}

First, the vacuum factors $\psi_0(u)$ and $\hat{\psi}_0(u)$ are 
\begin{equation}
\psi_0(u)\equiv1+ \frac{\sigma_3\psi_0}{u} \qquad\textrm{and} \qquad \hat{\psi}_0(u)\equiv1+\frac{\hat{\sigma_3}  \hat{\psi}_0}{u} \ .
\end{equation}
The bi-module $(\lambda, \lambda^{\star})$ contributes to both charges functions:
\begin{equation}\label{xforpsi}
\begin{aligned}
&\begin{cases}
\psi_{\lambda}(u)=\prod_{\blacksquare\in\lambda}\psi_{\,\blacksquare}(u)\\
\hat{\psi}_{\lambda}(u)=\prod_{\blacksquare\in \lambda}\hat{\psi}_{\,\blacksquare}(u)
\end{cases} \quad \textrm{with}  \quad \begin{cases}
\psi_{\,\blacksquare}(u)\equiv\varphi_2(u-g(\blacksquare))\\
\hat{\psi}_{\,\blacksquare}(u)\equiv\varphi^{-1}_2(u-\hat{g}(\blacksquare))
\end{cases}
\end{aligned}
\end{equation}
and similarly for $(\rho, \rho^{\star})$:
\begin{equation}\label{xbarforpsi}
\begin{aligned}
&\begin{cases}
\psi_{\hat{\rho}}(u)=\prod_{\overline{\blacksquare}\in \hat{\rho}}\psi_{\, \overline{\blacksquare}}(u)\\
\hat{\psi}_{\hat{\rho}}(u)=\prod_{\overline{\blacksquare}\in \hat{\rho}}\hat{\psi}_{\, \overline{\blacksquare}}(u)
\end{cases} \quad \textrm{with} \quad \begin{cases}
\psi_{\, \overline{\blacksquare}}(u)\equiv \varphi^{-1}_2(u-g(\overline{\blacksquare}))\\
\hat{\psi}_{\, \overline{\blacksquare}}(u)\equiv\varphi_2(u-\hat{g}(\overline{\blacksquare})) \ , 
\end{cases}
\end{aligned}
\end{equation}
The boxes in $\mathcal{E}$ (i.e.\ at the left corner)  only contributes to $\psi(u)$
\begin{equation}\label{eforpsi}
\psi_{\mathcal{E}}(u)=\prod_{\square\in\mathcal{E}}\psi_{\,\square}(u)
 \qquad \textrm{with} \qquad
\psi_{\,\square}(u)\equiv\varphi_3(u-h(\square)) \ . 
\end{equation}
Those in $\hat{\mathcal{E}}$ (i.e.\ at the right corner)  only contributes to $\hat{\psi}(u)$
\begin{equation}\label{hateforpsi}
\hat{\psi}_{\hat{\mathcal{E}}}(u)=\prod_{\hat{\square}\in\mathcal{E}}\hat{\psi}_{\,\hat{\square}}(u)
 \qquad \textrm{with} \qquad
\hat{\psi}_{\,\hat{ \square } }(u)\equiv\varphi_3(u-\hat{h}(\hat{\square}))\ . 
\end{equation}

\subsubsection{$(P,\bar{P})$ charge functions}

The form of the $(P,\bar{P})$ charge function was given in (\ref{XLambda}). 
Note that unlike ($\psi,\hat{\psi}$) function, which only see the left and right corner, respectively,  the $(P,\bar{P})$ function contain four types of contributions, since they control the action of the gluing operators, thus see both corners at the same time. 
We will only be able to derive these four contributions in Section.\ \ref{sec:residue}.

\section{OPE between corner operators and gluing operators}
\label{sec:pole}

The following two sections are devoted to deriving the OPEs of the glued algebra, by demanding that it has a faithful representation on the set of twin plane partitions. 
The strategy is summarized in Section~\ref{sec:strategyoutline}. 
We will now outline the procedure in more detail. 

Now we focus on (\ref{Oform}) again
\begin{equation}\label{Oform1}
	\mathcal{O}(w) |\Lambda\rangle = \sum_{i} \frac{O[\Lambda \rightarrow \Lambda'_i]}{w-w^*_{i}} |\Lambda_{i}'\rangle
\end{equation}
There are two aspects to (\ref{Oform1}) : pole and residue. 
The poles determine the set of final state $|\Lambda'\rangle$, and are relatively easy to fix --- it only requires the knowledge of charge functions $(\Psi_{\Lambda}(u),\hat{\Psi}_{\Lambda}(u))$ for arbitrary $\Lambda$ and the OPEs between the charge operators $(\psi(z) , \hat{\psi}(z))$ and $\mathcal{O}(w)$. 
To determine the residue, one needs to consider successive action of more than one $\mathcal{O}$'s and adjust the residue such that it's compatible with existing OPEs relations. 

For the procedure to work, it is important that there is a one-to-one correspondence between the pole $w^{*}_i$ and $|\Lambda_{i}'\rangle$. 
This is guaranteed by their respective one-to-one correspondence with charge functions 
\begin{equation}
w^{*}_i \quad \longleftrightarrow \quad (\bm{\Psi}_{\Lambda'_i}(z) , \hat{\bm{\Psi}}_{\Lambda'_i}(z))\quad \longleftrightarrow \quad
|\Lambda_{i}'\rangle \end{equation}
First, for each twin plane partition, one can easily compute its charge function $(\bm{\Psi}_{\Lambda}(z) , \hat{\bm{\Psi}}_{\Lambda}(z))$, explained in section.\ \ref{sec:TPPchargepsi}.\footnote{
See section 4.3 of \cite{Gaberdiel:2018nbs}, which we suitably generalize here to the case of two additional parameters.} 
Moreover, their structure is such that given a pair of charge functions, one can immediately see whether it  corresponds to an allowed twin plane partition configuration or not.\footnote{
By ``allowed plane partition", we mean there shouldn't be box outside the room, or floating in the air, or more than one box in one position, etc.
}

To proceed, we need the OPE between  $(\psi(z) , \hat{\psi}(z))$ and $\mathcal{O}(w)$: in general form these are 
\begin{equation}
\begin{cases}\begin{aligned}
\psi(z) \, \mathcal{O}(w)&\sim \phi[\mathcal{O}](z-w) \,  \mathcal{O}(w)\, \psi(z) \\
\hat{\psi}(z) \, \mathcal{O}(w)&\sim \hat{\phi}[\mathcal{O}](z-w) \,  \mathcal{O}(w)\, \hat{\psi}(z) 
\end{aligned}
\end{cases}
\end{equation}
The OPE coefficients are just the charges of various (leading) configurations created or destroyed by these operators.
For creation operators:
\be\label{charge1}
\begin{aligned}
\phi[e](u) &\equiv \varphi_3(u)\,, \quad  \phi[\hat{e}](u)\equiv 1\,, \quad \phi[x](u) \equiv \varphi_2(u)\,, \quad\phi[\bar{x}](u)\equiv \varphi^{-1}_2(-u-\sigma_3 \psi_0)
\\
\hat{\phi}[e](u)&\equiv 1\,, \quad \hat{\phi}[\hat{e}](u)\equiv \hat{\varphi}_3(u) \,,\quad
\hat{\phi}[x](u)\equiv \hat{\varphi}_2^{-1}(-u-\hat{\sigma}_3\hat{\psi}_0)\,, \quad \hat{\phi}[\bar{x}](u)\equiv \hat{\varphi}_2(u) 
\end{aligned}
\end{equation}
Those for the annihilation operators are the inverse of the one for the corresponding creation operators.

Applying the charge operators $(\psi(z) , \hat{\psi}(z))$ to
(\ref{Oform1}) gives
\begin{equation}
\begin{cases}
\begin{aligned}
\bm{\Psi}_{\Lambda'_i}(z) =    \phi[\mathcal{O}](z-w^{*}_i) \bm{\Psi}_{\Lambda}(z)\\
\hat{\bm{\Psi}}_{\Lambda'_i}(z) =    \hat{\phi}[\mathcal{O}](z-w^{*}_i) \bm{\hat{\Psi}}_{\Lambda}(z)
\end{aligned}
\end{cases}
\end{equation}
Since the initial state $|\Lambda\rangle$ corresponds to an allowed twin plane partition, one can easily compute its charge function $(\bm{\Psi}_{\Lambda}(z),\bm{\hat{\Psi}}_{\Lambda}(z))$.
Since we also know the OPE function $( \phi[\mathcal{O}](\Delta),  \hat{\phi}[\mathcal{O}](\Delta))$, one can always adjust the $w^{*}$ such that the final state's charge function $(\bm{\Psi}_{\Lambda'_i}(z),\hat{\bm{\Psi}}_{\Lambda'_i}(z))$ would correspond to an allowed twin plane partition.\footnote{
If such a pole $w^{*}$ does not exist, it means $\mathcal{O}$ annihilates the initial state  $\Lambda$.}
The set of admissible poles, over which the sum in (\ref{Oform1}) runs, are all those that make $\Lambda_i'$ and admissible twin plane partition.

The next step is to fix the residue $O_i[\Lambda \rightarrow \Lambda'_i]$, which depends both on the operator $\mathcal{O}_i$ and the twin-plane-partition $\Lambda$.
This is done by demanding compatibility both with the OPE between $\mathcal{O}_i$ and  $(\psi(z) , \hat{\psi}(z))$ and with 
the $\mathcal{O}_i\cdot \mathcal{O}_j$ OPE.\footnote{
See section 6.1. of \cite{Gaberdiel:2018nbs}, which will be slightly generalized in Section~7.2. in this paper to allow for the two additional parameters.
}

\subsection{Action of gluing operators on generic twin plane partitions: pole structures}
\label{sec:glueactiongeneric}

Recall that a generic initial twin plane partition state $\Lambda$ consists of four parts $(\lambda, \hat{\rho}, \mathcal{E}, \hat{\mathcal{E}})$, see Section~\ref{sec:TPPcharge}. 
Just like $e,f$ and $\hat e,\hat f$ act as creation/annihilation operators on $\mathcal{E},\hat{\mathcal{E}}$, there is a similar interpretation for the action of gluing operators on $\lambda,\hat\rho$ (see Sections \ref{eq:heisenberg-intuition} and \ref{eq:heisenberg-intuition-bis}).
Motivated by this, we consider the following ansatz for the actions of $x(w)$ and $\bar{x}(x)$ on a generic twin plane partition $|\Lambda\rangle$:
\begin{equation}\label{xansatz}
\begin{aligned}
x(w)|\Lambda\rangle &=\sum_{\blacksquare\in \textrm{Add}(\lambda)}\frac{
\Big[ \textrm{Res}_{u=p_{+}({\blacksquare})}  \textbf{P}_\Lambda(u) \Big]^{\frac{1}{2}} 
}{w-p_+(\blacksquare)}|[\Lambda+\blacksquare]\rangle+ \sum_{\overline{\blacksquare}\in \textrm{Rem}(\hat{\rho})}\frac{
\Big[ \textrm{Res}_{u=p_{-}(\overline{\blacksquare})} \textbf{P}_\Lambda(u) \Big]^{\frac{1}{2}}
}{w-p_-(\overline{\blacksquare})}|[\Lambda-\overline{\blacksquare}]\rangle  \ , \\
\bar{x}(w)|\Lambda\rangle &=\sum_{\overline{\blacksquare}\in \textrm{Add}(\hat{\rho})} \frac{
 \Big[ \textrm{Res}_{u=\bar{p}_{+}(\overline{\blacksquare})} \, \bar{\textbf{P}}_\Lambda(u) \Big]^{\frac{1}{2}}
}{w-\bar{p}_+(\overline{\blacksquare})}|[\Lambda+\overline{\blacksquare}]\rangle
+ \sum_{\blacksquare\in \textrm{Rem}(\lambda)} \frac{
 \Big[ \textrm{Res}_{u=\bar{p}_{-}({\blacksquare})} \, \bar{\textbf{P}}_\Lambda(u) \Big]^{\frac{1}{2}} 
}{w-\bar{p}_-({\blacksquare})}| [\Lambda-{\blacksquare}]\rangle  \ , 
\end{aligned}
\end{equation}
For $x$ the sum is over all positions where a box $\blacksquare$ can be added to the Young diagram $\lambda$, or removed from the Young diagram $\hat{\rho}$, and similarly for $\bar{x}$.
We also introduced the notation $[\Lambda+\blacksquare]$ to denote the resulting twin plane partition configuration from adding $\blacksquare$ to $\Lambda$, and sum over all possibilities that result in consistent twin-plane partitions. 
The action of the corresponding annihilation operators works similarly:
\begin{equation}\label{yansatz}
\begin{aligned}
y(w)|\Lambda\rangle &= \!\!\! \sum_{\blacksquare\in \textrm{Rem}(\lambda)}
\frac{ \Big[ \textrm{Res}_{u=p_{+}({\blacksquare})} \, \textbf{P}_\Lambda(u) \Big]^{\frac{1}{2}}
}{w-p_+(\blacksquare)} \, |[\Lambda-\blacksquare]\rangle
+ \!\!\! \sum_{\overline{\blacksquare}\in \textrm{Add}(\hat{\rho})}\frac{\Big[ \textrm{Res}_{u=p_{-}(\overline{\blacksquare})} \, \textbf{P}_\Lambda(u) \Big]^{\frac{1}{2}}}{w-p_-(\overline{\blacksquare})}|[\Lambda+\overline{\blacksquare}]\rangle  \ , \\
\bar{y}(w)|\Lambda\rangle &= \!\!\!\sum_{\overline{\blacksquare}\in \textrm{Rem}(\hat{\rho})}
\frac{ \Big[ \textrm{Res}_{u=\bar{p}_+(\overline{\blacksquare})} \, \bar{\textbf{P}}_\Lambda(u) \Big]^{\frac{1}{2}}
}{w-\bar{p}_+(\overline{\blacksquare})}
\, |[\Lambda-\overline{\blacksquare}]\rangle
+ \!\!\!\sum_{\blacksquare\in \textrm{Add}(\lambda)} 
\frac{\Big[ \textrm{Res}_{u=\bar{p}_{-}({\blacksquare})} \, \bar{\textbf{P}}_\Lambda(u) \Big]^{\frac{1}{2}}
}{w-\bar{p}_-({\blacksquare})}|[\Lambda+{\blacksquare}]\rangle  \ , 
\end{aligned}
\end{equation}

We will now proceed by fixing the set of admissible final states for a given $\Lambda$, i.e. we will derive the poles for the action of gluing operators.
Working our way up from the simplest configurations, we will eventually derive the poles for a  fully generic $\Lambda$.

\subsubsection{Action of $x$ and $\bar{x}$ on $|\square\rangle$ and $|\hat\square\rangle$}

The action of $x$ and $\bar{x}$ on one existing box $|\square\rangle$ (or one hatted box $|\hat\square\rangle$), does not depend on the conformal dimension of $x$ (i.e.\ $h=1+\rho$) or the framing parameter $p$. 
In fact the analysis turns out to be just identical to the one in \cite{Gaberdiel:2018nbs}. 
Here we summarize the result, a detailed derivation can be found in Appendix ~\ref{app:xxbaronbox} 
\begin{equation}\label{xonefinal}
\begin{aligned}
& x(w) |\square\rangle \sim \frac{1}{w-h_2} \, |\blacksquare+\hat{\square}_{\, \textrm{top}}\rangle  \qquad\quad \bar{x}(w) |\hat{\square}\rangle \sim \frac{1}{w-h_2} \, |\overline{\blacksquare}+{\square}_{\, \textrm{top}}\rangle  \\
&x(w) |\hat{\square}\rangle \sim \frac{1}{w} \, |\blacksquare+\hat{\square}_{\, \textrm{0}}\rangle  \qquad \qquad \qquad
 \bar{x}(w) |\square\rangle \sim \frac{1}{w} \, |\overline{\blacksquare}+{\square}_{\, 0}\rangle  
\end{aligned}
\end{equation}

\subsubsection{Action of $x$ on $ |\blacksquare\rangle$}\label{sec:x-x-on-vac}

The next simplest one is
\be
\begin{aligned}
x(z) \, |\blacksquare\rangle
& \sim  \, \sum_{i} \frac{1}{z-z^{*}_i} \, |\Phi^{xx}_{i}\rangle \ ,
\end{aligned}
\end{equation}
The charge functions of the resulting state $|\Phi^{xx}_{i}\rangle$ are
\begin{equation}\label{psixx}
|\Phi^{xx}_{i}\rangle:\qquad 
\begin{cases}
\begin{aligned}
\bm{\Psi}_{\Lambda'}(u) &= \psi_0(u)\cdot \varphi_2(u-z^{*}_i) \cdot \varphi_2(u) \\
\hat{\bm{\Psi}}_{\Lambda'}(u) &=\hat{\psi}_0(u)\cdot \hat\varphi^{-1}_2(-(u-z^{*}_i)-\hat \sigma_3\hat{\psi}_0)\cdot \hat \varphi^{-1}_2(-u-\hat \sigma_3\hat{\psi}_0)  \ . 
\end{aligned}
\end{cases}
\end{equation}
One might expect that $|\Phi^{xx}_{i}\rangle$ be
one of the two configurations corresponding to two infinite rows of boxes, with charge functions
\begin{equation}\label{eq:boxbox-1-bosonic}
|\blacksquare\blacksquare_{1}\rangle:\qquad 
\begin{cases}
\begin{aligned}
\bm{\Psi}_{\blacksquare\blacksquare_{1}}(u) &= \psi_0(u)\cdot \varphi_2(u) \cdot \varphi_2(u-h_1) \\
\hat{\bm{\Psi}}_{\blacksquare\blacksquare_{1}}(u) &=\hat{\psi}_0(u)\cdot \hat \varphi^{-1}_2(-u-\hat \sigma_3\hat{\psi}_0) \cdot \hat \varphi^{-1}_2(-u-\hat h_3-\hat \sigma_3\hat{\psi}_0) \ ,
\end{aligned}
\end{cases}
\end{equation}
and
\begin{equation}\label{eq:boxbox-3-bosonic}
|\blacksquare\blacksquare_{3}\rangle:\qquad  
\begin{cases}
\begin{aligned}
\bm{\Psi}_{\blacksquare\blacksquare_{3}}(u) &= \psi_0(u)\cdot \varphi_2(u) \cdot \varphi_2(u-h_3) \\
\hat{\bm{\Psi}}_{\blacksquare\blacksquare_{3}}(u) &=\hat{\psi}_0(u)\cdot \hat \varphi^{-1}_2(-u-\hat \sigma_3\hat{\psi}_0) \cdot \hat \varphi^{-1}_2(-u-\hat h_1-\hat \sigma_3\hat{\psi}_0) \ .
\end{aligned}
\end{cases}
\end{equation}
Now to match with the first case one should take 
\begin{equation}z_i^* = h_1 = -\hat h_3
\end{equation} which is possible if $h_i$ and $\hat h_i$ are related as in (\ref{eq:h-hhat-p}) with $p=1$. The second possibility is instead realized if one takes 
\begin{equation}
z_i^* = h_3 = -\hat h_1
\end{equation} 
which means $p=-1$ in (\ref{eq:h-hhat-p}). Thus we have 
\begin{equation}\label{eq:xx-0-p1}
x(w)\cdot x(z) |\emptyset\rangle \sim 
\begin{cases}
\begin{aligned}
\frac{1}{z}\frac{1}{w-h_1} |\blacksquare\blacksquare_1\rangle\qquad \qquad & p=1 \\
0\qquad \qquad  \qquad\qquad&p=0\\
\frac{1}{z}\frac{1}{w-h_3} |\blacksquare\blacksquare_3\rangle \qquad \qquad & p=-1
\end{aligned}
\end{cases}
\end{equation}

On the face of it, (\ref{eq:boxbox-1-bosonic}) seems to describe two rows next to each other along $x_1$ and two walls on the hatted side placed next two each other along the $\hat x_3$ direction.
Therefore these charge functions would seem to describe the box configurations where we have the conjugate of the \emph{transpose} of $\blacksquare\blacksquare_{1}$ (and similarly with $\blacksquare\blacksquare_{3}$ for charges in (\ref{eq:boxbox-3-bosonic})). 
This may appear to pose a problem since we have seen in (\ref{eq:char-decomp}) that on the $\hat{\cal Y}$ side the conjugate representation (as opposed to the conjugate of the transpose) should appear.

However let us recall that in (\ref{eq:axes-conventions}) we argued that the notion of transpose (or not transpose) along the common direction should be understood  with respect to an appropriate choice of coordinates on either side.
When $p=\pm1$ one should compare shapes of asymptotic partitions $(Y,Y^\star)$ through the identifications $\hat x_1 \sim x_3$ and $\hat x_3 \sim x_1$ which explains the naive appearance of a transpose (also see discussion of coordinate functions (\ref{ghatdefx}) and (\ref{ghatdef})).
In any case, for $p=\pm 1$ we have checked that the conformal dimensions of the above states are such that the character calculation (\ref{eq:char-decomp}) comes out correctly, following the procedure outlined in Section \ref{sec:vacuum-char-analysis-bose}. 

\subsubsection{Minimal buds for creation of $\blacksquare$ by $x$}
\label{sec:minimalBud}

Above we have analyzed the result of repeated action by $x$, and found that multiple $\blacksquare$ are only created if $p=\pm1$, and can only be stacked along certain directions summarized in (\ref{eq:xx-0-p1}).
Now we will discuss how to create other configurations, such as a configuration with two $\blacksquare$ when $p=0$, or $|\blacksquare\blacksquare_3\rangle$ when $p=1$.

The main idea is to repeat the analysis of Section \ref{sec:x-x-on-vac}, but replacing the initial configuration $|\blacksquare\rangle$ with a configuration $|\blacksquare+\square\cdots\square_i\rangle$ corresponding to $s_i$ extra boxes stacked along the infinite row, and displaced by $h_i$ for $i=1,3$.
The analysis is somewhat tedious but straightforward and can be found in Appendix \ref{sec:minimal-buds-appendix}.
The upshot is that $x$ will create a second row displaced by $h_3$, if the number of extra boxes is $s_3=2$ and $p=1$
\be\label{eq:x-bb-ww3-partial}
\begin{split}
	x(z)  |\blacksquare + {\square}{\square}_3\rangle  
	&  \ \supset \ 
	   \frac{(\#)}{z-h_3-2h_2}\,   |\blacksquare\blacksquare_3\rangle   
	\qquad ({p=1})
\end{split}
\ee
Likewise it turns out that the new row will be displaced by $h_1$ with $s_1=2$, if $p=-1$.
For $p=0$, the same analysis reveals the possibility to create a new row displaced either by $h_1$ or by $h_3$, respectively with $s_1=1$ or $s_3=1$.
These results have the following interpretations.

For $p=1$ we can only stack rows along the $x_3$ direction provided there is a ``bud'' of length two placed next to the initial row, in the $x_3$ direction. Then the second row will attach to the end of this bud.  (no bud is necessary to stack rows along the $x_1$ direction in this case)

The situation is similar for $p=-1$; in this case one needs a length-two bud displaced in the $x_1$ direction in order to create row stacked on the first one along the $x_1$ direction.
(no bud is necessary to stack rows along the $x_3$ direction in this case)

In the fermionic case $p=0$, the situation is more symmetric. 
The analysis carried out in \cite{Gaberdiel:2018nbs} applies essentially unaltered by the shifting parameter $\rho$, and leads to the conclusion that in this case one needs a bud of length one to support the second row, \emph{both} for stacking two rows along $x_1$ and for stacking them along $x_3$. 
This is compatible with the nilpotency-on-vacuum
property $x\cdot x \, |\emptyset\rangle\sim 0$ seen in (\ref{eq:xx-0-p1}), and the fermionic nature of $x$, since in the absence of a bud would lead to a disallowed twin plane partition.

While the appearance of these requirements on minimal buds may seem somewhat opaque from the viewpoint of the algebra, recall that there is an intuitive geometric explanation for all three cases, see Section \ref{sec:geometry}. 
In fact, let us also recall that a more precise prediction on the length of buds was also found from the character analysis of Section \ref{sec:vacuum-char-analysis-bose}. 
The length-one buds for $p=0$ can be seen in the leading $q$-powers of the $y^2$-coefficient in (\ref{eq:fermionic-char-buds}), while the length-two and length-zero for $p=\pm1$ can be seen in (\ref{eq:bosonic-char-buds}).

\paragraph{Length of minimal bud at generic position.}
Extending the above analysis of charge functions to more general twin plane partitions, can be shown to lead to the following  general rule. 
In order to create 
an infinite row along the $x_2$ direction at $(x_1,x_3)=(m,n)$ one must first create a bud of length 
\be\label{eq:x-bud}
	\boxed{m(1-p) + n(1+p)} \ . 
\ee
For instance for $p=1$ and with an infinite row at $(m,n)=(0,0)$ the next row can be created at $(m,n)=(1,0)$ without a bud (length zero) by just acting with another $x$; or it can be created at $m=0,n=1$ with a bud of length two, by acting with $x\cdot e\cdot e$ (the action of $e\cdot e$ will provide the two boxes that make up the length-two bud). The situation is reversed for $p=-1$.

We should also mention that in general, multiple configurations can be created simultaneously, resulting in a superposition of states. For example, if $x(z)$ acts on 
$ |\blacksquare + {\square}{\square}_3\rangle  $ it creates two configurations
\be\label{eq:x-bb-ww3-complete}
\begin{split}
	x(z)  |\blacksquare + {\square}{\square}_3\rangle  
	& \sim 
	   \frac{(\#)}{z-h_3-2h_2}\,   |\blacksquare\blacksquare_3\rangle  
	+ \frac{(\#)}{z-h_1} \,    |\blacksquare\blacksquare_1 + {\square}{\square}_3\rangle  
	\qquad ({p=1})
\end{split}
\ee

\subsubsection{Action of $x$  on generic twin plane partitions}

Above we analyzed the creation of $\blacksquare$'s by acting with $x$ on generic twin plane partitions, and derived the length of minimal buds that are required for various $p$, at various positions $(m,n)$.
Now we consider two generalizations. 
First, we ask what happens if a bud is longer than required by (\ref{eq:x-bud}).
Second, we recall that acting with $x$ on a generic twin plane partition can have two effects, depending on the specific shape of the partition: it can either add a row on the unhatted side (and a wall on the hatted side), or it can remove a wall on the unhatted side (and therefore a row on the hatted side).

\paragraph{Adding}
In order for $x$ to be allowed to add an infinite row along the $x_2$ direction at $(x_1,x_3)=(m,n)$, first it should be possible to add a new box $\blacksquare$ at $(x_1,x_3)=(m,n)$ to the Young diagram $\lambda$. 
In addition, there needs to be a bud of at least $m(1-p) + n(1+p)$ boxes extending in the $x_2$ direction at $(x_1,x_3)=(m,n)$.
Accordingly, the poles are located at
\begin{equation}\label{p+f}
p^{(\ell)}_+(\blacksquare) \equiv g(\blacksquare)+  [m(1-p) + n(1+p)]h_2=h(\blacksquare) +\ell h_2 \ ,
\end{equation}
where $\ell$ is the number of additional boxes extending the bud of minimal length along the $x_2$ direction.
\smallskip

Again, the key idea is to analyze how the charge function behaves upon acting with $x$.
In appendix \ref{app:xxbaronbox} we carry out a detailed analysis for the case  $m=n=0$ and $\ell=1$. This turns out to give
\begin{equation}\label{eq:xonbox-main}
 x(w) |\square\rangle \sim \frac{1}{w-h_2} \, |\blacksquare+\hat{\square}_{\, \textrm{top}}\rangle \ ,
\end{equation}
implying that the extra box extending the minimal bud gets pushed atop the high-wall on the hatted side, as previously observed in the $p=0, \rho=1/2$ case by \cite{Gaberdiel:2018nbs} (the statement now holds also for $p=\pm1$, and any $\rho$).

More generally, if the minimal bud is extended along the $x_2$ direction by $\ell$ additional $\square$s, one finds:
\begin{equation}
x(w) |\Lambda+\textrm{min.bud}[\blacksquare] + \square\square\dots \square_{\ell} \rangle \sim \frac{1}{w-p^{(\ell)}_+(\blacksquare) } \, |\Lambda+\blacksquare+\hat{\square}\hat{\square}\dots \hat{\square}_{\ell \, \textrm{top}}\rangle \,.
\end{equation}

\paragraph{Removing}
For the removal part of the action, let us assume that $\overline{\blacksquare} \in \hat{\rho}$ is a 
``removable box" at $(\hat{x}_1,\hat{x}_3)= (\hat{m},\hat{n})$. 
From the viewpoint of the left Yangian, this is part of the wall. Let us assume there are $\ell\geq 0$ unhatted boxes on top of it. 

Before we apply $x$, the $\overline{\blacksquare}$
at $(\hat{x}_1,\hat{x}_3)= (\hat{m},\hat{n})$ and  these $\ell$ boxes together contribute to the $(\psi, \hat{\psi})$ charge functions by
\begin{equation}
\begin{cases}
\begin{aligned}
\bm{\Psi}(u): &\qquad   \varphi^{-1}_2(u-g(\overline{\blacksquare})) \prod_{j=0}^{\ell-1} \varphi_3 \bigl(u - g(\overline{\blacksquare})- j h_2   \bigr) \label{blackbar} \\
\hat{\bm{\Psi}}(u): &\qquad \hat \varphi_2(u-\hat{g}(\overline{\blacksquare}))  \ .
\end{aligned}
\end{cases}
\end{equation}
Note the following useful identity
\begin{equation}\label{magic}
\varphi_2(u-h_2) \, \varphi_3(u)  =  \varphi_2(u) \ .
\end{equation}
Applying (\ref{magic}) recursively, and also using
\begin{equation}\label{varphi2m}
\varphi_2(- u) = \varphi_2(u- h_2) \ , 
\end{equation}
the $\bm{\Psi}(u)$ part of (\ref{blackbar}) can be rewritten as
\begin{equation}
\begin{aligned}
\bm{\Psi}(u): &\qquad \varphi_2^{-1}\bigl(u - (g(\overline{\blacksquare}) +\ell h_2 )\bigr) \ .
\end{aligned}
\end{equation}

The effect of $x$ on the removable box $\overline{\blacksquare}$ can be directly seen from its contribution to the $(\psi, \hat{\psi})$ charge functions.  
An $x(z)$ action with the pole at 
\begin{equation}\label{p-f}
	p^{(\ell)}_-(\overline{\blacksquare})
	\equiv
	\hat{h}(\overline{\blacksquare}) + (\ell+1) h_2 -  \sigma_3  \psi_0 
	=g(\overline{\blacksquare})+\ell h_2  \ , 
\end{equation}
contributes to the $(\psi, \hat{\psi})$ charge function by 
\begin{equation}\label{xkilling}
\begin{cases}
\begin{aligned}
\bm{\Psi}(u): &\qquad   \varphi_2\bigl(u - (g(\overline{\blacksquare}) + \ell h_2 )\bigr) 
 \\
\hat{\bm{\Psi}}(u): &\qquad \hat \varphi_2^{-1}\bigl(-u + (g(\overline{\blacksquare}) +\ell h_2 )- \hat \sigma_3\hat{\psi}_0\bigr) \ .
\end{aligned}
\end{cases}
\end{equation}
Combining (\ref{blackbar}) and (\ref{xkilling}) gives
\begin{equation}\label{eq:charges-wall-removal}
\begin{cases}
\begin{aligned}
\bm{\Psi}(u): &\qquad   1 
 \\
\hat{\bm{\Psi}}(u): &\qquad \prod_{j=0}^{\hat{m}(1+p)+\hat{n}(1-p)+\ell+1+2\rho} \hat\varphi_3\bigl(u - \hat{g}(\overline{\blacksquare}) - j  h_2  \bigr) \ .
\end{aligned}
\end{cases}
\end{equation}
We see that applying $x(u)$ with the pole (\ref{p-f}) removes the  $\overline{\blacksquare}$ at $(\hat{x}_1,\hat{x}_3)= (\hat{m},\hat{n})$, and replaces it by $(\hat{m}(1+p) +\hat{n}(1-p)+\ell+2+2\rho)$ hatted boxes, which sit at positions $(\hat{m},0,\hat{n}), (\hat{m},1,\hat{n}),\ldots, (\hat{m},\hat{m}(1+p)+\hat{n}(1-p)+\ell+1+2\rho,\hat{n})$ in the right plane partition.

Note the appearance of the shifting parameter $\rho$ in the form $2\rho+2$ (when $\hat{m}=\hat{n}=\ell=0$). Since the conformal dimension of each box is that of $e,\hat e$, and equal to $1$, the appearance of $2\rho+2$ boxes upon removal of a wall is to be expected, since this matches the dimension of $x \bar x |\emptyset\rangle$. Here $\bar x$ creates a wall, and $x$ removes it, and each gluing operator has conformal dimension $\rho+1$ (recall eq. (\ref{eq:rho-def})).

\subsubsection{Action of $\bar{x}$  on generic twin plane partitions}
\label{sec:xbar-gen-states}
The action of $\bar x$ is essentially identical to that of $x$, upon exchanging the two sides (hatted and unhatted). This operator creates rows of hatted boxes (and corresponding walls of unhatted boxes) or deletes walls of hatted boxes (together with corresponding rows of unhatted boxes). Once again, for the action to produce an allowed twin plane partition, certain requirements on the initial partition must be met, such as the presence of ``buds''.

For example, one readily finds the following poles for repeated action on the ground state of the vacuum module
\begin{equation}\label{eq:xx-0-p1}
\bar{x}(w)\cdot \bar{x}(z) |\emptyset\rangle \sim 
\begin{cases}
\begin{aligned}
\frac{1}{z}\frac{1}{w-\hat{h}_3}|\overline \blacksquare\overline \blacksquare_{\hat 3}\rangle\qquad \qquad & p=1 \\
0\qquad \qquad  \qquad\qquad&p=0\\
\frac{1}{z}\frac{1}{w-\hat{h}_1} |\overline \blacksquare\overline \blacksquare_{\hat 1}\rangle
 \qquad \qquad & p=-1
\end{aligned}
\end{cases}
\end{equation}
implying that no buds (or length-zero buds) are required to stack two hatted rows along  $\hat x_3 \sim x_1$ if $p=1$ and along $\hat x_1\sim x_3$ if $p=-1$. 
The derivation of these results can be found in Appendix \ref{app:boxbar-symm}.

Non-zero buds are required along other directions, and buds are always required if $p=0$.
More generally, in order to create an infinite (hatted) row along the $\hat x_2$ direction at $(\hat x_1,\hat x_3)=(\hat m,\hat n)$, it turns out that one must first introduce a bud of length 
\be\label{eq:x-bar-bud}
	\boxed{\hat m(1+p) + \hat n(1-p)}\,.
\ee
An example of this result is derived by direct computation in Appendix \ref{app:boxbar-antisymm}.
As expected, this condition is ``transposed" relative to the that for the $x$-action, cf.\ eq.~(\ref{eq:x-bud}). This follows from  (\ref{eq:axes-conventions}) and (\ref{eq:star-def}).

Taking into account both the addition and removal action of $\bar x$, we have the following general formula
\begin{equation}\label{barxansatz}
\bar{x}(w)|\Lambda\rangle =\sum_{\overline{\blacksquare}\in \textrm{Add}(\hat{\rho})} \frac{
 \Big[ \textrm{Res}_{u=\bar{p}_{+}(\overline{\blacksquare})} \, \bar{\textbf{P}}_\Lambda(u) \Big]^{\frac{1}{2}}
}{w-\bar{p}_+(\overline{\blacksquare})}|[\Lambda+\overline{\blacksquare}]\rangle
+ \sum_{\blacksquare\in \textrm{Rem}(\lambda)} \frac{
 \Big[ \textrm{Res}_{u=\bar{p}_{-}({\blacksquare})} \, \bar{\textbf{P}}_\Lambda(u) \Big]^{\frac{1}{2}} 
}{w-\bar{p}_-({\blacksquare})}|[\Lambda-{\blacksquare}]\rangle  \ , 
\end{equation}
where again summation is understood over all positions where a $\overline{\blacksquare}$ can be added and where a $\blacksquare$ can be removed. Accordingly, the poles correspond to the following positions 
\begin{align}
\label{eq:xbar-p-pm}
\bar{p}^{(\ell)}_+(\overline{\blacksquare}) & =  \hat{h}(\overline{\blacksquare}) + \ell  h_2 \qquad\textrm{and}\qquad
\bar{p}^{(\ell)}_-({\blacksquare})  =\hat{g}(\blacksquare) + \ell  h_2  \ ,
\end{align}
In the first equation $\ell$ counts the number of hatted boxes extending the bud of length (\ref{eq:x-bar-bud}) along the $\hat{x}_2$ direction,
while in the second one $\ell$ counts the number of hatted boxes atop the right high-wall corresponding to $\blacksquare$. 

\subsection{Partially fixing OPEs of gluing operators with single-box operators}
\label{sec:OPEcornerGluepartial}

Having fixed the poles of operators $x$ and $\bar x$, we can now employ this information to partially fix OPEs of these operators with single-box operators $e,\hat e, \dots$. 
The strategy will be to deduce constraints on the OPE coefficients from the action on twin-plane partitions. 
After a few illustrative examples, we will state the general outcome of this analysis.

\subsubsection{The $e \cdot x$ OPE }\label{sec:e-x}

As a first basic example, let us begin by studying the OPE of $e \cdot x$:
\be
	e(z)  \cdot x(w) \sim G(\Delta) \ x(w) \cdot e(z) \,.
\ee
Here as usual $\Delta= z-w$. 
As in the case with $p=0$ studied in \cite{Gaberdiel:2018nbs}, this OPE can be constrained by considering the action of the two sides on various states.

\paragraph{Constraints from $|\emptyset\rangle$}
We begin by evaluating both sides on the vacuum, 
\be\label{A.2}
\begin{split}
	e(z) x(w) |\emptyset\rangle  
	& \sim e(z) \frac{1}{w} |\blacksquare \rangle
	 \sim \frac{1}{w} \frac{(\#)}{z-h_1} |\blacksquare  + {\square}_1\rangle
	+ \frac{1}{w} \frac{(\#)}{z-h_3} |\blacksquare  + {\square}_3\rangle
\end{split}
\ee
and
\be\label{A.3}
\begin{split}
	x(w)e(z) |\emptyset\rangle  
	& \sim x(w)
	\frac{1}{z} |{\square}\rangle  
	\sim \frac{1}{w-h_2}\frac{1}{z} |\blacksquare  +\hat {\square}_{\rm top}\rangle  \ .
\end{split}
\ee
Since neither of the states in (\ref{A.2}) appears in (\ref{A.3}), and vice versa, we learn 
that the denominator of  $G(\Delta)$ must contain the factors $z-h_i-w = \Delta-h_i$ for $i=1,3$, while its numerator must contain $(z - (w- h_2)) = (\Delta + h_2)$. Thus we conclude that $G(\Delta)$ must contain the factors 
\be\label{eq:e-x-OPE-coeff-incomplete}
	G(\Delta) \sim \frac{ (\Delta +   h_2) }{(\Delta - h_1)(\Delta - h_3)}\ .
\ee
This is exactly as for $p=0$. Also similar to that case, it turns out to be impossible to fix $G(\Delta)$ completely via this route. 
We will show later, using also information on the residues of $x(u)$ action on twin plane partitions, that $G(\Delta) = \varphi_2 (\Delta)$, again identical to the case of $p=0$.

\subsubsection{The $\hat{f} \cdot x$ OPE}\label{sec:hat-f-x}

As the next example, we consider the following OPE :
\be
	\hat f(z)  \cdot x(w) \sim \hat{\bar{H}}(\Delta) \ x(w) \cdot \hat f(z)  \ .
\ee
Using constraints from $|{\square}\rangle$, $|\blacksquare + \hat {\square}_{\rm top}\rangle$, and $|\blacksquare  + {\square}{\square}_3 + \hat {\square}_{\rm top}\rangle$
it is easy to show that $\hat{\bar{H}}(\Delta)$ must contain the factors\footnote{
For details of this computation see Appendix.\ \ref{sec:fhatx}.}
\be
	\hat{\bar{H}}(\Delta) \sim \frac{ (\Delta + \hat\sigma_3\hat\psi_0 + \hat  h_1) (\Delta + \hat\sigma_3\hat\psi_0 + \hat  h_3)}{(\Delta + \hat\sigma_3\hat\psi_0)}\ . 
\ee
This is almost the same as the case $p=0$ (except that now the hatted $\hat{h}_i$ appear). Once again, it is impossible to fix the OPE coefficient entirely via this route. 
We will show later that $\hat{\bar{H}}(\Delta) = \hat\varphi_2^{-1} (-\Delta - \hat\sigma_3\hat\psi_0)$, after fixing the residues of the $x$-action.

\subsubsection{General structures of OPEs}

One can thus continue in this manner and try to fix all the OPEs between the single-box operator (i.e.\ $(e,f)$ and $(\hat{e},\hat{f})$) and the gluing operators. One can first show that these OPEs relations are not all independent, see  Figs.~\ref{OPEbosonicxy-conj} and \ref{figOPEbosonicfull}. 
Namely, there are only four independent functions:\begin{equation}
G(\Delta) \qquad \qquad H(\Delta) \qquad\qquad  \bar{G}(\Delta) \qquad\qquad \bar{H}(\Delta)
\end{equation}
and the hatted functions can be obtained from the unhatted ones via
\begin{equation}
h_i \mapsto \hat{h}_i \qquad \qquad \psi_0\mapsto \hat{\psi}_0
\end{equation}

Using the methods outlined above in this section, we can only determine certain factors of these rational functions: 
\begin{equation}\label{Fourfunctions}
\begin{aligned}
&G(\Delta)= \frac{(\Delta+a) (\Delta +   h_2) }{(\Delta - h_1)(\Delta - h_3)} \qquad  \qquad H(\Delta)=\frac{\Delta}{\Delta+a}  \\
&\bar{G}(\Delta)=\frac{\Delta+\sigma_3 \psi_0-b}{\Delta+\sigma_3 \psi_0-h_2}  \qquad  \quad \qquad \bar{H}(\Delta)= \frac{ (\Delta + \sigma_3\psi_0 +  h_1) (\Delta + \sigma_3\psi_0 +  h_3)}{(\Delta + \sigma_3\psi_0)(\Delta + \sigma_3\psi_0-b)}
\end{aligned}
\end{equation}
where $a$ and $b$ are undermined.
Note that in writing these four functions in the form of (\ref{Fourfunctions}), we have used the fact that (1) each one is a  rational function whose numerator and denominator have the same degree, as demanded by their free field limits (see \cite{Gaberdiel:2017hcn} for the case $p=0,\rho=1/2$);\footnote{
The algebra is also expected to admit a free field construction when $p=\pm1$ (and for certain values of $\rho$), and it should therefore be possible to verify a similar condition. However we have not checked this explicitly. 
We take as a good sign the fact that eventually we find a rather natural expression, which is valid for all $p$, together with the fact that this expression complies with known free-field constraints for the case of $p=0$ and $\rho=1/2$.} (2) the creating and annihilating functions are related by\footnote{
Note that since it is possible to have expressions like $\frac{\Delta}{\Delta}$,  these constraints can only serve as consistency check, i.e.\ we still have to derive the four functions independently. 
For example, this cannot tell the difference between $H(\Delta)=\frac{\Delta-a}{\Delta-a}$. 
}
\begin{equation}
G(u)\cdot H(u)=\varphi_2(u) \qquad \textrm{and} \qquad \bar{G}(u)\cdot \bar{H}(u)=\varphi^{-1}_2(-u-\sigma \psi_0)
\end{equation}
which can be derived by applying the OPE relation $ [e(z)\,, f(w)]  \sim  - \frac{1}{\sigma_3}\, \frac{\psi(z) - \psi(w)}{z-w}$ (and the hatted version) on an arbitrary plane partition \cite{Gaberdiel:2017hcn}.
In the next section, we will show that $a=0$ and $b=h_2$. 

\section{OPEs among gluing operators}
\label{sec:residue}

In the previous section, we focused on the OPE between  gluing operators and corner operators, i.e.\ those in the two copies of the affine Yangian of $\mathfrak{gl}_1$. The strategy so far was to first determine the pole structure for the action of gluing operators on generic twin plane partitions, and then to apply the OPE relation (to be determined) to various cases, thus using the pole structures to constrain OPE coefficients. 

Although we were able to determine the pole structure for all gluing operators on generic twin plane partitions, this procedure did not entirely fix  the OPE coefficients. Instead, the result was to determine the presence of some zeroes and poles of the coefficients, leaving open the possibility that more factors may be present.

In this section we take the next step, by fixing the \emph{residues} of the action of gluing operators on twin plane partitions. 
This will enable us to fix the remaining freedom in the OPE relations between corner operators and gluing operators. 

Once we fully know how the gluing operators act on a generic twin plane partition, i.e.\ both the poles and residues, we can further use these to determine all the OPE relations between gluing operators. 

\subsection{General strategy}

To determine the OPEs between gluing operators, we can essentially proceed as in the case with $\rho=\frac{1}{2}$ and $p=0$, see \cite{Gaberdiel:2018nbs}. 
As we will see later, the shifting parameter $\rho$ does not change the structures of these OPEs.
However, since the framing parameter $p$ changes the statistics of gluing operators, its effect on OPEs is far less trivial.
For example, recall that regardless of the shifting parameter $\rho$, at $p=0$, the $x$ operator is fermionic.
However, for $p=\pm 1$, the gluing operators are bosonic, independent of $\rho$. Therefore we expect non-trivial self OPEs for gluing operators when $p=\pm 1$. 
Indeed these considerations follow intuitively from the fact that the framing modulus $p$ determines the relative orientation of the two plane partitions that, glued together, make up a twin plane partition.\footnote{
Recall the intuitive geometric arguments from Section \ref{sec:statistics} and the related discussion of characters in Section \ref{sec:vacuum-char-analysis-bose}.}

\subsubsection{Structure of OPEs between gluing operators}

For starters, let us make a basic observation on the general form of the OPEs.
On general grounds, we expect the triplet $(x, P, y)$ to behave in a way analogous to the triplet $(e,\psi,f)$. 
This concerns only relations within each triplet, and is based on the fact that in either case we have a Heisenberg-like algebra acting on 3d or 2d partitions.
(see the discussion in Section \ref{eq:heisenberg-intuition}).
In fact, this was the motivation behind the choice of ans\"atze (\ref{xansatz}) and (\ref{yansatz}).
But in the case of $p=0$ the gluing generators are fermionic, and therefore the triplet $(x, P, y)$ actually generates a Clifford algebra (this can be explicitly verified in the case $p=0,\rho=1/2$ by matching with the $\mathcal{N}=2$ ${\mathcal W}_\infty$ algebra \cite{Gaberdiel:2018nbs}).
To keep track of this we introduce a sign 
\be\label{eq:epsilon-p}
	\epsilon_p = \left\{
	\begin{array}{lr}
		-1 \qquad &  (p  = 0) \\
		1 \qquad &  (p  = \pm1) 
	\end{array}
	\right.
\ee
in the OPE relations, which take a form similar to (\ref{bosonicdef}):
\begin{equation}\label{Pxy}
\begin{aligned}
\begin{aligned}
P(z)\, x(w)&\sim S_p(\Delta) \ x(w)\ P(z)\\
P(z)\, y(w)&\sim S^{-1}_p(\Delta)\ y(w)\ P(z)
\end{aligned}
&\qquad
\begin{aligned}
x(z)\ x(w) &\sim \epsilon_p\, S_p(\Delta) \ x(w)\ x(z)\\
 y(z)\ y(w) &\sim  \epsilon_p\, S^{-1}_p(\Delta)\ y(w)\ y(z)\\
\end{aligned}\\
[x(z), y(w)]_{ \epsilon_p\,} & \sim \frac{P(z)-P(w)}{z-w}
\end{aligned}
\end{equation}
Namely, there is only one independent function $S_p(u)$ that determines the relations among the $(x,P,y)$ triplet. 
Here $[x,y]_{\epsilon_p}$ denotes the commutator for $p=\pm 1$ and the anti-commutator for $p=0$.
Note that the ``charge operators'' $P,\bar P$ are expected to be bosonic for all $p$, therefore we did not introduce a sign in their OPEs with the gluing operators.

One can also prove (\ref{Pxy}) starting from the ans\"atze (\ref{xansatz}) and (\ref{yansatz}) and using the fact that $\textbf{P}_\Lambda(u)$ is the eigenvalue of the charge function $P(u)$ on the twin plane partition $\Lambda$.
We omit the proof here since it goes essentially the same as the bosonic case $(e,\psi,f)$.
The only subtlety is that when taking square-roots in (\ref{xansatz}) and (\ref{yansatz}), one needs to select the branch that comply with the statistic factor $\epsilon_{p}$ in (\ref{Pxy}). 
Through the analogy with $(e,\psi,f)$, it is clear that $S_p$ plays the role of $\varphi_3$.

The barred version of (\ref{Pxy}) is obtained by putting bars on all the fields in (\ref{Pxy}):
\begin{equation}\label{fermionicbardef}
\begin{aligned}
\begin{aligned}
\bar{P}(z)\, \bar{x}(w)&\sim \bar{S}_p(\Delta) \ \bar{x}(w)\ \bar{P}(z)\\
\bar{P}(z)\, \bar{y}(w)&\sim \bar{S}^{-1}_p(\Delta)\ \bar{y}(w)\ \bar{P}(z)
\end{aligned}
&\qquad
\begin{aligned}
\bar{x}(z)\ \bar{x}(w) &\sim \epsilon_p \,\bar{S}_p(\Delta) \ \bar{x}(w)\ \bar{x}(z)\\
 \bar{y}(z)\ \bar{y}(w) &\sim  \epsilon_p \, \bar{S}^{-1}_p(\Delta)\ \bar{y}(w)\ \bar{y}(z)\\
\end{aligned}\\
[\bar{x}(z), \bar{y}(w)]_{ \epsilon_p } & \sim \frac{\bar{P}(z)-\bar{P}(w)}{z-w}
\end{aligned}
\end{equation}
Recalling that $\bar x$ can destroy rows created by $x$ and vice versa, leads to the following relation for the barred OPE coefficients
\begin{equation}\label{Spbar}
\bar{S}_p(\Delta)=S^{-1}_{p}(\Delta)\,.
\end{equation}
The relations are summarized  in Fig.~\ref{OPEfermionicxy}.

Note that for $p=0$ we can simply take 
\begin{equation}\label{eq:S-zero}
S_0(\Delta)\equiv1\,.
\end{equation}
With this definition, the relations (\ref{Pxy}) and (\ref{fermionicbardef}) reproduce the OPEs among gluing operators $\bm{g}=\{x,\bar x, y, \bar y\}$ obtained in \cite{Gaberdiel:2018nbs} for $p=0,\rho=1/2$.

We will next proceed to fix $S_p(u)$ for more general $p$ and $\rho$.

\subsubsection{Procedure in steps}

As anticipated, the strategy to fix the algebra among gluing generators will be to mimic the construction of the algebra of single-box operators on standard plane partitions.
In fact, recall that for the bosonic affine Yangian of $\mathfrak{gl}_1$, the OPE relations (\ref{bosonicdef}) and the action on the plane partitions (\ref{ppartpsi}) and (\ref{ppartef}) are not independent. 
One can derive one knowing the other.
In particular, the $\varphi_3$ function appears both in the $e\cdot e$ OPE  and in $\psi_{{\square}}(u)$ (the charge function that is also the $\psi\cdot e$ OPE coefficient). 

For the gluing operators, the situation is more complicated, due to the fact that (1) the charge function $(P(u), \bar{P}(u))$ also receive contribution from individual boxes at the two corners and (2) the gluing operators also interact with the corner operators. 

Therefore, the strategy should be to start by fixing the contribution of \emph{individual} boxes to $(P(u), \bar{P}(u))$, and later use this information to fill in the remaining pieces. This is the roadmap we will follow:
\begin{enumerate}
\item Fix ($P_{\square}$, $P_{\hat{\square}}$, $\bar{P}_{{\square}}$ $\bar{P}_{\hat{\square}}$) together with the remaining freedom in $(G, H, \bar{G}, \bar{H})$. 
The knowledge of $P_\square$ fixes the $P\cdot e$ and $P\cdot f$ OPEs (and similarly for $\hat e,\hat f$).
\item Fix $S_p(\Delta)$, which appears in the $P\cdot x$ and $x \cdot x$ OPE. 
\item Fix ($P_\blacksquare$, $\bar P_{\overline\blacksquare}$, $P_{\overline\blacksquare}$, $\bar P_\blacksquare$)
\item Fix remaining OPEs.
\end{enumerate}

\subsection{Fixing ($P_{{\square}}$, $P_{\hat{{{\square}}}}$, $\bar{P}_{{\square}}$ $\bar{P}_{\hat{{{\square}}}}$) and $(G, H, \bar{G}, \bar{H})$}
\label{sec:PboxandG}

\subsubsection{$P_{{\square}}$, $G$ and $H$}
\label{sec:721}

We begin by fixing the factor $P_{\square}$, by following the argument from \cite[Sec.\ 6.1]{Gaberdiel:2018nbs}, which goes through essentially unchanged. Using the ansatz (\ref{xansatz}), we consider the two states obtained by acting with $e,x$ on a generic twin plane partition $\Lambda$:\footnote{
Here we focus on the \emph{creation} action of $x$ only.}
\be
	e(z) x(w) |\Lambda\rangle 
	\quad\text{and}\quad 
	x(w) e(z) |\Lambda\rangle \,.
\ee
Comparing these, using the OPE relation of $e\cdot x$, we conclude that 
\be
\begin{split}
	&
	{
	\Big[ -  \frac{1}{\sigma_3} {\rm Res}_{w = h({\tiny \yng(1)})} \bm{\Psi}_{\Lambda+\blacksquare}(w) \Big]^{\frac{1}{2}}
	}
	{
		\Big[ \textrm{Res}_{u=p_{+}({\blacksquare})} \, \textbf{P}_\Lambda(u) \Big]^{\frac{1}{2}} 
	}
	\\
	& \qquad =
	G( h({\tiny \yng(1)}) -p_+(\blacksquare)) \ 
	{
		\Big[ \textrm{Res}_{u=p_{+}({\blacksquare})} \, \textbf{P}_{\Lambda + {\tiny \yng(1)}}(u) \Big]^{\frac{1}{2}} 
	}
	{
		\Big[ -  \frac{1}{\sigma_3} {\rm Res}_{w = h({\tiny \yng(1)})} \bm{\Psi}_{\Lambda}(w) \Big]^{\frac{1}{2}}
	} \ , 
\end{split}
\ee
implying that
\be\label{Pfromex}
\begin{split}
	P^{1/2}_{{\tiny \yng(1)}}(p_+(\blacksquare) ) 
		& = G^{-1}( h( {\tiny \yng(1)} )  - p_+(\blacksquare) ) \ \Big[  \psi_{\blacksquare}(h( {{\tiny \yng(1)}} ) ) \Big]^{1/2} \,.
\end{split}
\ee
Considering (\ref{Pfromex}) together with the partial result for $G$ (\ref{eq:e-x-OPE-coeff-incomplete}) and the constraint (\ref{Fourfunctions}) from the free field limit for $p=0$ and $\rho=\frac{1}{2}$ given in \cite{Gaberdiel:2018nbs}, we see that the most natural solution is
\begin{equation}\label{Pbox1}
P_{{\square}}(u)=\varphi^{-1}_2(-u+h({\square}))
\qquad \textrm{and}\qquad
G(\Delta)=\varphi_2(\Delta)
\end{equation}
Note that  the result is exactly the same as for  $p=0$, $\rho=1/2$ case in \cite{Gaberdiel:2018nbs}. 
Using (\ref{Fourfunctions}), this also fixes $H$:
\begin{equation}
H(\Delta)=\frac{\Delta}{\Delta}\,.
\end{equation}
Note that $H$ should not be confused with the identity, since the purpose of the factors appearing in this ratio (as for all OPE coefficients, in general) is to cancel poles on either side of an OPE. 
More importantly, when translating to algebraic relations in terms of modes, the exact factors in the numerator and denominator are crucial. (For how to derive modes relation from an OPE-like relation see section 2 of \cite{Gaberdiel:2018nbs}; for examples with gluing operators see section 5 of  \cite{Gaberdiel:2017hcn}.)

\subsubsection{$P_{\hat{{{\square}}}}$ and $\hat{\bar{H}}$ }

In order to fix $P_{\hat {\square}}$, we consider instead the OPE of $x$ with $\hat f$ on a generic twin plane partition $\Lambda$. By the same steps as above, we obtain the following relation
\be
\begin{split}
	&
	{
	\Big[ -  \frac{1}{\hat \sigma_3} {\rm Res}_{w = \hat h(\hat {\tiny \yng(1)})} \hat{ \bm{\Psi}}_{\Lambda+\blacksquare}(w) \Big]^{\frac{1}{2}}
	}
	{
		\Big[ \textrm{Res}_{u=p_{+}({\blacksquare})} \, \textbf{P}_\Lambda(u) \Big]^{\frac{1}{2}} 
	}
	\\
	&\qquad =
	\hat  H( h({\tiny \yng(1)}) -p_+(\blacksquare)) \ 
	{
		\Big[ \textrm{Res}_{u=p_{+}({\blacksquare})} \, \textbf{P}_{\Lambda - \hat {\tiny \yng(1)}}(u) \Big]^{\frac{1}{2}} 
	}
	{
		\Big[ -  \frac{1}{\hat \sigma_3} {\rm Res}_{w = \hat h(\hat {\tiny \yng(1)})} \hat {\bm{\Psi}}_{\Lambda}(w) \Big]^{\frac{1}{2}}
	} 
\end{split}
\ee
Again, we see that the most natural solution is 
\begin{equation}\label{Pbox2}
P_{\hat {\tiny \yng(1)}}(u)=\hat\varphi^{-1}_2(u - \hat h(\hat {\tiny \yng(1)})-\hat\sigma_3 \hat\psi_0)
\qquad
\textrm{and}
\qquad
\hat{\bar{H}}(\Delta)=\hat{\varphi}^{-1}_2(-\Delta-\hat{\sigma}_3\hat{\psi}_0)\,.
\end{equation}

\subsubsection{$\bar{P}_{{\square}}$ and ${\bar{H}}$}

Repeating the above procedure with operators $\bar x$ and $f$, we get 
\begin{equation}\label{Pbox3}
\bar{P}_{{\tiny \yng(1)}}(u)=\varphi^{-1}_2(u -  h({\tiny \yng(1)})-\sigma_3 \psi_0)
\qquad
\textrm{and}
\qquad
{\bar{H}}(\Delta)={\varphi}^{-1}_2(-\Delta-{\sigma}_3{\psi}_0)
\end{equation}
which is the unhatted analogue of (\ref{Pbox2}).
Using (\ref{Fourfunctions}), this also fixes $\bar{G}$:
\begin{equation}
\bar{G}(\Delta)=\frac{\Delta+\sigma_3 \psi_0-h_2}{\Delta+\sigma_3 \psi_0-h_2} \,.
\end{equation}

\subsubsection{$\bar{P}_{\hat{{{\square}}}}$ and  $\hat{G}$}

Finally, in order to fix $\bar{P}_{\hat{{{\square}}}}$ we consider the states obtained by acting with $x$ and $\hat e$ on a generic twin plane partition $\Lambda$. This leads to
\begin{equation}\label{Pbox4}
\bar{P}_{\hat{{\square}}}(u)=\hat{\varphi}^{-1}_2(-u+\hat{h}(\hat{{\square}}))
\qquad \textrm{and}\qquad
\hat{G}(\Delta)=\hat{\varphi}_2(\Delta)
\end{equation}
which is the hatted version of (\ref{Pbox1}).

\subsubsection{Full form of OPEs of gluing generators}

We have now fixed all OPE functions of the gluing generators with corner (single-box) operators. For later convenience we collect them here:
\begin{equation}
\begin{aligned}
G(\Delta)&=\varphi_2(\Delta) \\\
\bar{G}(\Delta)&=\frac{\Delta+\sigma_3 \psi_0-h_2}{\Delta+\sigma_3 \psi_0-h_2}  
\end{aligned}
\qquad
\begin{aligned}
H(\Delta)&=\frac{\Delta}{\Delta}  \\
\bar{H}(\Delta)&=\varphi^{-1}_2(-\Delta-\sigma_3 \psi_0)
\end{aligned}
\end{equation}
Analogous expressions for the hatted counterparts can be easily obtained by replacements of $h_i\to \hat h_i, \psi_0\to \hat \psi_0$.

\subsection{OPEs between $(P, \bar{P})$ and single box generators $\{\e,f,\hat{e},\hat{f}\}$}
\label{sec:OPEPe}

The single box contribution to the charge function $(\textbf{P}_{\Lambda}(z),\bar{\textbf{P}}_{\Lambda}(z))$ we obtained in the previous subsection immediately give the OPEs between $(P(z),\bar{P}(z))$ and the single box generators $\{e,f,\hat{e},\hat{f}\}$. 
For example, in order to fix the $P\cdot e$ OPE, apply  
\begin{equation}\label{Pe}
P(z)e(w)\sim B(z-w) \,e(w) P(z)
\end{equation} 
 on an arbitrary twin plane partition $|\Lambda \rangle$, which gives
\begin{equation}
\begin{aligned}
&\sum_{\square \in \textrm{Add}(\Lambda)} \frac{E(\Lambda\rightarrow \Lambda+\square)}{w-h(\square)} \textbf{P}_{\Lambda+\square}(z) |\Lambda+\square\rangle \\
&\qquad 
= \sum_{\square \in \textrm{Add}(\Lambda)} B(z-w) \frac{E(\Lambda\rightarrow \Lambda+\square)}{w-h(\square)} \textbf{P}_{\Lambda}(z) |\Lambda+\square\rangle
\end{aligned}
\end{equation}
Since this is a vector equation, it needs to hold for every $|\Lambda+\square\rangle
$, this gives
\begin{equation}
B(z-h(\square))=\frac{\textbf{P}_{\Lambda+\square}(z)}{\textbf{P}_{\Lambda}(z)}=P_{\square}(z)=\varphi^{-1}_2(-z+h(\square))
\end{equation}
where in the last step we have used the result $P_{\square}(z)$ (\ref{Pbox1}). 
This fixes the  OPE (\ref{Pe}) to be
\begin{equation}
P(z) \,e(w) \sim \varphi^{-1}_2(-\Delta) \,e(w)\,P(z)
\end{equation}
Repeating this exercise we eventually obtain all OPEs between $(P,\bar{P})$ and the single box generators $\{e,f,\hat{e},\hat{f}\}$: 
\begin{equation}
\begin{aligned}
&P(z) \,e(w) \sim \varphi^{-1}_2(-\Delta) \, e(w)\,P(z) \qquad \,\,\, \qquad P(z) \,f(w) \sim \varphi_2(-\Delta) \, f(w)\,P(z)\\
&P(z) \,\hat{e}(w) \sim \hat{\varphi}^{-1}_2(\Delta-\hat\sigma_3\hat\psi_0) \, \hat{e}(w)\,P(z) \qquad P(z) \,\hat{f}(w) \sim \hat\varphi_2(\Delta-\hat\sigma_3\hat\psi_0) \, \hat{f}(w)\,P(z)\\
&\bar{P}(z) \,e(w) \sim \varphi^{-1}_2(\Delta-\sigma_3\psi_0) \, e(w)\,\bar{P}(z) \qquad  \bar{P}(z) \,f(w) \sim \varphi_2(\Delta-\sigma_3\psi_0) \, f(w)\,\bar{P}(z)\\
&\bar{P}(z) \,\hat{e}(w) \sim \hat{\varphi}^{-1}_2(-\Delta) \, \hat{e}(w)\,\bar{P}(z) \qquad\,\,\, \qquad \bar{P}(z) \,\hat{f}(w) \sim \hat{\varphi}_2(-\Delta) \, \hat{f}(w)\,\bar{P}(z)\\
\end{aligned}
\end{equation}
these are collected in Figure~\ref{OPEfermionicPe}.

\subsection{Fixing self-OPEs of gluing operators}
\label{sec:selfOPEglue}

We now turn to fixing OPEs of gluing operators with themselves. 
As explained earlier, the OPE among the triplet $(x,P,y)$ only depends on one function, $S_p(u)$, see (\ref{Pxy}). 
Therefore we only need to fix the easiest one, i.e.\ the  $P\cdot x$ OPE. 

As always, we apply both sides of the OPE we wish to fix on a generic twin plane partition $\Lambda$, and use this to deduce constraints on the OPE coefficient. 
Let us consider applying 
\begin{equation}\label{PxOPE}
P(z)x(w)\sim S_p(\Delta) x(w) P(z)
\end{equation} 
on a generic twin plane partition $\Lambda$. This gives
\begin{equation}\label{PxonLambda}
\begin{aligned}
&\sum_{\blacksquare}\textbf{P}_{[\Lambda+\blacksquare]}(z) \,  \frac{X(\Lambda\rightarrow [\Lambda+\blacksquare])}{w-p_{+}(\blacksquare)} |[\Lambda+\blacksquare]\rangle\\
&\qquad \qquad \sim S_{p}(z-w) \sum_{\blacksquare} \frac{X(\Lambda\rightarrow [\Lambda+\blacksquare])}{w-p_{+}(\blacksquare)} \ \textbf{P}_{\Lambda}(z) |[\Lambda+\blacksquare]\rangle
\end{aligned}
\end{equation}
Here we use $[\Lambda+\blacksquare]$ to denote the resulting twin plane partition configuration from adding $\blacksquare$ to $\Lambda$, and sum over all possibilities. 
And we use 
\begin{equation}
X(\Lambda\rightarrow [\Lambda+\blacksquare]) \equiv \Big[ \textrm{Res}_{u=p_{+}({\blacksquare})} \, \textbf{P}_\Lambda(u) \Big]^{\frac{1}{2}}
\end{equation}
to denote the residue for $x$'s action on $|\Lambda\rangle$.
Since the relation (\ref{PxonLambda}) is a vector identity, i.e. it holds for each term in the sum, $X$ drops out and hence can only be fixed later, and for each term we have 
\begin{equation}
\frac{\textbf{P}_{[\Lambda+\blacksquare]}(z)}{\textbf{P}_{\Lambda}(z) }=S_p(z-p_{+}(\blacksquare))
\end{equation}

Let's start with the simplest twin plane partition, the vacuum $|\emptyset\rangle$. Since there is only one resulting state $|\blacksquare\rangle$, with coordinate function $g(\blacksquare)=0$, and the pole 
\begin{equation}
p_+(\blacksquare)=p^{(0)}_+(\blacksquare)=0 \qquad \textrm{for} \qquad g(\blacksquare)=0
\end{equation} we have
\begin{equation}
\frac{\textbf{P}_{\blacksquare}(u)}{\textbf{P}_{0}(u) }=S_p(u) \qquad \textrm{for} \qquad g(\blacksquare)=0
\end{equation}
From the decomposition of $\textbf{P}$ in (\ref{XLambda})
\begin{equation}
\textbf{P}_{0}(u)=P_0(u) \qquad \textrm{and}\qquad \textbf{P}_{\blacksquare}(u)=P_0(z)P_{\blacksquare}(u)
\end{equation}
we immediately have that for the first $\blacksquare$ one create from the vacuum
\begin{equation}\label{Pblacksquare0}
P_{\blacksquare}(u)=S_p(u) \qquad \textrm{for} \qquad g(\blacksquare)=0
\end{equation}
Note that this reduces to the trivial case of $P_{\blacksquare}(u)=-1$  for $p=0$, due to (\ref{eq:S-zero}), also see  \cite[Table 1]{Gaberdiel:2018nbs}.

Then we look at the next simplest twin plane partition, the single box state $|{\square}\rangle$. Recall from (\ref{eq:xonbox-main}) the action of $x$ on a single box, we have
\begin{equation}
[{\square}+\blacksquare]=\blacksquare+\hat{{\square}}_{\,\text{top}}
\end{equation}
and 
\begin{equation} 
p_+(\blacksquare)=p^{(1)}_+(\blacksquare)=h_2\qquad \textrm{for} \qquad g(\blacksquare)=0
\end{equation}
which gives 
\begin{equation}\label{Pbox}
\frac{\textbf{P}_{\blacksquare+\hat{{{\square}}}_{\text{top}}}(u)}{\textbf{P}_{{{\square}}}(u) }=S_p(u-h_2)\qquad \textrm{for} \qquad g(\blacksquare)=h({\square})=0
\end{equation}
On the other hand, we can also evaluate the l.h.s. of (\ref{Pbox}) directly using the decomposition (\ref{XLambda}), since we have already derived all four types of $P(u),\bar P(u)$ factors for single boxes and we have just obtained $P_\blacksquare(u)$ (valid only for the case $g(\blacksquare) = 0$, which applies to this case) in (\ref{Pblacksquare0}).
Namely
\begin{equation}
\begin{aligned}
\textbf{P}_{{\square}}(u)&=P_0(u) \cdot P_{{\square}}(u)=P_0(u) \cdot \varphi^{-1}_2(-u) \\
\textbf{P}_{\blacksquare+\hat{{\square}}}(u)&=P_0(u)\cdot P_{\blacksquare}(u) \cdot P_{\hat{{\square}}}(u)=P_0(u) \cdot S_p(u) \cdot \hat{\varphi}_2^{-1}(-u)
\end{aligned}
\end{equation}
With this we obtain for the l.h.s. of (\ref{Pbox})
\begin{equation}\label{Pbox2}
\frac{\textbf{P}_{\blacksquare+\hat{{{\square}}}_{\text{top}}}(u)}{\textbf{P}_{{{\square}}}(u) }=\frac{S_p(u)\hat{\varphi}_2^{-1}(-u)}{\varphi^{-1}_2(-u)}\qquad \textrm{for} \qquad g(\blacksquare)=h({\square})=0
\end{equation}
Comparing (\ref{Pbox}) and (\ref{Pbox2}), we find the key recursion relation:
\be\label{Srecursion}
	{\frac{S_p(u-h_2)}{S_p(u)} =  \frac{\varphi_2(-u)}{ \hat{\varphi_2}(-u)}\ .}
\ee

Note that $S_p(u)$ then enjoys the following inversion property
\be\label{Sinv}
{S_p^{-1}(u) = S_p(-u)   \ . }
\ee
In fact this property is necessary for consistency of the exchange symmetry of the two operators in the $x\cdot x$ OPE, where $S_p$ appears (and it may have been anticipated in this way).

As regards to the $P(z)$ charge field, it is natural to expect some relation between the charge of a single box and the charge of an infinite row, much like to the $\psi(z)$ charge.
In this vein, we should expect some relation between the OPE coefficients of $P\cdot x$ and $P\cdot e$, namely $S_p(u)$ and $\varphi_2^{-1}(-u)$. The former will play the role of $\varphi_2$ (in the context of  $\psi(z)$-charges) while the latter should play the role of $\varphi_3$. 
Recalling their relations (\ref{eq:infinite-product-before})-(\ref{psiudef0}) and (\ref{magic}), we can anticipate similar relations between $S_p$ and $\varphi_2^{-1}$. 
Using the inversion property, we immediately recognize (\ref{Srecursion}) as the analogue of (\ref{magic}) for the single-box OPE coefficients, since the factors on the right hand side correspond exactly to $\varphi_2^{-1}(-u)$ and $\hat \varphi_2^{-1}(-u)$. 
In fact the key recursion equation (\ref{Srecursion}) is just the statement that the charge of a row minus the charge of an $h_2$-shifted row is equal to the charge of a $\square$ at one end, minus the charge of a $\hat\square$ at the other end.
In other words, the charge of the shifted row is equal to that of the un-shifted row after removing the initial $\square$, and adding a $\hat\square$ at the end.
This is natural and should have been expected. 

Overall we find the following solution to (\ref{Srecursion}) and  (\ref{Sinv}):
\be\label{eq:S-p}
	\boxed{S_p(u) = \frac{u+\delta_p}{u-\delta_p} \ ,}
\ee
where 
\be\label{eq:delta-p}
	\delta_p = \left\{\begin{array}{ll}
	h_1 \qquad & \ (p=1) \\
	0 \qquad & \ (p=0) \\
	h_3 \qquad & \ (p=-1) \\
	\end{array}\right. \,.
\ee
The function $S_p$ then fixes all OPEs among the triplet $(x,P,y)$ in (\ref{Pxy}).
Then using $\bar{S}_p(u)=S^{-1}_p(u)$ (\ref{Spbar}), we also obtain the OPEs among the triplet $(\bar{x},\bar{P},\bar{y})$ in (\ref{fermionicbardef})

\subsection{Fixing ($P_\blacksquare$, $\bar P_{\overline\blacksquare}$, $P_{\overline\blacksquare}$, $\bar P_\blacksquare$)} 
\label{sec:Pblackbox}

Since we now know all four types of contributions to $(\textbf{P}_{\Lambda}(u), \bar{\textbf{P}}_{\Lambda}(u)) $ by single boxes, as well as the self OPE coefficient $S_{p}(u)$, we can derive the contribution of $\blacksquare$ and $\overline{\blacksquare}$ to $(\textbf{P}_{\Lambda}(u), \bar{\textbf{P}}_{\Lambda}(u))$.

\subsubsection{$P_\blacksquare$ and $\bar P_{\overline\blacksquare}$}

To compute the contribution of $\blacksquare$ to  $\textbf{P}_{\Lambda}(z)$, let us first consider applying the $P\cdot x$ OPE (\ref{PxOPE}) to an initial state defined by a twin plane partition
\begin{equation}
\Lambda_\textrm{initial}=\Lambda_0+\textrm{minimal bud}[\blacksquare]
\end{equation}
which has a minimal bud for some $\blacksquare$ at position $g(\blacksquare)$:
\begin{equation}
P(z)\, x(w)\, |\Lambda_\textrm{initial}\rangle \sim S_p(z-w) \,x(w) P(z)\, |\Lambda_{\textrm{initial}}\rangle
\end{equation}
Consider the final state in which $x$ creates the $\blacksquare$ and in the meantime eats the minimal bud for $\blacksquare$:
\begin{equation}
\Lambda_\textrm{final}=\Lambda_0+\blacksquare\,.
\end{equation}
This corresponds to the creation action of $x$ at the position determined by the pole 
\begin{equation}
w=p^{(0)}_+(\blacksquare)=h(\blacksquare)\,.
\end{equation}
Since the charge function $P(z)$ on the l.h.s. sees $|\Lambda_\textrm{final}\rangle$ and the one on r.h.s. see  $|\Lambda_\textrm{initial}\rangle$, we have
\begin{equation}
P_{\blacksquare}(z)= S_p(z-p^{(0)}_{+}(\blacksquare)) 
P_{\textrm{minimal bud}(\blacksquare)}(z)
\end{equation}
where we have cancelled all terms shared by both sides of equations.
From the derivation of the minimal bud (\ref{eq:x-bud}) and $P_{{\square}}(z)$ (\ref{Pbox})
\begin{equation}\label{Pblacksquare0}
\begin{aligned}
P_{\blacksquare}(z)&= S_p(z-p^{(0)}_{+}(\blacksquare)) \prod^{m(1-p)+n(1+p)-1}_{k=0}\varphi^{-1}_2(-z+g(\blacksquare)+k h_2) 
\end{aligned}
\end{equation}
This has a natural interpretation: when acting with $x$ to create a $\blacksquare$, the minimal bud is destroyed. As a result, the charge of the newly created row is that of $x$ (the factor $S_p$) minus the charges of single boxes in the bud (the product of $\varphi^{-1}_2(u) = G(u)^{-1}$). 
This result is perfectly consistent with what we would have anticipated directly from knowledge of the length of minimal buds and of the charge functions. 

We have obtained (\ref{Pblacksquare0})  assuming that there is a minimal bud in the position where the $\blacksquare$ is to be created, 
Now let's look at the generic initial state $\Lambda$.
If the bud is shorter than the minimal one, then $x$ will not create a row there. 
On the other hand, if the bud is longer by $\ell$ boxes, then
\begin{equation}\label{Pblacksquareell}
\begin{aligned}
P_{\blacksquare}(z)&= S_p(z-p^{(\ell)}_{+}(\blacksquare)) \prod^{m(1-p)+n(1+p)-1+\ell}_{k=0}\varphi^{-1}_2(-z+g(\blacksquare)+k h_2) 
\end{aligned}
\end{equation}
Using the shifting property (\ref{Srecursion}) together with the definition of the pole (\ref{p+f}), we can rewrite (\ref{Pblacksquareell}) in a form that depends entirely on the coordinate function $g(\blacksquare)$:
\begin{equation}\label{Pblacksquare}
\begin{aligned}
P_{\blacksquare}(z)&= S_p\bigl(z - g(\blacksquare) \bigr)\!\!  \prod^{m(1-p)+n(1+p)-1}_{k=0}\hat{\varphi}^{-1}_2(-z+g(\blacksquare)+k h_2)
\end{aligned}
\end{equation}
Note that this reduces to the result (\ref{Pblacksquare0}) for the first $\blacksquare$.

It is straightforward to repeat this exercise with $x$ replaced by $\bar x$, and switching unhatted quantities for hatted ones where suitable. This leads to 
\be
	\bar P_{\overline \blacksquare}(z)
	=
	 S_p\bigl( \hat g(\overline\blacksquare) -z\bigr)\!\!  \prod^{\hat m(1+p)+\hat n(1-p)-1}_{k=0}{\varphi}^{-1}_2(-z+\hat g(\overline \blacksquare)+k h_2) \,.
\ee

\subsubsection{$P_{\overline\blacksquare}$ and $\bar P_\blacksquare$}

So far we have only considered the part of the $x$-action where a $\blacksquare$ is added; the analysis is analogous for the removal part of the $x$-action. 

Let us apply the $P\cdot x$ OPE (\ref{PxOPE}) to an initial state corresponding to the twin plane partition
\begin{equation}
\Lambda_\textrm{initial}=\Lambda_0+\overline{\blacksquare}\ .
\end{equation}
We consider the final state in which $x$ kills the $\overline{\blacksquare}$, leaving $(\hat{m}(1+p) +\hat{n}(1-p)+\ell+2+2\rho)$ $\hat{{\square}}$'s in its wake:
\begin{equation}
\Lambda_\textrm{final}=\Lambda_0+\textrm{leftover}[\blacksquare]\ .
\end{equation}
This corresponds to the annihilation action of $x$, with pole
\begin{equation}
w=p^{(0)}_-(\overline{\blacksquare})=\hat{g}(\overline{\blacksquare}) \,.
\end{equation}
Using the same argument as in the previous subsection, we are led to
\begin{equation}
P_{\textrm{leftover}[\overline{\blacksquare}]}(z)= S_p(z-p^{(0)}_{-}(\overline{\blacksquare}) ) \cdot 
P_{\overline{\blacksquare}}(z)
\end{equation}
Finally, since we know the contribution to $P$ by the leftover $\hat\square$'s we can evaluate explicitly
\be
\begin{split}
	P_{\overline\blacksquare} (z)
&= S_p\bigl(g(\overline\blacksquare) - z\bigr) \times
 	\prod_{j=0}^{\hat m(1+p) + \hat n(1-p) +1+2\rho} \hat \varphi_2^{-1}(z - g(\overline\blacksquare)  + j h_2 ) \ ,
\end{split}
\ee
This expression is therefore $\ell$-independent, as it should be. 

A similar exercise, studying the removal-action of $\bar x$, leads to the following expression for $\bar{P}_{{\blacksquare}}$.
\be
	\bar{P}_{{\blacksquare}}(z)	
	=
	S_p\bigl(z - \hat g(\blacksquare) \bigr) \times
 	\prod_{j=0}^{ m(1-p) +  n(1+p) +1+2\rho}  \varphi_2^{-1}(z - \hat g(\blacksquare)  + j h_2  ) \ ,
\ee
As expected, this is related in a simple way to ${P}_{{\overline \blacksquare}}$, namely by replacing  hatted quantities with unhatted ones, and vice versa, and by transposing $(m,n)\to(n,m)$.

\subsection{Remaining OPEs among gluing generators}
\label{sec:remainingOPE}

Having fixed both poles and residues of the action by gluing generators, we now know exactly how they act on twin plane partitions. Using this, we can finally fix all remaining OPEs among these generators.

\subsubsection{The $x$ $\bar{x}$ OPE}\label{sec:xxbarOPEchecl}

We begin from the OPE relation 
\begin{equation}
x(z) \, \bar{x}(w) \sim  D(\Delta) \, \bar{x}(w) \, x(z) \ .
\end{equation}

\paragraph{Formal expression for $D(\Delta)$:}
Acting with the operators on each side on a generic twin plane partition $\Lambda$ gives among others the states 
\be
	x(z) \, \bar{x}(w) \, |\Lambda\rangle  \sim \frac{A}{ \bigl(z - {p}_+({\blacksquare})\bigr) \, \bigl(w - \bar{p}_+(\overline{\blacksquare})\bigr) } 
\, |[\Lambda + \overline{\blacksquare} + \blacksquare] \rangle  + \cdots \ , 
\ee 
and 
\be
	\bar{x}(w) \, x(z) \, |\Lambda\rangle  \sim \frac{B}{\bigl(w - \bar{p}_+(\overline{\blacksquare})\bigr) \, \bigl(z - {p}_+({\blacksquare})\bigr) } 
\, |[\Lambda + \blacksquare + \overline{\blacksquare}] \rangle + \cdots \ . 
\ee
Numerators can be computed explicitly from explicit knowledge of the residue functions. For example the numerator on the l.h.s. can be computed as follows:
\be
\begin{split}
	x(z) \bar x(w) |\Lambda\rangle 
	& = x(z) \frac{\({\rm Res}_{u=\bar p_+(\overline \blacksquare)} \bar {\bf P}_{\Lambda}(u) \)^{1/2}}{w - \bar p_+(\overline \blacksquare)}   |[\Lambda +\overline\blacksquare] \rangle  + \cdots\\
	& =   \frac{\({\rm Res}_{u= p_+( \blacksquare)}  {\bf P}_{\Lambda+\overline\blacksquare}(u) \)^{1/2}}{z -  p_+( \blacksquare)}   
		\frac{\({\rm Res}_{u=\bar p_+(\overline \blacksquare)} \bar {\bf P}_{\Lambda}(u) \)^{1/2}}{w - \bar p_+(\overline \blacksquare)}   |[\Lambda +\overline\blacksquare + \blacksquare]\rangle +\cdots \\
\end{split}
\ee
Recall that, although obscured by notation, ${\bf P}_{\Lambda+\overline\blacksquare}$ carries the contribution of $\hat \ell$ boxes on top of the wall created by $\bar x$. These arise by shifting $\hat\ell$ hatted boxes that extend the minimal bud where $\bar x$ acts (on the hatted side).
Another hidden contribution within creation of $\overline\blacksquare$, is the removal of a bud of $\hat m(1+p)+\hat n(1-p) + \hat \ell$ hatted boxes.
Similar considerations apply to the $\bar x \cdot x$ OPE, in this case we denote the number of shifting boxes with the integer $\ell$ (it is uncorrelated to $\hat\ell$).
Thus, we can write \be
	A^2 = C \, \Bigl({\rm Res}_{u= p_+( \blacksquare)}  \textbf{P}_{\Lambda+\overline\blacksquare }(u) \Bigr)   	        \Bigl({\rm Res}_{u=\overline  p_+(\overline \blacksquare)} \bar{\textbf{P}}_{\Lambda}(u) \Bigr)  \ , 
\ee
and similarly
\be
	B^2 = C \, \Bigl({\rm Res}_{u= \overline p_+( \overline \blacksquare)}  \bar{\textbf{P}}_{\Lambda+\blacksquare }(u) \Bigr)   
		\Bigl({\rm Res}_{u= p_+( \blacksquare)} \textbf{P}_{\Lambda}(u) \Bigr) \ , 
\ee
Equating the two sides we arrive the following formal expression for the $x\bar x$ OPE coefficient
\be
\begin{split}
	x(z) \, \overline x(w) 
	& = \frac{A}{B}  \, \overline x(w)\,  x(z) \\
	& = \(
	\frac{
		{\rm Res}_{u= p_+( \blacksquare)}  \textbf{P}_{\Lambda+\overline\blacksquare }(u) 
	}{
		{\rm Res}_{u= p_+( \blacksquare)} \textbf{P}_{\Lambda}(u)  
	}  
	\cdot
	\frac{
		{\rm Res}_{u=\overline  p_+(\overline \blacksquare)} \bar{\textbf{P}}_{\Lambda}(u) 
	}{
		{\rm Res}_{u= \overline p_+( \overline \blacksquare)}  \bar{\textbf{P}}_{\Lambda+\blacksquare }(u) 
	}\)^{1/2}
	\overline x(w) \, x(z) \ .
\end{split}
\ee

\paragraph{Evaluation of residue factors:}

Let's begin with the addition of $\overline \blacksquare$: the change in $\textbf{P}_\Lambda$ is due to the addition of $\overline\blacksquare$ and to the removal of the bud on the hatted side (where $\overline\blacksquare$ looks like an infinite row), as well as the shifting of the additional boxes to the other side at 
$h({\square}) = g(\overline\blacksquare)+ j h_2$ with $j = 0\dots \hat \ell-1$. Therefore
\be\label{eq:5.8}
\begin{split}
&	\frac{
		\textbf{P}_{\Lambda+\overline\blacksquare}(u) 
	}{
		\textbf{P}_{\Lambda}(u)  
	}  
	 = 
	\frac{P_{\overline\blacksquare}(u)\prod_{{\square} \, {\rm shifted}} P_{{\square}}(u) 
 }{\prod_{\hat{\square} \, {\rm bud}}  P_{\hat{\square}}(u)} \\[4pt]
	& = 
	\frac{
		S_p(g(\overline\blacksquare) - u) \, \prod_{k=0}^{\hat m(1+p) + \hat n(1-p) + 2\rho+1}
		\hat\varphi_2^{-1}(u - g(\overline\blacksquare) + k h_2)
	}{
		\prod_{k=0}^{\hat m(1+p) + \hat n(1-p)+\hat\ell-1} \hat\varphi_2^{-1}(u - \hat g(\overline\blacksquare) - \hat\sigma_3\hat\psi_0 - k h_2)
	} 
	 \cdot 
	\prod_{k=0}^{\hat \ell-1} \varphi_2^{-1}(-u +  g(\overline\blacksquare) + k h_2  )
	 \\[4pt]
	& = 
	S_p(g(\overline\blacksquare) - u) \, 
	\prod_{k=0}^{ 2\rho+1}    \hat\varphi_2^{-1}(u - g(\overline\blacksquare) + k h_2)  	\cdot \prod_{k=0}^{\hat\ell-1} 
	\frac{\varphi_2^{-1}(-u +  g(\overline\blacksquare) + k h_2  )}{\hat\varphi_2^{-1}(-u +  g(\overline\blacksquare) + k h_2 - (2+2\rho) h_2 )}
	\ ,
\end{split}
\ee
where we have used 
\be
	\hat g(\overline\blacksquare) = g(\overline\blacksquare) -  h_2(\hat m(1+p) + \hat n(1-p) ) + \sigma_3\psi_0 - h_2 \ .
\ee
This should then be evaluated at $u= p_+( \blacksquare)$. We can simplify this expression; indeed, since these are the only factors that involve $\hat{\ell}$, this parameter must disappear already from this expression.
After a bit of algebra we arrive at the following nice, and most importantly $\hat{\ell}$ independent, form
\be\label{eq:5.17}
\begin{split}
\left. \frac{\textbf{P}_{\Lambda+\overline\blacksquare}(u)}{\textbf{P}_{\Lambda}(u)}  \right|_{u=p_+( \blacksquare)} 
& = 
S_p \bigl(\bar{p}_+(\overline\blacksquare)- p_+( \blacksquare) + h_2 - \sigma_3 \psi_0 \bigr) \\[2pt]
& \quad 	\times \prod_{k=0}^{2\rho+1} 
\hat\varphi_2^{-1}\bigl(p_+( \blacksquare) -  \bar{p}_+(\overline\blacksquare)  + \sigma_3 \psi_0  - h_2 +k  h_2 \bigr) 
	\ .
\end{split}
\ee

The analysis for the other factor works similarly, the final result is 
\be\label{eq:5.28}
\begin{split}
\left. \frac{\bar{\textbf{P}}_{\Lambda+\blacksquare}(u)}{\bar{\textbf{P}}_{\Lambda}(u)}  \right|_{u=\bar p_+(\overline \blacksquare)} 
& = 
S^{-1}_p \bigl({p}_+(\blacksquare)- \bar p_+(\overline \blacksquare) + h_2 -\hat \sigma_3 \hat \psi_0 \bigr) \\[2pt]
& \quad 	\times \prod_{k=0}^{2\rho+1} 
\varphi_2^{-1}\bigl(\bar p_+(\overline  \blacksquare) -  {p}_+(\blacksquare)  + \hat\sigma_3 \hat \psi_0  - h_2 +k  h_2 \bigr) 
	\ .
\end{split}
\ee
Again, as expected, all dependence on ${\ell}$ eventually dropped out. 
\medskip

\paragraph{Explicit form of the OPE coefficient:}
It thus follows that the OPE coefficient is 
\be
\label{5.34}
\begin{split}
	\(\frac{A}{B}\)^2
	& = 
	\frac{ 
		S_p \bigl(\bar{p}_+(\overline\blacksquare)- p_+( \blacksquare) + h_2 - \sigma_3 \psi_0 \bigr) 
		}{
		S^{-1}_p \bigl({p}_+(\blacksquare)- \bar p_+(\overline \blacksquare) + h_2 -\hat \sigma_3 \hat \psi_0 \bigr) 
		}
	 \prod_{k=0}^{ 2\rho+1} 
	 \frac{ 
		\hat\varphi_2^{-1}\bigl(p_+( \blacksquare) -  \bar{p}_+(\overline\blacksquare)  + \sigma_3 \psi_0  - h_2 +k  h_2 \bigr) 
	  }{
		\varphi_2^{-1}\bigl(\bar p_+(\overline  \blacksquare) -  {p}_+(\blacksquare)  + \hat\sigma_3 \hat \psi_0  - h_2 +k  h_2 \bigr) 
	  } \\
	  & = 1\ .
\end{split}
\ee
where in the last step we used   the recursion property (\ref{Srecursion}) of $S_p(u)$ and the expression (\ref{eq:rho-def}) for the the shifting modulus  $\rho$.

The fact that the ratio turns out to be trivial is rather striking, given how involved the computation has been, but in fact it mimics the result for $p=0, \rho=1/2$ \cite{Gaberdiel:2018nbs}.
Nevertheless, the full OPE coefficient $D(\Delta)$ must be non-trivial since there are various poles that need to be cancelled.
Following in fact the steps as in  \cite[Section~5.3]{Gaberdiel:2018nbs}, the same argument goes through essentially unchanged. This means that   $D(\Delta)$ must contain the factors 
\be
D(\Delta) \sim \frac{(\Delta + h_2 - \hat{\sigma}_3 \hat{\psi}_0)}{(\Delta - h_2 + \sigma_3 \psi_0)} \ . 
\ee
On the other hand,  (\ref{5.34}) implies that generically $D(\Delta)\sim 1$. It follows that the pole structure must be completed as follows:
\be
D(\Delta) = \epsilon_p \, \frac{(\Delta + h_2 - \hat{\sigma}_3 \hat{\psi}_0)}{(\Delta - h_2 + \sigma_3 \psi_0)}\, 
\frac{(\Delta - h_2  + {\sigma}_3 {\psi}_0)}{(\Delta + h_2 - \hat\sigma_3 \hat\psi_0)} \ .
\ee

\subsubsection{Creation-annihilation OPE for gluing generators: $[x,y]$ and $[\bar{x},\bar{y}]$}

The OPE of the gluing creation generators $x,\bar x$ with the respective annihilation generators $y,\bar y$ was postulated in (\ref{Pxy}). 
One can check that this is indeed realized by the action of these operators on a generic twin plane partition $\Lambda$. Indeed it follows easily from the definitions (\ref{xansatz}) and (\ref{yansatz}) that 
\be
	\left(x(z)\, y(w) - \epsilon_p \,y(w) \, x(z)\right) |\Lambda\rangle \ \sim \ \frac{\textbf{P}_\Lambda(z) - \textbf{P}_\Lambda(w)}{z-w} |\Lambda\rangle\ ,
\ee
where the sign $\epsilon_p$ has been introduced by hand, according to the discussion around equation (\ref{eq:epsilon-p}), and $\textbf{P}_\Lambda(u)$ is the eigenvalue of the operator $P(u)$ on the state corresponding to the twin plane partition $\Lambda$, defined in (\ref{eigendef}) and computed by (\ref{XLambda}), and controls the action of $x$ and $y$ on $\Lambda$ via (\ref{xansatz}) and (\ref{yansatz}). 
A similar identity holds for the barred fields $\bar{x}$, $\bar{y}$, the analysis is essentially identical to the one for $x$, $y$.

\subsubsection{Creation-annihilation OPE for conjugate generators: $x\cdot\bar y$ and $\bar{x}\cdot{y}$}

A much less trivial result is to study the OPE of creation gluing generators $x,\bar x$ with the conjugate annihilation gluing generators $\bar{y},y$ respectively. 
The action on twin plane partitions offers a way to compute these coefficients. 
Using effectively  the same approach as in Section~\ref{sec:xxbarOPEchecl}, we consider the action of two such operators on a generic twin plane partition $\Lambda$: 
\be
\begin{split}
	x(z) \, \bar{y}(w) \, |\Lambda\rangle  
	& \sim \frac{
		\({\rm Res}_{u=p_+(\blacksquare)} \textbf{P}_{\Lambda - \overline\blacksquare}(u)\)^{\frac{1}{2}}
	}{ 
		\bigl(z - {p}_+({\blacksquare})\bigr) 
	} 
	\cdot
	\frac{
		\({\rm Res}_{u=\bar p_+(\overline \blacksquare)} \bar {\textbf{P}}_{\Lambda}(u)\)^{\frac{1}{2}}
	}{ 
		\bigl(w - \bar{p}_+(\overline{\blacksquare})\bigr) 
	} 
	\, |\Lambda - \overline{\blacksquare} + \blacksquare \rangle  + \cdots \ , 
	\\
	\bar{y}(w) \, x(z) \, |\Lambda\rangle  
	&\sim \frac{
		\({\rm Res}_{u=\bar p_+(\overline \blacksquare)} \bar{\textbf{P}}_{\Lambda + \blacksquare}(u)\)^{\frac{1}{2}}
	}{ 
		\bigl(w - \bar{p}_+(\overline{\blacksquare})\bigr) 
	} 
	\cdot
	\frac{
		\({\rm Res}_{u=p_+( \blacksquare)}  \textbf{P}_{\Lambda}(u)\)^{\frac{1}{2}}
	}{ 
		\bigl(z - {p}_+({\blacksquare})\bigr) 
	} 
	\, |\Lambda - \overline{\blacksquare} + \blacksquare \rangle  + \cdots \ . 
\end{split}
\ee
This implies that the $x\cdot \bar y$ OPE should be
\be
\begin{split}
	x(z) \overline y(w) 
	& \sim 
	\(\frac{
		{\rm Res}_{u=p_+(\blacksquare)} \textbf{P}_{\Lambda - \overline\blacksquare}(u)
	}{ 
		{\rm Res}_{u=p_+( \blacksquare)}  \textbf{P}_{\Lambda}(u)
	} \)^{\frac{1}{2}}
	\cdot
	\(\frac{
		{\rm Res}_{u=\bar p_+(\overline \blacksquare)} \bar{\textbf{P}}_{\Lambda}(u)
	}{ 
		{\rm Res}_{u=\bar p_+(\overline \blacksquare)} \bar{\textbf{P}}_{\Lambda + \blacksquare}(u)
	} \)^{\frac{1}{2}}
	\ \overline y(w) x(z)  \ .
\end{split}
\ee
We already computed the second factor in (\ref{eq:5.28}), while the first factor is related to (\ref{eq:5.17}) by
\be
	\frac{
		{\rm Res}_{u=p_+(\blacksquare)} \textbf{P}_{\Lambda - \overline\blacksquare}(u)
	}{ 
		{\rm Res}_{u=p_+( \blacksquare)}  \textbf{P}_{\Lambda}(u)
	}
	=
	\(\left. \frac{\textbf{P}_{\Lambda+\overline\blacksquare}(u)}{\textbf{P}_{\Lambda}(u)}  \right|_{u=p_+( \blacksquare)} \)^{-1} \ .
\ee
Furthermore, noting that $S_p\bigl({p}_+(\blacksquare)- \bar p_+(\overline \blacksquare) + h_2 -\hat \sigma_3 \hat \psi_0 \bigr)  $ can be rewritten as either 
\be
\begin{split}
	& 
	S_p^{-1} \bigl( \bar p_+(\overline \blacksquare) - {p}_+(\blacksquare) + h_2 - \sigma_3  \psi_0 \bigr) 
	\prod_{k=0}^{2\rho+1} \frac{\varphi_2^{-1}( {p}_+(\blacksquare)- \bar p_+(\overline \blacksquare) - h_2 + \sigma_3\psi_0 + k h_2  )}{ \hat\varphi_2^{-1}  ({p}_+(\blacksquare)- \bar p_+(\overline \blacksquare) - h_2 + \sigma_3\psi_0 + k h_2 )}
\\
	=& 
	S_p^{-1} \bigl( \bar p_+(\overline \blacksquare) - {p}_+(\blacksquare) + h_2 - \sigma_3  \psi_0 \bigr) 	\prod_{k=0}^{2\rho+1} \frac{\varphi_2^{-1}(  \bar p_+(\overline \blacksquare)- {p}_+(\blacksquare) - h_2 + \hat \sigma_3 \hat \psi_0 + k h_2  )}{ \hat\varphi_2^{-1}  ({p}_+(\blacksquare)- \bar p_+(\overline \blacksquare) - h_2 + \sigma_3\psi_0 + k h_2 )}\,,
\end{split}	
\ee
leads to the appearance of a perfect square within the square-root in the OPE coefficient. 
The resulting OPE is then
\be\label{eq:x-y-bar-OPE}
\begin{split}
	x(z) \cdot \overline y(w) 
	& \sim 
	\epsilon_p \,
	\Biggl[ S_p \bigl( \Delta  + \sigma_3  \psi_0 - h_2 \bigr) 
	\prod_{k=0}^{ 2\rho+1} 
		\hat\varphi_2\bigl(\Delta   + \sigma_3 \psi_0  - h_2 +k  h_2 \bigr) \Biggr]
	\   \overline y(w) \cdot x(z) \ , 
\end{split}
\ee
where, as always,  $\Delta = z-w$.
\medskip

A similar argument gives the OPE of $\overline x\cdot  y$. From the action on a generic twin plane partition $\Lambda$ we find
\be
\begin{split}
	\overline  x(z)  \cdot y(w) 
	& \sim 
	\(\frac{
		{\rm Res}_{u=\bar p_+(\overline \blacksquare)} \bar{\textbf{P}}_{\Lambda - \blacksquare}(u)
	}{ 
		{\rm Res}_{u=\bar p_+( \overline \blacksquare)} \bar{\textbf{P}}_{\Lambda}(u)
	} \)^{\frac{1}{2}}
	\cdot
	\(\frac{
		{\rm Res}_{u=p_+( \blacksquare)} \textbf{P}_{\Lambda}(u)
	}{ 
		{\rm Res}_{u= p_+( \blacksquare)}  \textbf{P}_{\Lambda +\overline  \blacksquare}(u)
	} \)^{\frac{1}{2}}
	\ y(w) \cdot \overline x(z)  \ .
\end{split}
\ee
Again the first factor is the inverse of (\ref{eq:5.28}) whereas the second factor is the inverse of (\ref{eq:5.17}). Therefore the coefficient is the same as the one for $x\cdot \bar y$ in (\ref{eq:x-y-bar-OPE}).
Specializing to $p=0, \rho=1/2$ these expressions reproduce the OPEs found in \cite{Gaberdiel:2018nbs}.

\subsection{Relation between $(\psi, \hat{\psi})$ and $(P,\bar{P})$ charge functions}\label{sec:P-psi-rel}

Having obtained the $P$-charges of all building blocks of a twin plane partition, we deduce from their explicit expressions (which are collected in Table \ref{tab2} for convenience) that
\begin{equation}
	\frac{\textbf{P}_\Lambda(u)}{P_0(u)} \frac{P_0(u+h_2)}{\textbf{P}_\Lambda(u+h_2)} 
	= 
	\frac{\bm{\Psi}_\Lambda(u)}{\psi_0(u)} 
	\frac{\hat{\bm{\Psi}}_\Lambda(u+ h_2 - \hat\sigma_3\hat\psi_0)}{\hat \psi_0(u+ h_2 - \hat\sigma_3\hat\psi_0)}
\end{equation}
as well as 
\begin{equation}
	\frac{\bar{\textbf{P}}_\Lambda(u)}{\bar P_0(u)} \frac{\bar P_0(u+h_2)}{\bar{\textbf{P}}_\Lambda(u+h_2)} 
	= 
	\frac{\bm{\Psi}_\Lambda(u+ h_2 - \sigma_3\psi_0)}{\psi_0(u + h_2 - \sigma_3\psi_0) } 
	\frac{\hat{\bm{\Psi}}_\Lambda(u)}{\hat \psi_0(u)}
\end{equation}
for any twin plane partition $\Lambda$. 
These relations imply a corresponding identity between actual charge operators $P(u),\bar P(u)$ and $\psi(u),\hat\psi(u)$, namely
\be\label{eq:P-psi-rel}
\begin{split}
	\frac{P(u)}{P_0(u)} \frac{P_0(u+h_2)}{P(u+h_2)} 
	&= 
	\frac{\psi(u)}{\psi_0(u)} 
	\frac{\hat \psi(u+ h_2 - \hat\sigma_3\hat\psi_0)}{\hat \psi_0(u+ h_2 - \hat\sigma_3\hat\psi_0)}
	\,,
	\\
	\frac{\bar P(u)}{\bar P_0(u)} \frac{\bar P_0(u+h_2)}{\bar P(u+h_2)} 
	&= 
	\frac{\psi(u+ h_2 - \sigma_3\psi_0)}{\psi_0(u + h_2 - \sigma_3\psi_0) } 
	\frac{\hat \psi(u)}{\hat \psi_0(u)} 
	\,.
\end{split}
\ee
In the case $p=0,\rho=1/2$, these identities reduce the the ones previously observed by \cite{Gaberdiel:2018nbs}. 

The content of these relations is the statement that $P, \bar P$ are not independent generators of the algebra, rather they can be  expressed in terms of $
\psi,\hat\psi$. 
The relation between the modes of these two sets of charge functions can be worked out recursively, starting from these identities. 

Recall the mode expansions of $\psi,\hat \psi$ given in (\ref{generating}) and by the counterpart replacing $\psi_j$ with $\hat\psi_j$. For $P,\bar P$ we take the following mode expansion:
\be
	P(z) =  1+\sum_{j\geq 0} \frac{P_j}{z^{j+1}}  
	\qquad \textrm{and}\qquad
	\bar P(z) = 1+ \sum_{j\geq 0} \frac{\bar P_j}{z^{j+1}} \,.
\ee
The vacuum contributions are:
\be
\begin{split}
	\psi_0(z) & = 1+\sigma_3 \frac{\psi_0}{z} \\
	\hat\psi_0(z) &= 1+ \hat\sigma_3  \frac{\hat\psi_0}{z} \\
\end{split}
\qquad \qquad
\begin{split}
	P_0(z) &=  \( 1+\sigma_3 \frac{\psi_0}{z}  \)  \(   1- \hat\sigma_3  \frac{\hat\psi_0}{z}  \) \\
	\bar P_0(z) &=   \( 1-\sigma_3 \frac{\psi_0}{z}  \)  \(   1+ \hat\sigma_3  \frac{\hat\psi_0}{z}  \) 
\end{split}	
\ee
Here we fixed $P_0(z)$ and $\bar P_0(z)$ by following the result for $p=0, \rho=1/2$ obtained in  \cite{Gaberdiel:2018nbs} based on a free field limit. Our expressions are obtained by natural promotion to hatted quantities (such as $\sigma_3\hat \psi_0\to \hat\sigma_3\hat \psi_0$) where appropriate.
It is then easy to recast (\ref{eq:P-psi-rel}) into relations for the modes, the first few are
\be\label{eq:P-psi-modes}
\begin{split}
	P_0 &= 
	\sigma _3 \psi _0-\hat{\sigma }_3 \hat{\psi }_0
	+
	\frac{1}{h_2} \( {\sigma _3 \psi _1}+{\hat{\sigma }_3 \hat{\psi }_1}\)
	\\
	P_1
	& =
	\frac{1}{2} \( {\sigma _3} \psi _1 - \hat{\sigma }_3 \hat{\psi }_1\)
	+ \frac{1}{2 h_2} \(\sigma _3 \psi _2 	+ \hat{\sigma }_3 \hat{\psi }_2 \)
	-\sigma _3 \hat{\sigma }_3 \psi _0 \hat{\psi }_0
	\\
	&
	+\frac{1}{2}\(\frac{\sigma _3 \psi _1 + \hat \sigma _3 \hat \psi _1}{h_2} \)^2 
	+
	\frac{\sigma _3^2 }{2 h_2} \psi _0 \psi _1
	+\frac{\sigma _3 \hat{\sigma }_3 }{h_2} \psi _0 \hat{\psi }_1
	-\frac{\hat{\sigma }_3 \sigma _3  }{h_2}\psi _1 \hat{\psi }_0
	-\frac{\hat{\sigma }_3^2 }{2 h_2} \hat{\psi }_0 \hat{\psi }_1
\end{split}
\ee
as well as
\be\label{eq:P-psi-modes-hat}
\begin{split}
	\bar P_0
	&=
	\hat{\sigma }_3 \hat{\psi }_0 - \sigma _3 \psi _0
	+
	\frac{1}{h_2} \( {\sigma _3 \psi _1}+{\hat{\sigma }_3 \hat{\psi }_1}\)
	\\
	\bar P_1
	& =
	\frac{1}{2} \( {\hat \sigma _3} \hat \psi _1 - {\sigma }_3 {\psi }_1\)
	+ \frac{1}{2 h_2} \(\hat \sigma _3 \hat \psi _2 	+ {\sigma }_3 {\psi }_2 \)
	-\sigma _3 \hat{\sigma }_3 \psi _0 \hat{\psi }_0
	\\
	&
	+\frac{1}{2}\(\frac{\sigma _3 \psi _1 + \hat \sigma _3 \hat \psi _1}{h_2} \)^2 
	+
	\frac{\hat \sigma _3^2 }{2 h_2} \hat \psi _0 \hat \psi _1
	+\frac{\sigma _3 \hat{\sigma }_3 }{h_2} \hat \psi _0 {\psi }_1
	-\frac{\hat{\sigma }_3 \sigma _3  }{h_2}\hat \psi _1 {\psi }_0
	-\frac{{\sigma }_3^2 }{2 h_2} {\psi }_0 {\psi }_1
\end{split}
\ee
These identities are valid for all three cases $p=0,\pm1$.

\section{Conclusion and discussion}\label{sec:sum}
\label{sec:summary}

We conclude by first summarizing the defining relations of our four-parameter family of algebras in subsection \ref{sec:OPErelations}, then discussing open questions and future directions.

\subsection{Summary of OPE relations}\label{sec:OPErelations}

We first summarize the main results of this paper by collecting explicit expressions for all OPEs of generators of our two-parameter family of algebras. 
It is important to stress that we obtained these relations by constructing a non-trivial (faithful) representation on twin-plane-partitions. In particular, this establishes self-consistency of  the following set of relations.

\subsubsection{Charges of gluing generators}

The OPE relations of the $x$ and $\bar{x}$ generators with $\psi$ and $\hat{\psi}$ are\footnote{
As throughout the paper,  ``$\sim$" means true up to terms that are regular at either $z=0$ or $w=0$. }
\begin{equation}
\begin{split}
\psi(z) \, x(w)  &\sim  \varphi_2(\Delta) \, x(w) \,\psi(z)   \\
\hat{\psi}(z) \, \bar{x}(w)  &\sim  \hat\varphi_2(\Delta) \, \bar{x}(w)\, \hat{\psi}(z)  
\end{split}
\qquad
\begin{split}
\hat{\psi}(z) \, x(w)  &\sim  \hat\varphi^{-1}_2(-\Delta-\hat \sigma_3\hat{\psi}_0) \, x(w) \,\hat{\psi}(z) \\
\psi(z) \, \bar{x}(w)  &\sim  \varphi^{-1}_2(-\Delta-\sigma_3\psi_0) \, \bar{x}(w)\, \psi(z)   
\end{split}
\end{equation}
Although these expressions are  formally very similar to those derived in \cite{Gaberdiel:2017hcn, Gaberdiel:2018nbs}, they differ through the relation between $\hat h_i, \hat\psi_0$ and $h_i,\psi_0$, which depends on $p$ and on $\rho$. 
Similarly the charges for the $y$ and $\bar{y}$ generators  are
\begin{equation}\label{psiFBy0} 
\begin{aligned}
\psi(z) \, y(w)  &\sim  \varphi_2^{-1}(\Delta) \, y(w) \,\psi(z)   \qquad
\hat{\psi}(z) \, y(w)  \sim  \hat\varphi_2(-\Delta-\hat\sigma_3\hat{\psi}_0) \, y(w) \,\hat{\psi}(z)\\
\hat{\psi}(z) \, \bar{y}(w)  &\sim \hat \varphi_2^{-1}(\Delta) \,\, \bar{y}(w) \,\hat{\psi}(z) \qquad
\psi(z) \, \bar{y}(w)  \sim  \varphi_2(-\Delta -\sigma_3 {\psi}_0 ) \,\, \bar{y}(w) \,\psi(z)  
\end{aligned}
\end{equation}
These relations are summarized in Fig.~\ref{OPEeverybody2}.

\subsubsection{OPEs of gluing generators with single-box operators}

OPEs of $x$ with $e,\hat{e}$ and $f,\hat{f}$ are
\be
\begin{aligned}
e(z) \, x(w) & \sim \varphi_2(\Delta)\, x(w) \, e(z) \ ,  
 \  &
f(z) \, x(w) & \sim \frac{\Delta}{\Delta} \, x(w) \, f(z) \ , \\
\hat{e}(z) \, x(w) & \sim \frac{(\Delta+\hat \sigma_3\hat{\psi}_0-  h_2)}{(\Delta+\hat \sigma_3\hat{\psi}_0- h_2)} \, x(w) \, \hat{e}(z) \ , 
 \  &
\hat{f}(z) \, x(w) & \sim \hat \varphi_2^{-1}(-\Delta - \hat \sigma_3 \hat{\psi}_0)\, x(w) \, \hat{f}(z) \ .
\end{aligned}
\ee
We stress again that the OPE coefficient  $\Delta / \Delta$ should not be misunderstood as trivial, since the translation to algebraic relations in terms of modes depends crucially on the exact factors in the numerator and denominator.

For the corresponding annihilation operator $y$ we found
\begin{equation}
\begin{aligned}
e(z)\, y(w) & \sim \frac{\Delta}{\Delta} \, y(w)\, e(z)  \ ,
&\quad
f(z)\, y(w) & \sim \varphi_2^{-1}(\Delta) \, y(w)\, f(z)  \ , 
\\
\hat{e}(z)\, y(w) & \sim \hat \varphi_2(-\Delta-\hat \sigma_3\hat{\psi}_0) \, y(w)\, \hat{e}(z)  \ ,
&\quad
\hat{f}(z)\, y(w) & \sim \frac{(\Delta+\hat \sigma_3 \hat{\psi}_0 -  h_2) }{(\Delta +\hat \sigma_3 \hat{\psi}_0 -  h_2) }   \, y(w)\, \hat{f}(z)  \ .
\end{aligned}
\end{equation}

The OPE relations for the conjugate gluing creation operators $\bar{x}$  are related to the above ones by symmetry
\be
\begin{aligned}
e(z) \, \bar{x}(w) & \sim  \frac{(\Delta+\sigma_3{\psi}_0-h_2)}{(\Delta+\sigma_3 {\psi}_0-h_2)}  \, \bar{x}(w) \, e(z) \ ,  
& \quad 
f(z) \, \bar{x}(w) & \sim \varphi_2^{-1}(-\Delta - \sigma_3{\psi}_0 ) \, \bar{x}(w) \, f(z) \ , \\
\hat{e}(z) \, \bar{x}(w) & \sim\hat \varphi_2(\Delta) \, \bar{x}(w) \, \hat{e}(z) \ , 
& \quad
\hat{f}(z) \, \bar{x}(w) & \sim \frac{\Delta}{\Delta}   \, \bar{x}(w) \, \hat{f}(z) \ ,
\end{aligned}
\ee
and likewise for the corresponding annihilation operator $\bar{y}$, 
\be
\begin{aligned}
e(z)\, \bar{y}(w)  & \sim  \varphi_2(-\Delta - \sigma_3\psi_0) \,\, \bar{y}(w) \, e(z) \,,
\quad & 
f(z)\, \bar{y}(w)  & \sim  \frac{(\Delta + \sigma_3 \psi_0 - h_2)}{(\Delta + \sigma_3 \psi_0 - h_2)}  \bar{y}(w) \, f(z) \,,
\\
\hat{e}(z) \, \bar{y}(w)  & \sim \frac{\Delta}{\Delta} \,\, \bar{y}(w)\, \hat{e}(z) \,.
\quad&
\hat{f}(z) \, \bar{y}(w)  & \sim \hat\varphi_2^{-1}(\Delta) \,\, \bar{y}(w)\, \hat{f}(z) \ .
\end{aligned}
\ee
These relations are schematically summarized by the thick red arrows in Figures.~\ref{OPEbosonicxy-conj} and~\ref{figOPEbosonicfull}. 

The OPEs between $(P,\bar{P})$ and the single box generators $\{e,f,\hat{e},\hat{f}\}$ are
\begin{equation}
\begin{aligned}
&P(z) \,e(w) \sim \varphi^{-1}_2(-\Delta) \, e(w)\,P(z) \qquad \,\,\, \qquad P(z) \,f(w) \sim \varphi_2(-\Delta) \, f(w)\,P(z)\\
&P(z) \,\hat{e}(w) \sim \hat{\varphi}^{-1}_2(\Delta-\hat\sigma_3\hat\psi_0) \, \hat{e}(w)\,P(z) \qquad P(z) \,\hat{f}(w) \sim \hat\varphi_2(\Delta-\hat\sigma_3\hat\psi_0) \, \hat{f}(w)\,P(z)\\
&\bar{P}(z) \,e(w) \sim \varphi^{-1}_2(\Delta-\sigma_3\psi_0) \, e(w)\,\bar{P}(z) \qquad  \bar{P}(z) \,f(w) \sim \varphi_2(\Delta-\sigma_3\psi_0) \, f(w)\,\bar{P}(z)\\
&\bar{P}(z) \,\hat{e}(w) \sim \hat{\varphi}^{-1}_2(-\Delta) \, \hat{e}(w)\,\bar{P}(z) \qquad\,\,\, \qquad \bar{P}(z) \,\hat{f}(w) \sim \hat{\varphi}_2(-\Delta) \, \hat{f}(w)\,\bar{P}(z)\\
\end{aligned}
\end{equation}
these are collected in Figure~\ref{OPEfermionicPe}.

\subsubsection{Mutual OPEs of gluing generators}\label{sec:mutbos}

Among the gluing operators $x,y$ and the charge operator $P$ we found the following relations
\begin{equation}
\begin{aligned}
\begin{aligned}
P(z)\, x(w)&\sim S_p(\Delta) \ x(w)\ P(z)\\
P(z)\, y(w)&\sim S^{-1}_p(\Delta)\ y(w)\ P(z)
\end{aligned}
&\qquad
\begin{aligned}
x(z)\ x(w) &\sim {\epsilon_p}\, S_p(\Delta) \ x(w)\ x(z)\\
 y(z)\ y(w) &\sim {\epsilon_p}\,  S^{-1}_p(\Delta)\ y(w)\ y(z)\\
\end{aligned}\\
[x(z), y(w)]_{\epsilon_p} & \sim \frac{P(z)-P(w)}{z-w}
\end{aligned}
\end{equation}
Where 
\be
	{S_p(u) = \frac{u+\delta_p}{u-\delta_p} \ ,}
	\qquad
	\delta_p = \left\{\begin{array}{ll}
	h_1 \qquad & \ (p=1) \\
	 0 \qquad & \ (p=0) \\
	h_3 \qquad & \ (p=-1) \\
	\end{array}\right. \,.
\ee
Analogous relations hold for the triplet $(\bar x,\bar y, \bar P)$, with $\bar{S}_p(u)=S^{-1}_p(u)$ (\ref{Spbar}).
In addition we obtained OPE relations between barred and un-barred operators:
\begin{equation}
\begin{aligned}
	x(z) \, \bar{x}(w) &\sim {\epsilon_p}\, \frac{(\Delta + h_2 - \hat{\sigma}_3 \hat{\psi}_0)}{(\Delta - h_2 + \sigma_3 \psi_0)}\, 
\frac{(\Delta - h_2  + {\sigma}_3 {\psi}_0)}{(\Delta + h_2 - \hat\sigma_3 \hat\psi_0)}  \, \bar{x}(w) \, x(z) \ ,\\
	y(z) \, \bar{y}(w)  &\sim {\epsilon_p}\, \frac{(\Delta + h_2 - \hat{\sigma}_3 \hat{\psi}_0)}{(\Delta - h_2 + \sigma_3 \psi_0)}\, 
\frac{(\Delta - h_2  + {\sigma}_3 {\psi}_0)}{(\Delta + h_2 - \hat\sigma_3 \hat\psi_0)} \, \, \bar{y}(w) \, y(z)  \ ,
\end{aligned}
\end{equation}
as well as
\be
\begin{split}
	x(z) \cdot \overline y(w) 
	& \sim 
	{\epsilon_p}\, \Biggl[ S_p \bigl( \Delta  + \sigma_3  \psi_0 - h_2 \bigr) 
	\prod_{k=0}^{ 2\rho+1} 
		\hat\varphi_2\bigl(\Delta   + \sigma_3 \psi_0  - h_2 +k  h_2 \bigr) \Biggr]
	\   \overline y(w) \cdot x(z) \ , 
	\\
	\overline  x(z)  \cdot y(w) 
	& \sim 
	{\epsilon_p}\, \Biggl[ S_p \bigl( \Delta  + \sigma_3  \psi_0 - h_2 \bigr) 
	\prod_{k=0}^{ 2\rho+1} 
		\hat\varphi_2\bigl(\Delta   + \sigma_3 \psi_0  - h_2 +k  h_2 \bigr) \Biggr]
	\ y(w) \cdot \overline x(z)  \ .
\end{split}
\ee
See thick red lines in Figs.~\ref{OPEfermionicxy}.

\subsubsection{Charge operators $P, \bar P$ and $\psi, \hat \psi$}

We also ascertained that $P,\bar P$ are not independent operators, but can be expressed entirely in terms of charge operators $\psi, \hat\psi$, shown in eqs.~(\ref{eq:P-psi-rel}). 
It is straightforward to mode-expand the fields in (\ref{eq:P-psi-rel})  and derive the algebraic relations for their modes; the first few are given in (\ref{eq:P-psi-modes}) and (\ref{eq:P-psi-modes-hat}). 

\subsection{Geometric interpretation of framing and shifting via BPS partition function}
\label{sec:BPS-interpretation}

In deriving the OPE relations of the  2-parameter family of VOAs, we found that certain properties, such as the length of ``minimal buds'' and the related restrictions on the choice of framing $p$ have a suggestive interpretation in terms of the geometry of toric Calabi-Yau threefolds. 
In section \ref{sec:geometry} it was observed that $p$ is related to a choice of Calabi-Yau of the form $\CO(-s_3) \oplus \CO(-s_1) \to \IP^1$, with $(s_1, s_3)$ (subject to $s_1+s_3=2$) related to $p$ by (\ref{eq:p-s-i-relation}).

In this vein, we can also deduce a geometric interpretation for $\rho$ and $y$ (the fugacity associated with gluing operators that appeared in vacuum characters (\ref{chi0}) and (\ref{eq:vac-char-bose})). 
The vacuum character for $p=0$ given in (\ref{chi0}) can be rewritten in the following form
\be\label{eq:fermionc-char-DT}
\begin{split}
	&
	\left[
		M(q)^{-2} \ 
		\prod_{n\geq 1} (1-Q\, q^{n})^{n}  
		\prod_{n\geq m} (1- Q^{-1} \, q^{n})^{n} 
	\right]
	\ \times \ 
	\left[
		\prod_{n\geq 1}(1-Q^{-1} q^{n+2\rho})^{-1}
	\right]^{2\rho}
	\\[6pt]
	= &
	Z_{BPS}(Q,q) \ \times \ \left[Z_{\textrm{free boson}}\right]^{2\rho} \,,
\end{split}
\ee
where $M(q)$ is the MacMahon function, and we adopted the change of variables
\be
	Q = -y \, q^{\rho}\,, \qquad m=2\rho + 1\,.
\ee
The first factor in (\ref{eq:fermionc-char-DT}) is strongly reminiscent of the partition function of (framed) Donaldson-Thomas invariants of the conifold \cite{Szendroi:2007nu, Jafferis:2008uf, Aganagic:2009kf}. In that context, $Q=e^{-t}$ is the complexified K\"ahler parameter, whereas $m$ is related to a choice of chamber in the moduli space of stability conditions for framed wall-crossing, \emph{i.e.} to a choice of $B$-field.
It is known that framed wall-crossing happens for integer shifts of $B$ (cf. for example \cite[eq. (4.6)]{Aganagic:2009kf}), this matches nicely with our constraint that $\rho$ be quantized by half-integers.
 
One might even hope that the glued algebra is the BPS algebra  
of type IIA string compactified on the conifold. 
However, we do not know how to interpret the mismatch with the DT partition function, i.e\ the  second factor of (\ref{eq:fermionc-char-DT}), which appears to count $2\rho$ free bosons with charge $Q^{-1}q^{2\rho}$. 
A possible interpretation might be that our algebra includes a $\mathfrak{u}(1)^{\oplus\, 2\rho}$ that can be decoupled.
Recall that in the construction of the $\mathcal{N}=2$ affine Yangian, which corresponds to $p=0$ and $\rho=\frac{1}{2}$, we first need to tensor a $\mathfrak{u}(1)$ to the $\mathcal{W}^{\mathcal{N}=2}_{\infty}$ algebra. 
It could be that the $\mathfrak{u}(1)$ is only needed to make the resulting algebra more (left-right) symmetric, but otherwise it is not essential to the algebra and can be easily decoupled from the final algebra.
It would be interesting to check if for generic $\rho$, there are indeed $2 \rho$ $\mathfrak{u}(1)$s that can be naturally decoupled and if so work out the decoupled algebra.
We hope to resolve these puzzles in future work. 

\subsection{Open problems}

There are certainly a number of questions raised by our work, that may serve as inspiration for future developments.

First, we emphasize again that although the gluing construction via twin plane partitions is very efficient in constructing new algebras of affine Yangian type, it is highly non-trivial to rewrite them in terms of the $\mathcal{W}$ algebra basis. 
Recall that even for the isomorphism between affine Yangian of $\mathfrak{gl}_1$ and UEA of $\mathcal{W}_{1+\infty}[\lambda]$, the detailed translation is highly non-local and can only be obtained order by order in spin.\footnote{
For the $\mathcal{N}=2$ version of the isomorphism, even though the $\mathcal{W}$ algebra is already known and is just $\mathcal{N}=2$ $\mathcal{W}_{\infty}[\lambda]$ algebra, we haven't completely obtained the full translation explicitly. 
}
It would be interesting and useful to translate 
our four-parameter family of VOAs (labeled by $c,\lambda,\rho,p$) to the $\mathcal{W}$ algebra basis.
In particular, one may ask if any of the algebras that we constructed is isomorphic to known $\mathcal{W}$ algebras. 
One might also ask whether  some of the algebra in this family is related to some of the known affine Yangian algebra, e.g.\ with the shifted affine Yangian of $\mathfrak{gl}_2$, which can be obtained as the rational limit of quantum toroidal algebra of $\mathfrak{gl}_2$.\footnote{
For quantum toroidal algebra of $\mathfrak{gl}_n$ see \cite{Feigin1204, Feigin:2013fga} and for rational limit see \cite{Bershtein1512}.}
Another important question is whether we need Serre relations beyond those from the two bosonic subsectors.

We would also like to generalize the gluing construction even further. 
For example, how to glue along larger $(p,q)$-web diagrams? 
In particular, what novelties arise in the algebra (or in plane partitions) when dealing with a diagram with loops? 
Further, is it possible to use gluing operators that transform as $({\square}, {\square})$ and $(\overline{{\square}}, \overline{{\square}})$ w.r.t.\ the two corner affine Yangians?

Finally, we expect that this gluing technique works equally well for quantum toroidal algebras. 
Namely, since quantum toroidal algebra of $\mathfrak{gl}_1$ also has a natural action on plane partitions, one can glue quantum toroidal algebra of $\mathfrak{gl}_1$ along $(p,q)$-web diagrams using the constraints from the action on plane partitions glued appropriately.  
This could provide new examples and help check/prove the recent conjecture of \cite{Feigin:2018bkf}. 
We hope to be able to answer some of these questions in future work. 

\section*{Acknowledgements}

We thank Matthias Gaberdiel and Miroslav Rap\v{c}ak for initial collaboration on this project, and for many discussions. 
We also thank Bryce Bastian, Stefan Hohenegger, Mauricio Romo and Francesco Sala for helpful discussions. 
WL thanks support from the Thousand Talent Program, Max-Planck Partergruppen fund, and NSFC 11875064; she is also grateful for the hospitality of ETH Zurich, MPI-AEI, and ESI workshop ``Higher spin and holography" during various stages of this project. 
The work of Pietro Longhi is supported by a grant from the Swiss National Science foundation. 
He also acknowledges the support of the NCCR SwissMAP that is also funded by the Swiss National Science foundation.
PL gratefully acknowledges hospitality from the Simons Center for Geometry and Physics at Stony Brook University, the Kavli IPMU at the University of Tokyo, and the Institut de Physique Nucl\'eaire de Lyon at Universit\'e Claude Bernard during completion of this work. 

\appendix

\section{Some details on computations with gluing generators}
\label{sec:app1}

The purpose of this appendix is to provide the derivation of some key formulae that appear in the main body of the paper, while also illustrating how to perform computations with twin plane partitions for generic $p$ and $\rho$.

\subsection{Minimal buds of length two for $p=\pm1$}\label{sec:minimal-buds-appendix}

Here we show how to derive the fact that minimal buds must have length 2 (or multiples of 2, by extension of this reasoning) when $p=\pm1$.
For this purpose we consider the creation of a minimal row by acting with $x$ on the vacuum $|\emptyset\rangle$, and then creation of a next-minimal row.

As we have seen in Section \ref{sec:x-x-on-vac}, acting simply with $x\cdot x$ would create two rows stacked along the ``symmetric'' direction, in the sense of (\ref{eq:axes-conventions}). 
That would correspond to stacking the two rows along the direction with a bud of length zero. 
Therefore in order to probe the stacking of rows along the ``anti-symmetric'' direction we need to apply $e$ generators in between the two $x$'s. 

Let us first consider  the case where we generate one additional box in between, 
\begin{equation}
\begin{aligned}
x(z)\cdot e(w)\cdot x(v) |\emptyset\rangle 
&\sim  \frac{1}{v} \, x(z) \left[\frac{1}{w-h_1}|\blacksquare+{\tiny\yng(1)_{\, 1}}\rangle+\frac{1}{w-h_3} |\blacksquare+{\tiny\yng(1)_{\, 3}}\rangle\right]   
\end{aligned}
\end{equation}
With the ansatz 
\begin{equation}
x(z) |\blacksquare+{\tiny\yng(1)_{\, i}}\rangle \sim\sum_{j} \frac{1}{z-z^{*}_{i,j}} \, |\Phi^{xex}_{i,j}\rangle \ , \qquad \hbox{where $i=1,3$.}
\end{equation}
The charge functions of the final state $ |\Phi^{xex}_{i,j}\rangle$ are
\begin{equation}
|\Phi^{xex}_{i,j}\rangle:\qquad 
\begin{cases}
\begin{aligned}
\bm{\Psi}(u) &= \psi_0(u)\cdot \varphi_2(u-z^{*}_{ i,j})\cdot\varphi_3(u-h_i)\cdot \varphi_2(u) \\
\hat{\bm{\Psi}}(u) &=\hat{\psi}_0(u)\cdot \hat \varphi^{-1}_2(-(u-z^{*}_{i,j})-\hat \sigma_3\hat{\psi}_0)\cdot \hat \varphi^{-1}_2(-u-\hat \sigma_3\hat{\psi}_0) \ , 
\end{aligned}
\end{cases}
\end{equation}
for $i=1,3$.
We found that for each $i$, there is \emph{no} pole that makes this charge compatible with $|\blacksquare\blacksquare_3\rangle$ (for $p=1$). On the other hand one finds that 
\be
\begin{split}
	x(z)\cdot e(w)\cdot e(u)\cdot x(v) |\emptyset\rangle 
	&\sim \frac{1}{v}  \frac{1}{u - h_3}  \frac{1}{w - h_3 - h_2} \frac{1}{z-h_3-2h_2}\, x(z)  |\blacksquare\blacksquare_3\rangle  
\end{split}
\ee
For $p=1$ we can therefore only stack rows along the $x_3$ direction provided there is a ``bud'' of length two placed next to the initial row, in the $x_3$ direction. Then the second row will attach to the end of this bud. 

The analysis is similar for $p=-1$; in this case one needs a length-two bud displaced in the $x_1$ direction in order to create row stacked on the first one along the $x_1$ direction.

\subsection{Action of $x$ and $\bar{x}$ operators on one existing box}
\label{app:xxbaronbox}
Consider 
\begin{equation}
 x(w) |\square\rangle \sim \sum_{i} \frac{1}{w-w^{*}_i} |\Phi^{xe}\rangle \ ,
\end{equation}
where the sum runs over all possible poles $w^*_i$ for which  $|\Phi^{xe}_{i}\rangle$ is an allowed twin-plane-partition configuration. 
The $(\bm{\Psi}(u),\hat{\bm{\Psi}}(u))$ charge functions of $|\Phi^{xe}_{i}\rangle$ are 
\begin{equation}\label{chargexeapp}
|\Phi^{xe}_{i}\rangle:\qquad 
\begin{cases}
\begin{aligned}
\bm{\Psi}(u) &= \psi_0(u)\cdot \varphi_2(u-w^{*}_i)\cdot \varphi_3(u) \\
\hat{\bm{\Psi}}(u) &=\hat{\psi}_0(u)\cdot \hat\varphi^{-1}_2(-(u-w^{*}_i)-\hat\sigma_3\hat{\psi}_0) \ . 
\end{aligned}
\end{cases}
\end{equation}
Given that $|\Phi^{xe}_{i}\rangle$ is created by a single $x$ generator, it should be an $e$ or $\hat{e}$ descendant of $|\blacksquare\rangle$. 
The charge functions (\ref{chargexeapp}) allows us to compute 
the conformal dimension of $|\Phi^{xe}_{i}\rangle$, using 
\be\label{Ltotal}
L_0^{\rm total} = L_0 + \hat{L}_0 \ , \qquad L_0 = \frac{1}{2}\psi_2 \ , \qquad \hat{L}_0 = \frac{1}{2} \hat{\psi}_2 \ , 
\ee
(see eq.\ (3.5) of \cite{Gaberdiel:2017dbk}). 
The result is 
\begin{equation}\label{2plusrho} 
L_0^{\rm total} =2+\rho =1+h_{\blacksquare} \ ,
\end{equation}
independent of the position of the pole $w^*_i$; in the second ``=" we have used $(\ref{eq:rho-def})$.

The conformal dimension (\ref{2plusrho}) means that the state $|\Phi^{xe}_{i}\rangle$ must be a single box excitation of $|\blacksquare\rangle$.
And there are four possible candidates:
$|\blacksquare + {\square}_1\rangle$, $|\blacksquare + {\square}_3\rangle$, $|\blacksquare + \hat {\square}_0\rangle$, $|\blacksquare + \hat{\square}_{\rm top}\rangle$, with charge functions
\begin{equation}
\begin{aligned}
&|\blacksquare + {\square}_j\rangle :\qquad  
\begin{cases}
\begin{aligned}
\bm{\Psi}(u) &= \psi_0(u)   \varphi_2(u)   \varphi_3(u-h_j) \\
\hat{\bm{\Psi}}(u)&=\hat\psi_0(u) \hat\varphi_2^{-1}(-u-\hat\sigma_3\hat\psi_0)
\end{aligned}
\end{cases} \qquad \qquad j=1,3\\
&|\blacksquare + \hat {\square}_0\rangle:\qquad 
\begin{cases}
\begin{aligned}
\bm{\Psi}(u)& = \psi_0(u)   \varphi_2(u)    \\
\hat{\bm{\Psi}}(u) &=\hat\psi_0(u) \hat\varphi_2^{-1}(-u-\hat\sigma_3\hat\psi_0) \hat \varphi_3(u)
\end{aligned}
\end{cases}\\
&| \blacksquare + \hat{\square}_{\rm top}\rangle:\quad  
\begin{cases}
\begin{aligned}
\bm{\Psi}(u)&= \psi_0(u)   \varphi_2(u)    \\
\hat{\bm{\Psi}}(u) &=\hat\psi_0(u) \hat\varphi_2^{-1}(-u-\hat\sigma_3\hat\psi_0) \hat \varphi_3(u+\hat\sigma_3 \hat\psi_0 - h_2) \ .
\end{aligned}
\end{cases}
\end{aligned}
\end{equation}

Exploring the different possibilities, one can show that there is only one consistent value for $w^{*}$, namely 
\begin{equation}\label{wpole}
w^{*}=h_2\ . 
\end{equation}
Indeed, using (\ref{magic}) 
and similarly for hatted functions, we find that for the pole (\ref{wpole}) 
\begin{equation}
|\Phi^{xe}\rangle:\qquad 
\begin{cases}
\begin{aligned}
\bm{\Psi}(u) &= \psi_0(u)\cdot \varphi_2(u)\\
\hat{\bm{\Psi}}(u) &=\hat{\psi}_0(u)\cdot \hat \varphi^{-1}_2(-u-\hat \sigma_3\hat{\psi}_0)\cdot\hat \varphi_3(u+\hat \sigma_3\hat{\psi}_0- h_2) \ , 
\end{aligned}
\end{cases}
\end{equation}
therefore 
\begin{equation}\label{xonefinal}
 x(w) |\square\rangle \sim \frac{1}{w-h_2} \, |\blacksquare+\hat{\square}_{\, \textrm{top}}\rangle \ . 
\end{equation}

%This result has a simple geometric interpretation: since the starting configuration is a single box in the left corner, the only natural way in which one can add an infinite row is to start this infinite row at the position $(x_1,x_2,x_3) = (0,1,0)$, thus leading to the pole (\ref{wpole}). 
%Furthermore, it is suggestive that one extra box sticks out at the right corner, and thus leads to an $\hat{e}$ descendant of $|\blacksquare\rangle$. 
Similarly we find 
\begin{equation}
\begin{aligned}
&x(w) |\hat{\square}\rangle \sim \frac{1}{w} \, |\blacksquare+\hat{\square}_{\, \textrm{0}}\rangle\,,  \quad 
 \bar{x}(w) |\square\rangle \sim \frac{1}{w} \, |\overline{\blacksquare}+{\square}_{\, 0}\rangle\,, \quad
\bar{x}(w) |\hat{\square}\rangle \sim \frac{1}{w-h_2} \, |\overline{\blacksquare}+{\square}_{\, \textrm{top}}\rangle  \ .
\end{aligned}
\end{equation}

\subsection{Partially fixing the $\hat{f}\cdot x$ OPE}
\label{sec:fhatx}

Here we give the detailed procedure of fixing $\hat{f}\cdot x$ OPE 
\begin{equation}
	\hat f(z)  \cdot x(w) \sim \hat{\bar{H}}(\Delta) \ x(w) \cdot \hat f(z)  \ , 
\end{equation}
by applying the OPE relations on various twin plane partition configurations.

\subsubsection{Constraints from $|{{\square}}\rangle$}
We calculate 
\be
	\hat f(z) x(w) |{\square}\rangle  
	\sim \hat f (z) \frac{1}{w-h_2} |\blacksquare +\hat {\square}_{\rm top} \rangle
	\sim \frac{1}{w-h_2} \frac{1}{z+\hat \sigma_3\hat\psi_0  -  h_2} |\blacksquare \rangle
\ee
and 
\be
	x(w)\hat f(z) |{\square}\rangle  
	\sim 0 \ . 
\ee
Thus the denominator of $\hat{\bar{H}}(\Delta)$ must contain the factor $(\Delta + \hat\sigma_3\hat\psi_0)$.

\subsubsection{Constraints from $|\blacksquare + \hat {{\square}}_{\rm top}\rangle$}

In the following we shall consider, for definiteness, the case $p=1$. First we note that 
\be
	 x(w) |\blacksquare + \hat {\square}_{\rm top} \rangle  
	\sim 0 \ , 
\ee
since $x(w) |\blacksquare\rangle$ creates $|\blacksquare\blacksquare_1\rangle$, and this would correspond to a high-wall with two missing rows at the top, for which the presence of $\hat {\square}_{\rm top}$ is not be allowed (since the row beneath it has been removed by $x$). On the other hand, 
\be
\begin{split}
	x(w)\hat f(z)  |\blacksquare + \hat {\square}_{\rm top} \rangle  
	 & \sim \frac{1}{w-h_1} \frac{1}{z + \hat \sigma_3 \hat\psi_0 - h_2}|\blacksquare\blacksquare_1 \rangle   \ ,
	 \\
\end{split}
\ee
where in the second step we have used (\ref{eq:xx-0-p1}).
Therefore the numerator of $\hat{\bar{H}}(\Delta)$ must contain the factor $(\Delta + \hat\sigma_3\hat\psi_0 -  h_2 + h_1)$ (for $p=-1$ we just need to replace $h_1$ by $h_3$ here). Using the relation between $h_i$ and $\hat h_i$ for $p=1$ this can be rewritten as $ (\Delta + \hat\sigma_3\hat\psi_0 + \hat h_1)$. 

\subsubsection{Constraints from $|\blacksquare  + {{\square}}{{\square}}_3 + \hat {{\square}}_{\rm top}\rangle$ }

Let us fix again $p=1$, then we observe that
\be
	\hat f(z) x(w) |\blacksquare  + {\square}{\square}_3 + \hat {\square}_{\rm top} \rangle  
	\sim 0
\ee
due to the fact that $x(w) |\blacksquare  + {\square}{\square}_3 \rangle$ would either create $|\blacksquare\blacksquare_1  + {\square}{\square}_3 \rangle$ or $|\blacksquare\blacksquare_3\rangle$ by (\ref{eq:x-bb-ww3-complete}). 
The latter case would correspond to a high-wall with two missing rows at the top (as in the example given in  (\ref{eq:axes-conventions}) upon exchanging $x_1\leftrightarrow x_3$ and $\hat x_1\leftrightarrow \hat x_3$ since there $p=-1$), for which the presence of $\hat {\square}_{\rm top}$ is not allowed, since the whole row beneath it has been removed by $x$. 
The former case would correspond to two parallel high walls, where the second one is created closer to the origin. Then the hatted box would end up sitting on the high wall that is further from the origin, which is also not allowed (see also discussion at the end of section 4.6.1 in \cite{Gaberdiel:2018nbs}). On the other hand,
\be
\begin{split}
	x(w)\hat f(z)   |\blacksquare  + {\square}{\square}_3 + \hat {\square}_{top} \rangle  
	 & \sim \frac{1}{w-h_1} \frac{1}{z + \hat \sigma_3 \hat\psi_0 - h_2}|\blacksquare\blacksquare_1+ {\square}{\square}_3    \rangle  
	 \\
	  & + (\#) \frac{1}{w-h_3-2h_2} \frac{1}{z + \hat \sigma_3 \hat\psi_0 - h_2}|\blacksquare\blacksquare_3  \rangle  \ , 
\end{split}
\ee
where in the second step we have used (\ref{eq:x-bb-ww3-complete}). 
Therefore the numerator of $\hat H(\Delta)$ must contain the factor $(\Delta + \hat\sigma_3\hat\psi_0 +  h_2 + h_3) $. Using the relation between $h_i$ and $\hat h_i$ for $p=1$ this can be rewritten as $ (\Delta + \hat\sigma_3\hat\psi_0 + \hat  h_3)$.

Combining these constraints we therefore find that $\hat{\bar{H}}(\Delta)$ must contain the factors 
\be
	\hat{\bar{H}}(\Delta) \sim \frac{ (\Delta + \hat\sigma_3\hat\psi_0 + \hat  h_1) (\Delta + \hat\sigma_3\hat\psi_0 + \hat  h_3)}{(\Delta + \hat\sigma_3\hat\psi_0)}\ . 
\ee
This is as for $p=0$ (except that now the hatted $\hat{h}_i$ appear).

\subsection{The action of $\overline x$ on generic states}\label{sec:x-bar-gen-states}

So far we concentrated on the action of $x$. For illustration purposes, here we explore the action of $\overline\blacksquare$.

\subsubsection{Bosonic behaviour: symmetric stacking of $\overline\blacksquare$}\label{app:boxbar-symm}

We consider the ``symmetric action'' of these bosonic operators, by acting twice with $\bar{x}$ on the vacuum:
\be
\begin{aligned}
	\bar{x}(z) \cdot \bar{x}(w) \, |\emptyset\rangle 
	& \sim 
	\frac{1}{w} \, \bar{x}(z) \, |\overline\blacksquare\rangle
	& \sim \frac{1}{w} \, \sum_{i} \frac{1}{z-z^{*}_i} \, |\Phi^{\overline x\overline x}_{i}\rangle \ ,
\end{aligned}
\end{equation}
whose charges are 
\begin{equation}\label{psixbxb}
	|\Phi^{\overline x\overline x}_{i}\rangle:\qquad 
	\begin{cases}
	\begin{aligned}
		\bm{\Psi}(u) 
			&= \psi_0(u) \cdot  \varphi^{-1}_2(-(u-z^{*}_i)- \sigma_3 {\psi}_0)\cdot  	\varphi^{-1}_2(-u- \sigma_3 {\psi}_0)  \ ,\\
		\hat{\bm{\Psi}}(u) 
			&=  \hat{\psi}_0(u)\cdot \hat\varphi_2(u-z^{*}_i) \cdot \hat \varphi_2(u) \ . 		
	\end{aligned}
	\end{cases}
\end{equation}
One might expect that $|\Phi^{\overline x\overline x}_{i}\rangle$ should match 
one of the two configurations corresponding to 
\begin{equation}
|\overline\blacksquare\overline\blacksquare_{\hat 1}\rangle:\qquad 
\begin{cases}
\begin{aligned}
\bm{\Psi}(u) &= \psi_0(u) \cdot \varphi^{-1}_2(-u- \sigma_3 {\psi}_0) \cdot  \varphi^{-1}_2(-u-  h_3- \sigma_3 {\psi}_0) \\
\hat{\bm{\Psi}}(u) &=\hat{\psi}_0(u)\cdot \hat \varphi_2(u) \cdot \hat \varphi_2(u-\hat h_1)
\ ,
\end{aligned}
\end{cases}
\end{equation}
and
\begin{equation}
|\overline\blacksquare\overline\blacksquare_{\hat 3}\rangle:\qquad  
\begin{cases}
\begin{aligned}
\bm{\Psi}(u) &= \psi_0(u) \cdot \varphi^{-1}_2(-u- \sigma_3 {\psi}_0) \cdot  \varphi^{-1}_2(-u-  h_1- \sigma_3 {\psi}_0) \\
\hat{\bm{\Psi}}(u) &=\hat{\psi}_0(u)\cdot \hat \varphi_2(u) \cdot \hat \varphi_2(u-\hat h_3)
\ ,
\end{aligned}
\end{cases}
\end{equation}
To match with the first case one should take $z_i^* = \hat h_1 = - h_3$, this is possible provided that $h_i,\hat h_i$ are related as in (\ref{eq:h-hhat-p}) with $p=-1$. The second possibility is realized if one takes $z_i^* = \hat h_3 = - h_1$, which means $p=1$ in (\ref{eq:h-hhat-p}). Thus altogether we find 
\be
	\boxed{p=1} : \qquad
	\bar{x}\cdot \bar{ x} |\emptyset\rangle \sim |\overline \blacksquare\overline \blacksquare_{\hat 3}\rangle
\ee
\be
	\boxed{p=-1} : \qquad
	\bar{ x}\cdot\bar{x} |\emptyset\rangle \sim |\overline \blacksquare\overline \blacksquare_{\hat 1}\rangle
\ee
By symmetry one would then also expect that the conformal dimensions add up correctly (in order to lead to eq.~(\ref{eq:char-decomp}). We have also checked this explicitly. 

\subsubsection{Antisymmetric stacking of $\overline\blacksquare$}\label{app:boxbar-antisymm}

Just like for the analysis of the $x$-action, we have also studied the action of $\bar{x}$ on higher descendant states. For example, for $p=1$ we find 
\be
\begin{split}
	\overline x(z)\cdot \hat e(w)\cdot \hat e(u)\cdot \overline x(v) |\emptyset\rangle 
	&\sim \frac{1}{v}  \frac{1}{u - \hat h_3}  \frac{1}{w - \hat h_3 -  h_2} \frac{1}{z-\hat h_1-2 h_2}\, \overline x(z)  |\overline\blacksquare\overline\blacksquare_{\hat 1}\rangle  \ .
\end{split}
\ee
This means that, in order to stack rows along the $\hat x_1$ direction when $p=1$, one has to first create a ``bud'' of length two placed next to the initial row, in the $\hat x_1$ direction. Then the second row will attach to the end of this bud.

Likewise for $p=-1$ we need to create a length-two bud placed next to the initial row, displaced in the $\hat x_3$ direction. Then acting with another $x$ will create a second row stacked on the first one along the $\hat x_3$ direction. 

\bibliographystyle{utphys}
\bibliography{yangian}

\end{document}